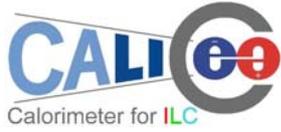

# Report to the Calorimeter R&D Review Panel

# CALICE Collaboration
# June 2007

# 1 Executive Summary

The CALICE collaboration is undertaking a major program of R&D into calorimetry for the ILC. It has members from many institutes and is represented in all three major ILC regions. The collaboration has a coordinated approach and, due to its size, is able to efficiently share its expertise across many different detector studies.

The main direction of the collaboration R&D is to study particle flow (PFA) calorimetry, software compensation and individual particle reconstruction. As such, the studies are concentrating on fine granularity calorimeters with a high degree of longitudinal segmentation. These studies include comparison of simulation models with data to measure their degree of agreement, the technical issues of building a detector optimised for PFA calorimetry, and development of algorithms for software compensation and particle flow reconstruction.

The collaboration is studying several technologies for electromagnetic and hadronic calorimeters with a goal of comparing their performance in terms of ILC physics requirements. By performing this work within a common framework, a meaningful comparison of the various proposed detectors can be made.

Two major electromagnetic calorimeter prototypes have already been built and exposed to beam within the last year. A major dataset of around 100 million events was accumulated, covering a wide range of energies and incident angles with both electrons and pions. The data from these beam tests are being analysed and first results are now available. Calibration and correction procedures are still at too preliminary a stage to quote final resolutions, but the detector performance is in reasonable agreement with simulation expectations.

The first hadronic calorimeter prototype was also run in the beam tests last year. A significant amount of data was acquired, both with this calorimeter alone and combined with the electromagnetic calorimeter and again, preliminary results from the analysis of these data are now ready. The first stage was to understand the response to electromagnetic interactions and compare to simulation. Having verified the level of agreement of the detector response, then measurements of hadronic showers become meaningful and results on these studies are now becoming available.

These tests will continue over at least the next 18 months to incorporate several other electromagnetic and hadronic detector technologies and so allow comparisons between them. These will be done with identical mechanical structures where possible so as to keep systematic differences under control.

The large size of the CALICE collaboration allows a high degree of shared experience, effort and equipment. Many of the various detector technologies use the same on- and off-detector readout electronics. They all share the same DAQ and online software system. The offline software is also common to all studies. To allow rapid and easy access to data throughout the collaboration, data management and processing are widely based on grid tools.

Beyond the beam tests, ILC-like prototypes are planned to address the mechanical, technical and integration issues of a realistic detector. These will again include several electromagnetic and hadronic calorimeter technologies. They will also use a DAQ system which would meet the requirements of such a system for the ILC.

We believe an appropriate evaluation of calorimeter technologies for the ILC can only be made by a direct comparison of the options within a well-controlled environment. Such an approach results in a high degree of commonality, leading to a more efficient use of resources, both in terms of funding and effort. We also believe this broad program of work is best coordinated by the experts involved directly in the studies. The CALICE collaboration has already made significant progress towards fulfilling these goals and aims to complete them over the next few years.

# 2 Overview of the CALICE Collaboration

## 2.1 Introduction

The CALICE collaboration is undertaking a major program of R&D into calorimetry for the ILC, directed towards the design of an ILC calorimeter optimised for both performance and cost. It now has over 200 members from 41 institutes worldwide including all three major ILC regions, and is by far the largest group studying calorimetry for the ILC. New groups continue to join CALICE and these have recently included institutes in India, Canada and Spain.

The main direction of the collaboration R&D is to study particle flow (PFA) calorimetry and software compensation. This is a very broad area and there are many proposals for ways to build a calorimeter optimised for such a concept. Really determining which is the "best" solution would be a difficult task without a systematic and fair comparison of the options. One critical aspect of CALICE is that it is a large enough collaboration that it can study a significant number of these options and can ensure that they are compared in a meaningful way. Specifically, CALICE covers both electromagnetic calorimeters (ECAL) and hadronic calorimeters (HCAL). However, within these broad ranges, it also covers digital (also called binary) and analogue readout for both types of calorimeter. For the ECAL, CALICE groups are studying silicon and scintillator sensitive layers, while for the HCAL, the collaboration is looking at several scintillator and gas detectors. There is a common mechanical converter structure for comparing both the analogue and digital HCALs and similarly there will be a common structure for the comparing the analogue and digital silicon ECALs. Furthermore, the collaboration also includes R&D into tail catcher and muon tagger (TCMT) detectors, which should allow an optimisation of the calorimeter design in terms of the HCAL interaction length compared to cost.

There are two main R&D directions within CALICE; the "physics prototypes" and the "technical prototypes". These have very different aims. The physics prototypes are being used to acquire large data samples in beam tests so that the agreement of the various available physics simulations can be checked. This will allow the design and optimisation of ILC detectors based on similar technologies with a high degree of confidence in the simulation. In contrast, the technical prototypes are designed to be a first attempt at building semi-realistic "ILC-like" calorimeter modules. These are intended to give crucial information on the real integration issues and constraints of building such a calorimeter.

The size of the collaboration means it can benefit enormously, in terms of efficiency, effort and cost, from sharing many aspects of the R&D work. Specifically, for the physics prototypes, the design, and indeed physical boards, of the readout electronics is common for the silicon ECAL, scintillator ECAL, scintillator HCAL and TCMT and there are plans to use them further for the one of the gas HCALs. The DAQ online software and raw data format has been common for all detectors used in the beam tests and future physics prototype beam tests will also use these collaboration-wide tools. These data all have a common offline format, specifically LCIO, which allows shared analysis techniques throughout the collaboration. This common offline structure is also used for ILC physics and global detector studies, allowing the output of the beam test studies to be efficiently transferred into detector optimisation work.

Longer term, the technical prototypes will also have large amounts of commonality. The front-end, on-detector electronics for several of the detectors will be implemented in ASICs all designed in a similar fashion. Furthermore, the second generation DAQs for use with the technical prototypes will have a common basis. This will again allow a single set of off-detector readout electronics, control/timing and online software to be used for all such tests. For analysis of the data, again a common offline software format will be used for all detectors.

We believe this coordination not only makes CALICE highly efficient in terms of cost and effort, but also allows the collaboration, possibly uniquely, to make meaningful and systematic comparisons between the various calorimeter technology choices for the ILC.

## 2.2 Physics prototype program

The goal of producing a calorimeter which is capable of delivering the best PFA and software compensation performance requires detailed simulation studies. However, particularly for hadronic interaction simulations, the simulation models can vary substantially between each other. Figure 2-1 shows an example of these differences, specifically a comparison of the shower radius for 10GeV pions in the HCALs, using various simulation physics models. This shows that there are differences of up to 60% between the models and that both scintillator and gas calorimeters have these uncertainties. This is a significant issue in any calorimeter design work.

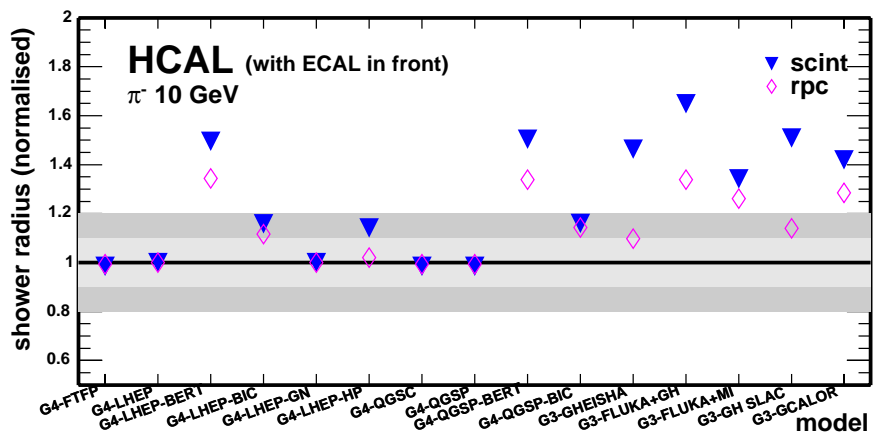

**Figure 2-1: Radius of showers from 10GeV pions in both scintillator (blue triangles) and gas (pink diamonds) HCALs as a function of various simulation models.**

To overcome this problem, CALICE is conducting a major physics prototypes program, which involves testing the two main ECAL technologies along with at least two types of HCAL technology in electron and hadron beams. Since the degree of simulation agreement will depend on the technology, this must involve calorimeter prototypes with detector technologies and materials close to those proposed. This will allow differentiation between the many available models so as to use one which agrees well with data and hence enable the optimisation of the ILC calorimeters to proceed with confidence. An additional aim of the physics prototypes program is to gain experience of operating real calorimeter systems with the proposed technology and to measure their performance. This is considered a crucial intermediate step to designing a larger scale detector, particularly as many of the proposed technologies are novel. The physics prototype beam test campaign started at DESY with an electron beam in May 2006 and continued at CERN with electron and hadron beams in the summer of 2006. Further data were taken at DESY in March 2007. For the future, a second CERN run is scheduled for summer 2007 and a run will take place at FNAL on the same timescale. Following this, the focus will move to FNAL by the end of 2007. There will be beam tests with electron and hadron beams at FNAL at various periods throughout 2008.

Historically, the first ECAL physics prototype was the silicon-tungsten sampling calorimeter with analogue readout. It consists of 30 layers of silicon wafers interspersed between tungsten sheets. Each wafer layer contains a 3×3 array of silicon wafers, each containing 36 1×1cm$^2$ diode pads. There are around 10,000 channels in total occupying a volume of approximately (18cm)$^3$. The ECAL assembly bas been paced by the silicon wafer production and there were delays in this production during 2005 and 2006. However, these have now been overcome and the total required number of functional wafers is now effectively in hand. Around 80% of these have now been assembled into detector planes and the rest are expected to be completed by July 2007. The readout is through a custom on-detector ASIC and VME readout boards. The prototype was exposed to beams at both DESY and CERN in 2006 when approximately half and two-thirds complete, respectively. The upcoming CERN and FNAL beam tests in 2007/8 will be with the completed calorimeter.

The second ECAL physics prototype is a scintillator-tungsten sampling calorimeter, again with analogue readout. It consists of 27 layers of scintillator strips, with each layer covering an $18 \times 18 cm^2$ area. The scintillator strips are $1 \times 4.5 cm^2$ in their transverse dimensions and each layer contains a $4 \times 18$ array of strips, for a total of around 2,000 channels. This calorimeter uses silicon photomultipliers (SiPM) to detect the scintillator light. In addition, it uses the same on-detector electronics as the scintillator HCAL (see below) and the same readout electronics as the silicon ECAL. This reuse of existing equipment and expertise illustrates well the power of a large collaboration like CALICE. A test stack of one-quarter size (the full depth of 27 layers, but with only a $9 \times 9 cm^2$ area in each layer) was assembled and tested in a beam at DESY in March 2007. The rest of the physics prototype is under construction and will be completed in January 2008 for the FNAL beam tests in that year.

The scintillator HCAL is a sampling calorimeter with 38 layers of steel absorber sheets instrumented with scintillator tiles which have analogue readout using SiPMs. Each layer has a $96 \times 96 cm^2$ area and the total volume is approximately $(1m)^3$. The tiles are of varying sizes, with the highest granularity central region using $3 \times 3 cm^2$ tiles, increasing to $12 \times 12 cm^2$ for the outermost tiles. This HCAL has around 8,000 channels in total. The readout from the SiPM is through a custom on-detector board (which was reused by the scintillator ECAL). The HCAL then uses the same VME readout boards as both ECALs. The HCAL assembly has been mainly paced by the manufacturing rate of the SiPMs used to read out the tiles. These are now all in hand and the HCAL will be fully assembled by May 2007. Around two-thirds of the layers were installed in time for the CERN beam tests in summer 2006. The completed detector will take data at CERN and FNAL in 2007/8.

The digital HCAL is a binary readout gas-steel sampling calorimeter. The sensitive layers will be mainly resistive plate chambers (RPC) although as the tests progress, one or more layers, up to the complete stack if production goes smoothly, will be replaced with gas electron multiplier (GEM) or micro-megas detectors. This will give invaluable experience of operating all these detectors and will allow their performance in terms of crosstalk, noise rates, etc, to be measured in realistic operating conditions. In all cases, the pads will be $1 \times 1 cm^2$, giving around 350,000 channels, each reading one bit. The readout electronics for the HCALs are based on on-detector, custom-designed ASICs which can run with ILC-like timing as well as in triggered mode as needed for the beam test. As such, they already fulfil many of the requirements of a second-generation technical prototype readout chip. These ASICs can be used with all the gas technologies and so will allow the different detector types to be tested in a very similar environment. The off-detector readout will be through VME, simplifying the integration into the common DAQ and online software system. As one of the main aims of the beam tests is to compare the performance of the various HCAL options, the same absorber structure and tail catcher as for the scintillator HCAL will be used for all cases, so as to eliminate any spurious differences which might otherwise arise. Hence, the digital HCAL will also be around $(1m)^3$ in volume. A test with layers of a smaller area, but still using the full readout chain, will take place at FNAL in summer 2007 and this will incorporate both RPC and GEM detectors. The full-size version of the gas HCAL physics prototype should be completed by the middle of 2008 and will then take data at FNAL.

The HCALs are complemented by a TCMT detector consisting of 96cm of iron instrumented with 16 layers of 5mm×5cm scintillator strips, which tag shower leakage and detect muons. It has a total of around 300 channels. The scintillator strips use the same SiPM readout as the scintillator HCAL and also have the same downstream readout electronics, again demonstrating the efficiencies of collaborative R&D. The TCMT detector was completed in 2006 and took data in the CERN beam tests that year. It will also participate in the beam tests at CERN and FNAL in 2007/8.

## 2.3 Technical prototype program

The technical prototype program is growing rapidly and has expanded significantly in the last year. The aim of this effort is to understand the real-world issues of building a calorimeter for the ILC. Hence, the technical prototypes are being designed to be a "best guess" of how such a calorimeter module might look in terms of mechanical structure, cooling, sensitive layers, on-detector electronics and off-detector readout. As such, they will give an indication of the likely integration issues and so will be invaluable for the eventual detector technical designs.

These "ILC-like" calorimeter modules will be tested in beam when complete and this part of the program is scheduled to start in 2009. However, unlike the physics prototypes, the aim would not be to accumulate a high statistics sample for comparison with simulation, but to assess the functionality (signal/noise, leakage, timing resolution, etc.) of the modules. As such, they do not need to be fully populated. This effort started with the EUDET grant, which provided funding to help build technical prototypes of the silicon ECAL and the scintillator HCAL. The expansion of the gas HCAL effort to include European groups has enabled work to start on a gas HCAL technical prototype also. There are no firm plans for a scintillator-tungsten ECAL technical prototype at this stage.

The ECAL technical prototype will be mechanically the size of a LDC current design detector ECAL module. This is specifically 1/40 of the total barrel calorimeter, being approximately $1\times1.5m^2$ in area and 30 layers deep. It will be partially equipped with silicon wafers; the middle third of the module will have 28 of the 30 layers equipped for 1/7 of their length while the remaining 2 layers of the 30 will be fully populated. The module is scheduled to be completed in 2008 and exposed to beams from mid 2009 onwards. In addition, as the tests progress, several layers will be replaced with digital ECAL MAPS sensors, allowing a controlled comparison between them in a very similar environment.

In a similar way, the scintillator HCAL technical prototype will also represent a fraction of 1/32 of the barrel HCAL of a calorimeter close to the LDC design. It would correspond to a half-octant of the half-barrel, with an area of approximately $1\times2m^2$ and a depth of 40 layers, although again, not all will be instrumented. It is scheduled to be complete on the same timescale as the ECAL and the two will be tested together. The gas HCAL technical prototype study started more recently and so is less well-defined at this stage. It is likely to be similar in size to the scintillator HCAL, around 1/36 of a barrel HCAL detector.

The on-detector readout for the technical prototypes will use second-generation ASICs. These will be different for the various detectors but will be designed with a common basis, such that they will have a very similar back-end and DAQ interface. This means they will share significant parts of the control and readout designs, with obvious benefits in terms of debugging, upstream interfaces and reduced design effort. They will be able to operate within an ILC-like timing structure and will not require an external trigger. They will also buffer data with no deadtime during an ILC-like beam spill and then read out between spills. Furthermore, the ASICs will be embedded on the detectors and will have the ability to be powered down to reduce the heat load before the next spill. As such, they will incorporate effectively all the features which would be needed for an eventual ILC readout chip. One novel feature of the ASIC used for the HCAL is that it will incorporate a TDC which will allow O(ns) timing of hits, to allow for non-prompt neutron identification.

The off-detector control, timing and readout will be with a prototype DAQ system which would in principle be capable of operating at the ILC. It will be common to all detectors, with the system-dependent differences restricted to a well-defined interface to each. The system will be triggerless, as expected with most DAQ concepts at the ILC. It will also be based on the concept of no off-detector custom electronics, with only commercial components being used; in particular, there will be none of the traditional crates used in previous HEP experiments. This will give invaluable experience in the real applications of these DAQ ideas.

# 3 The ECAL based on scintillator

## 3.1 Introduction

At ILC, the most important role of the Electro-magnetic Calorimeter (ECAL) is to identify photons in jets, since final states are dominated by jets coming from quarks and gluons produced in the decays of electroweak bosons and Higgs particles. Most photons are produced by the two-photon decay of neutral pions in the jets. The opening angle of two photons from the neutral pion decay becomes smaller and smaller with increasing pion energy, as shown in Figure 3-1.

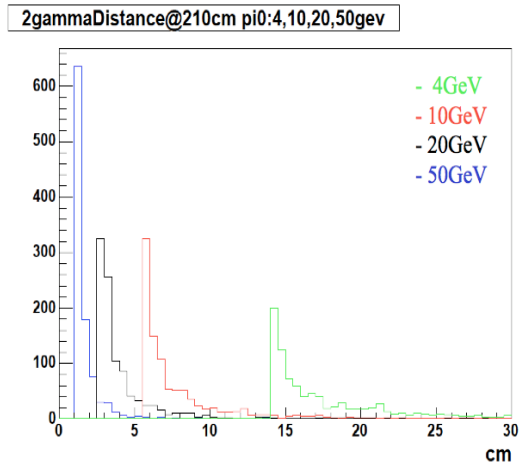

**Figure 3-1: Distance distribution between two photons from neutral pion decays at the ECAL sitting 210 cm from the IP.**

An ECAL granularity of around 1cm will allow the resolution of photons from the decay of neutral pions with energies up to around 50 GeV. It will also allow the efficient separation of charged and neutral energy deposits when used in conjunction with the tracking detector, an essential feature for Particle Flow Algorithms. To achieve such a highly segmented ECAL, we adopted scintillator as active material. The scintillator is a robust, stable, easy to assemble, and reliable material of which we have extensive experience. A sandwich structure is indispensable to measure the EM shower shape in the longitudinal direction. To reduce the Molière radius of the EM shower, we chose to use Tungsten absorber. In order to read out the scintillation light from the scintillator, we employ a wavelength shifting fibre to make the detector more homogeneous along the fibber direction. Our previous ECAL prototype with 1 cm x 20 cm strips showed fairly uniform response to minimum ionizing particles (MIP) [NIM A 557 (2006) 460-478]. One of the results of angle resolution is shown in Figure 3-2.

This long strip scintillator detector also showed good tracking performance in a single particle environment. The scintillation photons must be converted to an analogue signal. This photon detector must have a very small size, and must operate in a strong magnetic field. We have been developing a Multipixel APD to satisfy these requirements. These detectors have greatly improved photon counting capability with respect to previously available devices, and will be useful in many different light detection applications.

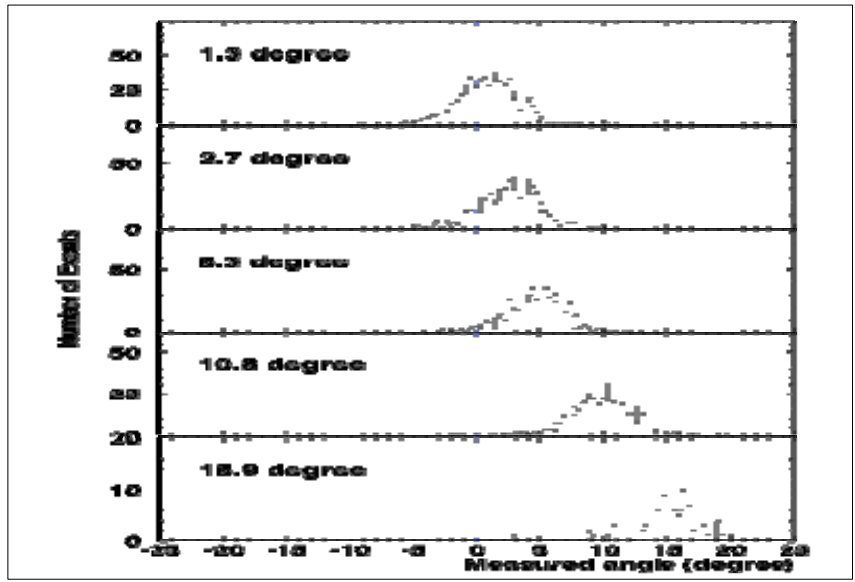

**Figure 3-2:** The angle resolution of long strip ECAL prototype.

## 3.2 Detector configuration

The ILC-ECAL consists of 1 cm x 4 cm x 0.2 cm strips, aligned orthogonally in adjacent layers. This reduces the number of readout channels with respect to a granularity of 1cm x 1cm, while keeping the same effective granularity. The mean Moliere radius of this ECAL detector is 29 mm which means that about 5 strips in each layer will receive electrons in a typical E-M shower. The thickness of the Tungsten absorber material is 3 mm at 90 degrees to the beam direction, as shown in Figure 3-3. At the end of every strip, a photon sensor is attached to receive the scintillation light via the WLS fiber.

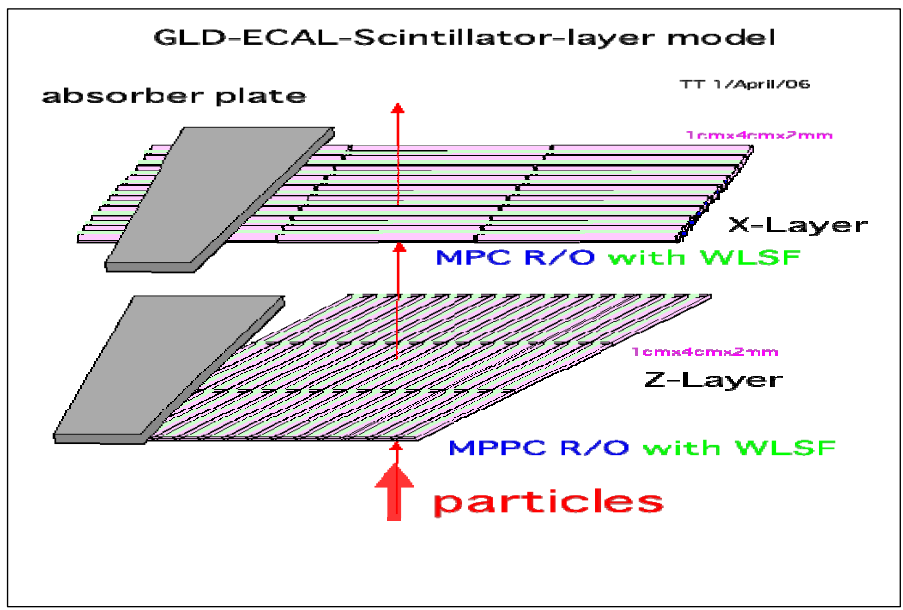

**Figure 3-3:** Scintillator ECAL model of orthogonal oriented short strips with tungsten absorbers.

## 3.3 Development items

### 3.3.1 Photon sensor (MPPC)

The new photon sensor is named "MPPC" (Multi Pixel Photon Counter), which we are developing together with the Hamamatsu Photonics Company. It is a silicon photon sensor, designed to operate at room temperature. The size of the photon detection area is 1mm x 1mm, which contains 40 x 40 small pixels of 25 x 25 microns, giving a total of 1600 pixels, which are connected in parallel. Each pixel operates in the limited Geiger Mode region, giving an ON/OFF signal of with a gain of about $3 \times 10^5$, depending on the bias voltage. The number of photons in the detector is proportional to the pulse height of the output signal for low light input. The basic understanding of this sensor and various improvements are in progress. A current MPPC which has dimensions of 3 mm x 4 mm x 1.3 mm, currently the smallest plastic package of the Hamamatsu product. In Figure 3-4, we show the response of a typical MPPC as a function of the applied bias voltage.

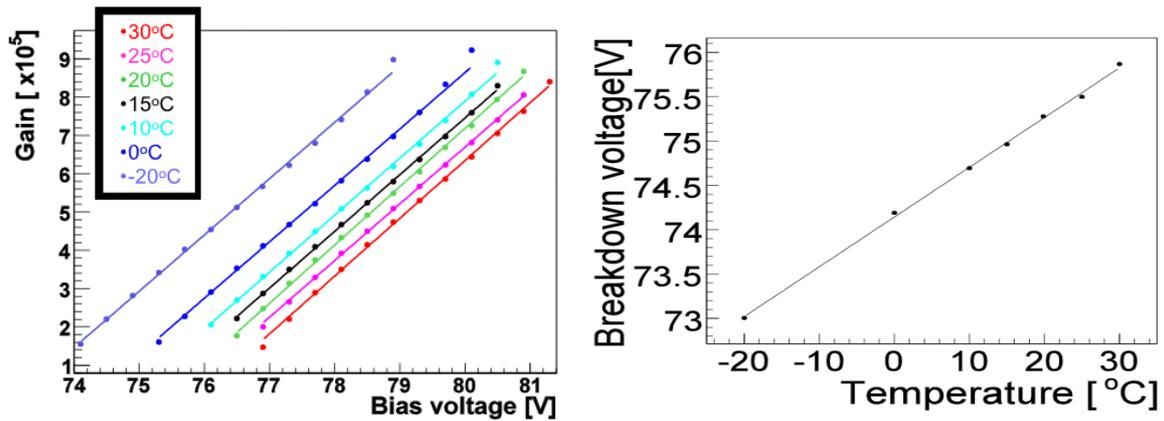
.

**Figure 3-4 Left: The MPPC amplification gain is plotted as a function of the bias voltage at various temperatures. Right: The MPPC breakdown voltage as a function of the temperature, which is extracted from the left figure.**

This sample of MPPCs has very uniform characteristics, as can be seen in Figure 3-5 (left), where we show the relation between the gain and the over-voltage (defined as applied bias voltage - breakdown voltage) for a large sample of MPPCs. In this figure, there are two colours which correspond to the two different delivery times.

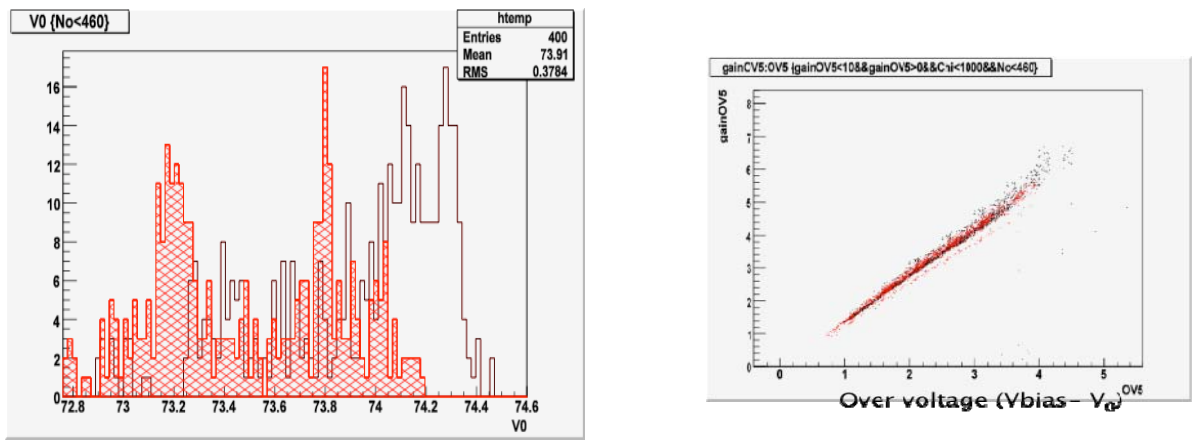

**Figure 3-5 Left: The variation of breakdown voltage with temperature is 56mV/degree for this particular MPPC. Right: The distribution of Vb for 809 MPPCs.**

Once a pixel in a photon sensor has fired, there is a possibility to have signal at the neighbouring pixel. It is called as cross talk. Even in the darkness, the MPPC may have signals, although the signal is equivalent with a single pixel firing. If two or more bigger signals are seen, they must be due to the cross talk. The cross-talk probability distribution for 468 MPPCs is shown in Figure 3-6, which shows fairly uniform behaviour as a function of the over voltage. This cross-talk probability is expected to be reduced by future sensor improvements.

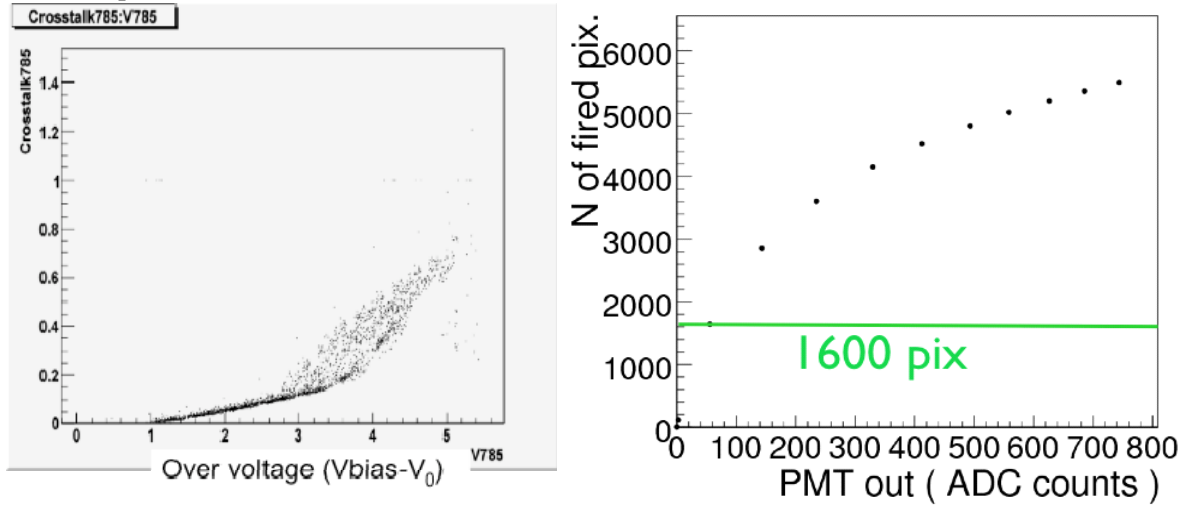

**Figure 3-6: Left: The cross talk probability distribution for 463 MPPCs as a function of the over-voltage. Right: MPPC linearity measured by LED system as a function of input light measured by a photo-multiplier.**

In Figure 3-6, we have unexpectedly good linearity than the number of pixels which is 1600 in this particular MPPC. This measurement has been carried out with a 100 ns gated ADC, which leads an idea of rapid recovery of pixels in the photon sensor. To ensure the linearity of the ECAL system, we need to increase the number of pixels in a MPPC. Further sensor development is required to increase the number of pixels.

### 3.3.2 Scintillator and its fabrication

We are studying the ECAL using scintillator from different scintillator producers. One is commercially available from the Kuraray company which makes a casting scintillator plate. We have machined this scintillator plate to produce the mega-strip structure whose cross section is shown in Figure 3-7 where one can find a hole for the fiber and grooves to prevent light leak to the neighboring strips.

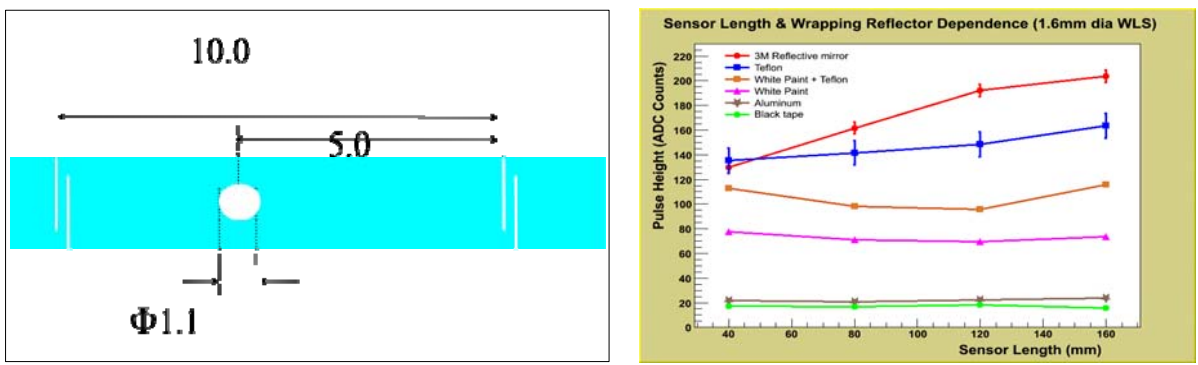

**Figure 3-7 Left: Cross section of mega-strip scintillator. Right: The light amount received with different reflector materials.**

The other scintillator is developed by Kyungpook University. It is manufactured using an extrusion method, which allows an outer light shield of TiO2 and the central hole (for the WLS) to be incorporated directly into the long scintillator strips. They are also trying to produce a mega-strip using an extrusion method.

### 3.3.3  Reflector

We have studied how the detected light depends on what material is used to cover the scintillator; the results are shown in Figure 3-7. The 3M radiant mirror film [http://www.3m.com/] is seen to give the largest signals. At present this is not commercially available, but the company says it will soon produce a similar reflector.

### 3.3.4  Wavelength shifting fiber

We have been studying the light collection from the scintillator. In order to reduce the non-uniform response of the scintillator, we set a light collection fiber in a straight hole at the center of the scintillator. This fibre contains Y11 wavelength shifter, and absorbs scintillation light (which has a wavelength of about 400 nm) and emits it at a wavelength of 550 nm. Although this emission wavelength is a little longer than the most efficient wavelength for the MPPC, we have measured a signal of 23 photo electrons (p.e.) from a 1 cm x 4.5cm x 0.3cm scintillator strip with a WLSF using a MPPC for a minimum ionizing particle (MIP). The number of photon electron distribution by a MIP is plotted in Figure 3-8.

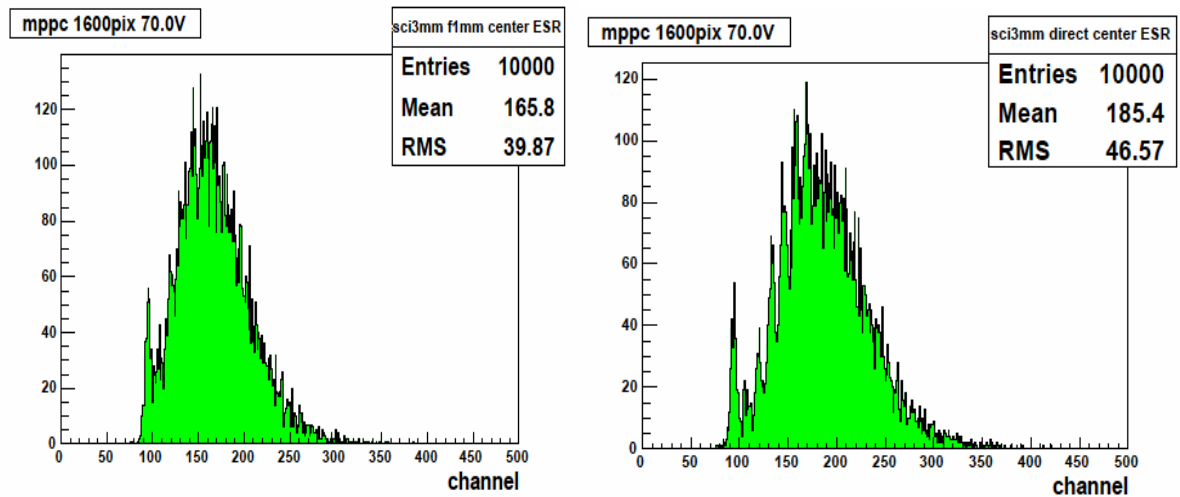

**Figure 3-8 Left: The energy deposit distribution by a mip with a wave length shifting fibre read out. Right: The energy deposit distribution by a mip without a wave length shifting fibre via direct read out.**

The peak efficiency of the MPPC is at a wavelength of around 400nm, so we tried to directly attach the MPPC to the scintillator, without a WLSF.  While reducing the number of photo electrons to a half of that with the WLSF, it shows a narrow distribution of the number of detected photons. This approach makes the detector construction easier, since we do not need to carefully align the MPPC to a WLSF.

### 3.4  The prototype detector and its performance

A prototype ECAL detector was constructed at Shinshu University and tested at the positron beam line at DESY.It consists of 26 layers of scintillator, each with 18 scintillator strips, interleaved with 3.5mm thick super-hard Tungsten plate. The strips are aligned in orthogonal directions in successive layers, as shown in Figure 3-9.

At each end of each scintillator strip, we made a small rectangular hole (4 mm x 1.3mm) to hold the MPPC. We have constructed three types of detector, each with 13 layers: Kuraray scintillator with and without the WLSF, and one of the extruded Korean scintillator. We prepared three types of ECAL combining these three detector types. The MPPC signals are read out by the electronics system of the AHCAL (Analog Hadron calorimeter of scintillator tile with SiPM read out).

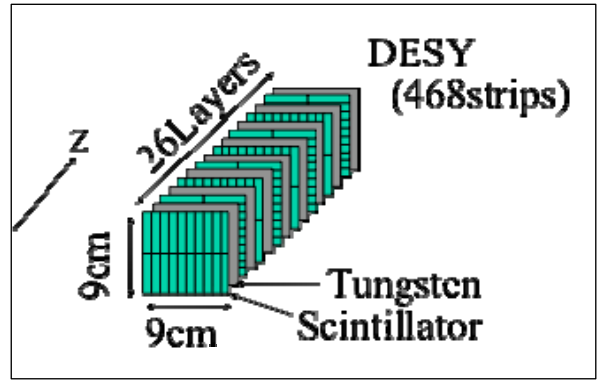

**Figure 3-9: A schematic view of scintillator ECAL prepared for the DESY beam test.**

## 3.5 Beam test at DESY

The detector performance was tested at the DESY synchrotron. Positrons with energies energies of between 1 and 6 GeV were injected onto the ECAL at various points. The detector was calibrated by removing the tungsten plates, and firing positrons at the centre of each scintillator strip position. The breakdown voltage of each MPPC was measured at the laboratory prior to the beam test. The bias voltages of the photon sensors were set 2.5 V higher than the breakdown voltage in the case of the strips with WLSF, and 3.5 V higher for the the other detectors.

Some preliminary results of the beamtest are presented here.The basic performance of this ECAL is shown in Figure 3-10, where we show the linearity and energy resolution when using various calibration methods which we are presently investigating. The homogeneity of a strip along its length is shown in Figure 3-11 for the three types of scintillator.

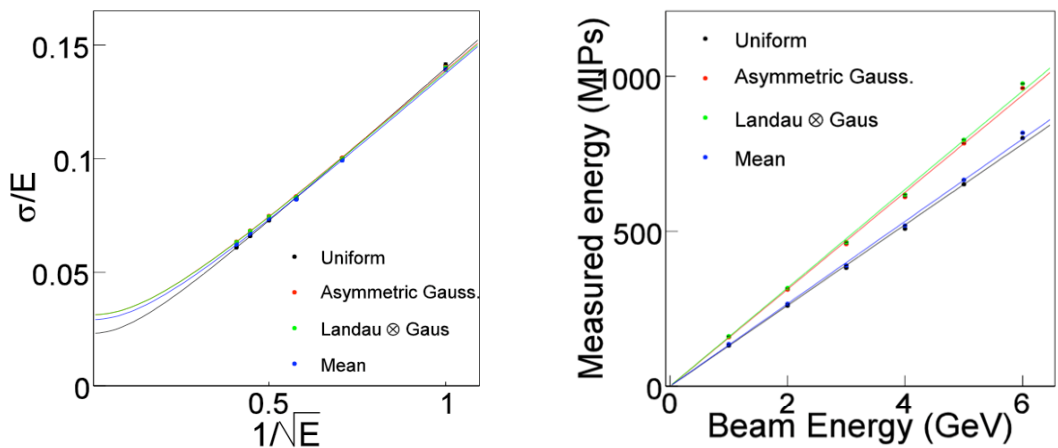

**Figure 3-10 Left: The energy resolution of the prototype scintillator ECAL detector tested at DESY beam with different calibrations. Right: The linearity of the ECAL with different calibration methods.**

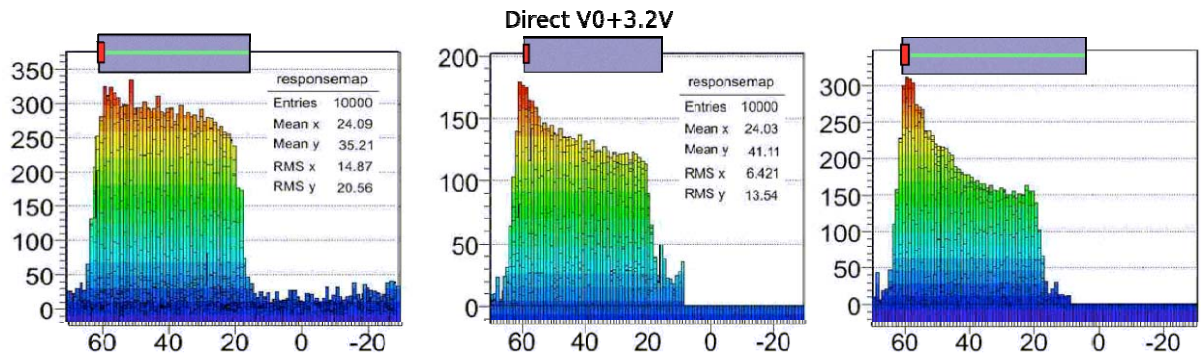

**Figure 3-11: Homogeneity of a scintillator strips. Left: Kuraray scintillator with a WLSF. Middle: Kuraray without WLSF. Right: KNU scintillator.**

## 3.6 Future plans and milestones

The prototype ECAL detector of 468 MPPCs has been successfully constructed and tested. Its performance has been measured and shows reasonable characteristics. Moreover, we are planning to build a larger detector, 4 times bigger than the current prototype. Using the next prototype, we will test neutral pion reconstruction at the pion beam line at Fermilab, where we will place a target in front of the ECAL module. The pi- + neutron -> pizero + proton charge exchange process pizero producing process will be employed to identify neutral pions using this ECAL module in 2008.

We will further improve the scintillator and the photon sensor. The MPPC photon sensor should have a smaller cross talk rate; we will work on this together with the Hammamatsu Photonics. The Korean scintillator will consist of extruded mega-strips with an embedded WLS fiber. This new production procedure is simpler and should improve the response uniformity.

## 3.7 Remaining issues

We have not yet verified the possibility of neutral pion reconstruction using a detector simulation, although we plan to test this using real pi-zero data at Fermilab in 2008. Only one third of hadrons interact in the ECAL, so we will need to use a HCAL system to fully measure hadron and eventually jet energies. The best combination of ECAL/HCAL detectors is not yet known.

# 4 The ECAL based on silicon

## 4.1 Introduction

At the future e+e− linear collider, the detector must be able to perform a very efficient particle flow reconstruction. One of the ways to perform it consists in instrumenting calorimeters with a high degree of tracking capabilities, making possible a reliable track-shower association (TCA) and energy estimation. To optimize the TCA and the energy resolution, both the electromagnetic (ECAL) and hadronic (HCAL) calorimeters must be put inside the coil. Figure 4-1 shows the general geometry of the calorimeter system, which avoids any blind region in the middle of the shower development.

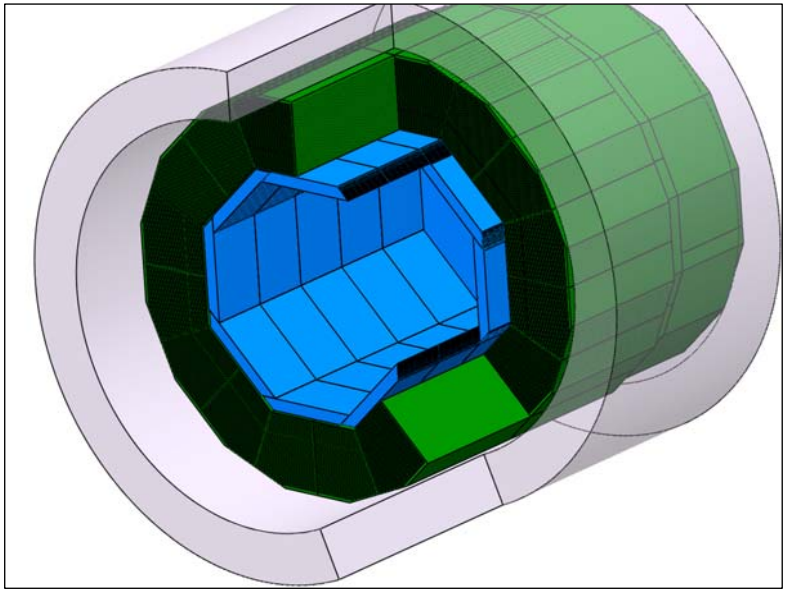

**Figure 4-1: General view of the calorimeter system internal to the coil. The thin ECAL is represented in blue.**

Therefore, the available space for both calorimeters is constrained by the internal radius of the coil and a compact ECAL is mandatory for shower separation giving in addition enough space for a large number interaction lengths in the HCAL. In addition, it has been shown that a "tracking calorimeter" could be the right choice to optimize the performance of the pattern recognition, since it produces a 3D view of the shower development inside the calorimeter. Such a calorimeter must have a large segmentation with small readout pad size and a good shower to shower separation, i.e. a small Molière radius but also a small hadronic shower spatial dispersion. Basic construction units of the electromagnetic calorimeter are modules, mechanical stiffness of which is ensured by a carbon fiber skeleton, minimizing dead region. A detailed version of the proposed ECAL geometry is shown in Figure 4-2.

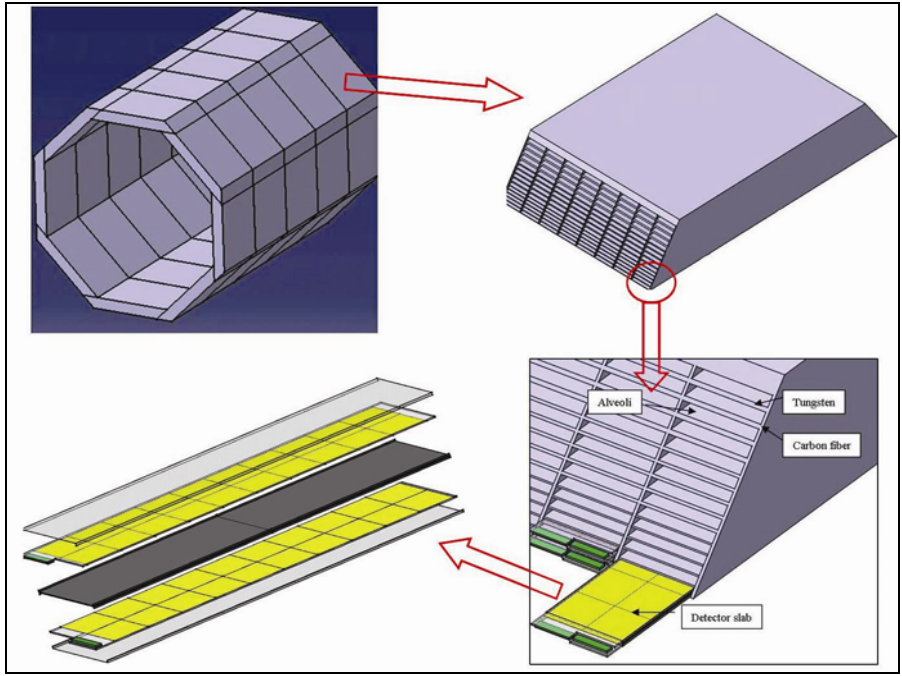

**Figure 4-2: Cross section of the barrel part of the detection system (top left) and basic construction elements of the electromagnetic calorimeter, with at the end (bottom left) the basic detector element, called detector slab.**

## 4.2 The electromagnetic calorimeter concept

All these considerations led to choose high Z number and density material for the radiator going naturally for the tungsten, which has a Molière radius $r_M$ of 9mm and a radiation length ($X_0$) of 3.5 mm. For the active part of the device, silicon PIN diodes seem perfect apart from their cost. The individual e.m. energy resolution needed and the future cost evolution will drive the area of silicon and therefore the number of layers. Many studies have been performed with 40 layers, as given in the TESLA TDR [1], while a prototype under construction will have only 30 layers. The pad size comes from a compromise between
1. the density of extraction lines of the readout,
2. the electronics cost with the total number of channels,
3. the effective Molière radius of the calorimeter.

A pad size of 1x1cm$^2$ has been adopted for the first studies and first prototype, but the second generation prototype as well as the recent studies use 5x5 mm².

The basic detection units are detector slabs (Figure 3.2) which consist of a stiffening H-structure from carbon fiber material embedding a tungsten sheet overlaid by silicon pad sensors. The entire slab is shielded by an aluminum envelope and is slit into the module. As shown in Figure 4-3, for the final project, the only possible geometry design locates the VFE chip inside the detector. That will be possible only if

1. the power dissipation is low enough,
2. the duty cycle reduce sufficiently the power dissipation. If not, a cooling system has to be designed in this thin device,
3. A 500 GeV electromagnetic shower crossing the VFE chip does not disturb it too strongly.

All these points are part of the CALICE R&D program. Ensuring the behavior of the first ECAL prototype in the test beam, for the prototype, the VFE chips are located outside, on the external part of the PCB.

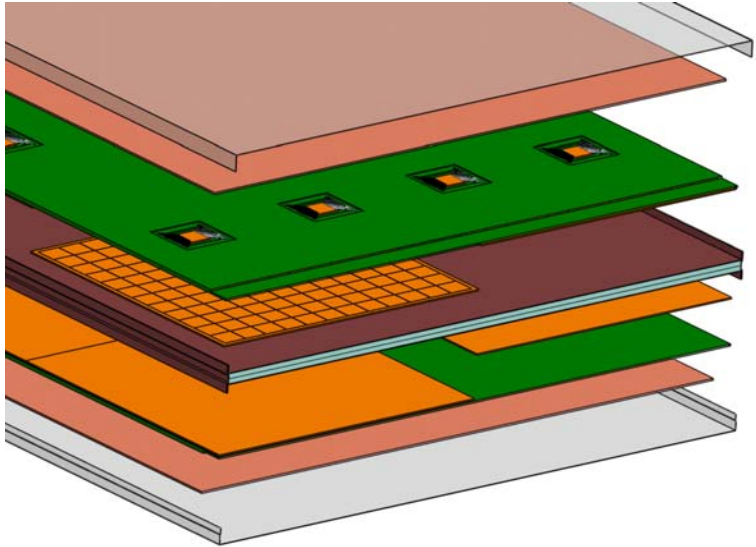

**Figure 4-3: Schematic view of a possible detector slab. Green is the PCB in which the VFE are included, the silicon matrices (orange) are glued on PCB. The tungsten sheet (light blue) is wrapped in carbon fibre (brown). An overall protection sheet in aluminium (grey) surrounds everything.**

## 4.3  Current status of the project

The agenda of the project concerns also the construction of a second generation of prototype, this time as close as possible to the final detector. This second generation is partially funded by the EU program EUDET. The test with hadron beams is foreseen to be done together with the different HCAL options of CALICE.

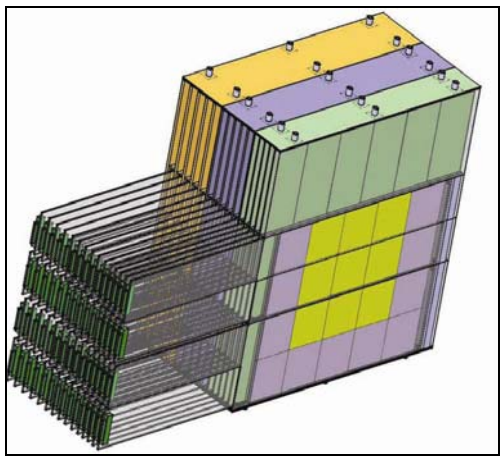 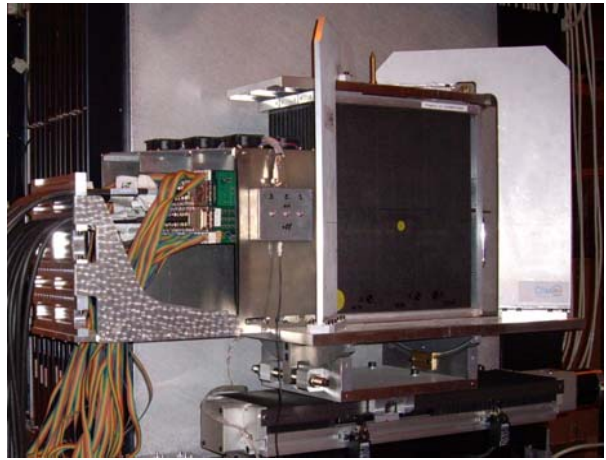

**Figure 4-4 Left: Schematic view of the first generation prototype. Right: ECAL prototype at CERN test beam area in 2006.**

## 4.4  The prototype in test beam

### 4.4.1  General description

The ECAL prototype mimics the final project with tungsten wrapped in carbon fiber, with 30 layers and a pad size of $1 \times 1 \text{cm}^2$ but the very front end (VFE) electronics is located outside the device, on the same PCB on which the silicon wafers are glued. The schematic view of the prototype is shown in Figure 4-4, with different shade of grey for the three stacks, each one with different tungsten thickness. This choice ensures a good resolution at low energy, due to the thin tungsten in the first stack, and a good containment of the e.m. shower due to a tungsten thickness 3 times larger in the third stack. The overall thickness is about 20 cm or $24X_0$. The structure is realized by carbon fiber wrapping half of the tungsten sheet, leaving free slots between each tungsten sheet, called alveoli. In these free slots, detector slabs are slit. The detector slab consists of 2 active readout layers surrounding one tungsten sheet. The active layer is made of PCB 14 layers (2.1mm) and 500 microns high resistivity silicon wafers.

### 4.4.2 The active device

The wafers are cut in matrices of 62x62 mm with 36 pads, while the space reserved for the guard ring is about 1mm. It must be noted that there is only one set of guard ring per matrix. The production, made in Russia, managed by the MSU group, and in Czech Republic, managed by the IOP- ASCR, is of very high quality, with a typical leakage current less than a few nA/pad for all the pads, apart from a few per cent of the production, where 1 or 2 pads among the 36 of a matrix go up to 20 nA/pad. The matrix will work in overdepleted mode around 250V bias voltage. The prototype is still under completion with silicon ratio about 50% from Russia and 50% from Prague. This prototype will be completed in June 2006.

The connection to the PCB is realized by conductive glue for each pad, with an AC coupling mode of the readout. For the final project, it is considered to use amorphous silicon deposited directly on the pad, to provide the resistance and capacitance. This is the object of an R&D program inside the CALICE-ECAL groups. For the prototype, the AC coupling is realized using discrete components put directly on the PCB, before the amplification in the VFE chip.

### 4.4.3 The very front end electronics

The PIN diodes are read by a VFE chip developed by the LAL-Orsay group. It consists of a preamplifier, a shaper with two possible gains and a multiplexer. This chip treats 18 channels, with low noise and high dynamic range. It provides a linearity of 0.2%, a signal uniformity better than 2% and less than 0.2% of cross talk. These values are measured over the 600 MIPS dynamical range. The PCB with the silicon matrix and the VFE on the side are shown on Figure 4-5. The schematic view of the VFE is also presented in Figure 4-5.

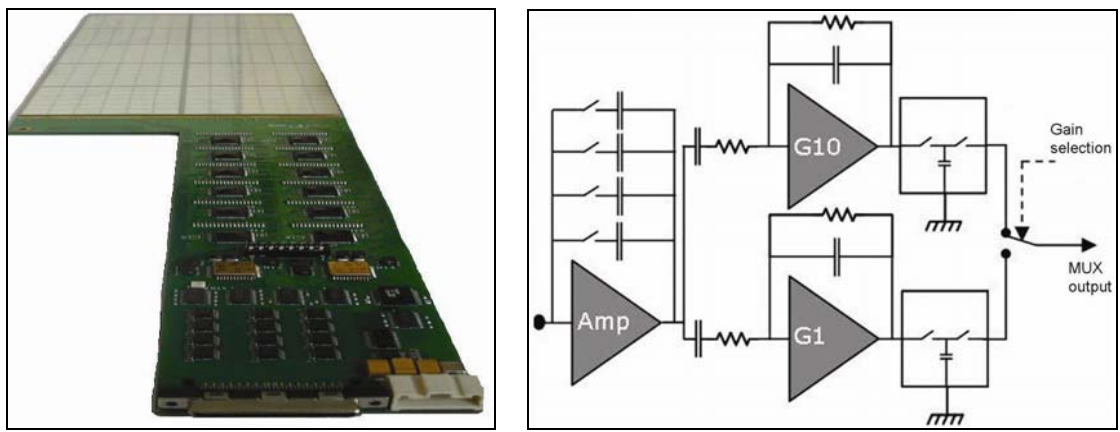

**Figure 4-5 Left: PCB equipped with silicon matrices. Right: VFE schematic used for the prototype and VFE chip.**

### 4.4.4 Assembly

The wafers are glued on the PCB by an automat controlled by a PC. The glue is a conductive glue from Epotek. Four years after the assembling, no problem is observed. The process has been studied to be useable down to a pad size of 5x5 mm².

### 4.4.5 First test of the device

One of the important advantages of the silicon is the response stability with time and temperature. In order to check that, the detector is intercalibrated with dedicated muon runs, by fitting the m.i.p. peak. This intercalibration is then used for all the year run period. After run analysis leads to observe, after 2 years of running, (DESY 2005 and CERN 2006) a number of dead cells of the order few per mill. The signal over noise observed as the mip peak to the standard deviation of the noise is about 8, with some dispersion from PCB to PCB, but also wafer to wafer. Most of the problems of common noise are related to the connexion at the PCB end toward the CALICE readout card (CRC) VME readout card.

Analysing the runs taken at CERN with electrons from 10 to 50 GeV/c, the detector shows a very good behaviour, and the overall performance seems well described by Geant4 simulation. It must be noted however some problems under investigation which have to be solved for the future.
1) The readout electronics shows some common mode on the pedestal, during the data taking
2) A very strong signal on one pixel, induced a common pedestal shift in the same wafer
3) The floating guard ring induced a cross talk at long distance between pads.
4) A loss of about 20% of signal in the guard ring region, even with a staggering of 1.2 mm layer to layer.
These problems should be cleared off for the second generation prototype already under design.

## 4.5 The second generation prototype

A second generation of prototype, as close as possible to the final detector is under design now. The goal is to have this prototype in test beam in 2009.

### 4.5.1 The new wafers

The silicon wafers will have a new design of the guard ring and a pixel size of 5x5 mm² and depending of the cost, wafers of 6" would be used, which reduces the ratio of the guard ring region over the pixel active region. Quality test of delivery are under development in LLR and Clermont LPC, able to deal with the new geometry. The wafers them self are under design studies at Prague, IOP-ASCR.

### 4.5.2 The new VFE chip

On the readout side, a new VFE ASIC will be directly mounted on the PCB, without packaging, and ion the opposite side of the silicon matrix. This new ASIC will include the ADC and the storage of the digital signal during the spill (similar but longer to the one during the bunch train). It will also have a good linearity up to 3000 mips, which is needed by the final project.

### 4.5.3 The new PCBs

One group of the collaboration is performing longer term stability studies of the gluing process. The assembling process could be, a priori, easily transferred to industry for the final project, through the use of stichable PCB assembled before to build the detector slab made of two layers above and below a tungsten sheet, the overall wrapped in carbon fiber. In order to minimize the loss in the case of something goes wrong during the gluing and assembling processes, the unit size of PCB will just be about 20x20 cm. Since the VFE ASIC being very close to the active silicon wafer, the number of layer in the PCB can be reduced, and the PCB thickness could go down to about 800 μm .

### 4.5.4 The new mechanics

The new module will be as close as possible to the module for the final detector. The new prototype is therefore expected to be a kind of module "zero" for the final detector. For a question of cost, this prototype will have the full length of a barrel module but only half of the width, 3 instead of 7 alveoli. We therefore will have this time a prototype 1.6 m long, about 54cm wide and about 18cm thick. The number of layers is 30, with only two stacks for what concerns the tungsten thickness (optimising the energy resolution at low and medium energy). The alveoli will be 18 cm wide, which optimises the use of 4" as well as 6" wafers. Figure 4-6 shows a schematic view of this new module prototype.

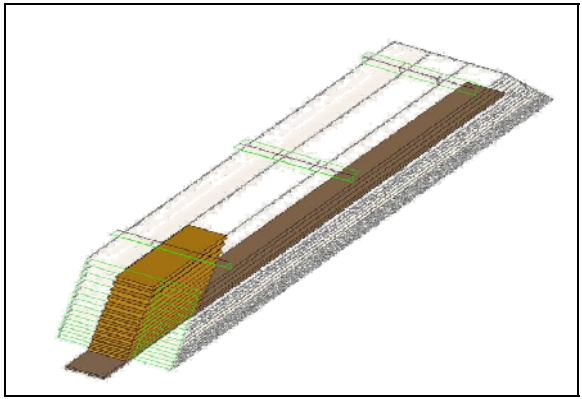

**Figure 4-6: Schematic view of the new ECAL prototype, with silicon coverage over a slab of full length and a region of full depth**

## 4.6 Schedule of the beam test

The tests of the ECAL prototype foreseen in 2007 at CERN period are the following:
- ➢ Detector completed – mid of 2007, test of hadron shower with full prototype
- ➢ Test of the optimised layers staggering, impact on energy resolution
- ➢ Test of the ASIC VFE in high energy em shower

In the year 2008, the EUDET prototype will be designed and finalized before a construction which must lead to having this new prototype in test beam before the end of 2009.

## 4.7 Alternative option for the silicon active device : MAPS

The studies of silicon-tungsten ECALs include work on a digital ECAL where the cells have binary read out. Since such threshold readout does not retain information on the number of particles passing through the cell, then to avoid a non-linear response, the probability of more than one particle hitting the cell should be low. The high density of particles in an EM shower therefore requires the cells to be very small. A typical shower density for the highest energy EM showers expected at the ILC is 100 MIPs/mm$^2$ which corresponds to an average of 1MIP in a $100 \times 100 \mu m^2$ area. Hence, cell sizes of this order or lower are required. This implies an extremely large number of cells for an ECAL and hence requires a very high level of readout integration.

The digital ECAL design being studied within CALICE is implemented using a Monolithic Active Pixel Sensor (MAPS) which integrates the sensor and the readout into the same silicon wafer.
MAPS can be implemented using a standard CMOS process, available at many foundries around the world, and so is more readily available than the high resistivity silicon process needed for the diode pads wafers. This should allow any large-scale production to be sourced from multiple vendors and so is likely to be significantly cheaper per unit area than a diode pad wafer production.

MAPS are sensitive to charged particles through charge liberation in an epitaxial layer under the electronics circuit layer. This charge diffuses in the epitaxial layer and is subsequently absorbed by n-well structures in the circuit layer. N-well collection diodes are used to absorb this charge and serve as the inputs to charge-sensitive circuitry on the sensor. The charge-sensitive circuit then discriminates the level of charge collected and the output is stored on-sensor. However, unless the circuit design can be restricted to otherwise only use p-well diodes, then the circuit n-well diodes will also absorb charge, leading to a significant loss of signal.

To overcome this problem, a novel "INMAPS" process has been developed which isolates the circuit n-well diodes from the epitaxial layer using a "deep p-well" layer between the two.

### 4.7.1 Project status

A first design of such a MAPS device has been completed and was submitted for fabrication using the INMAPS process in April 2007. It will be available in July for testing before a second sensor is fabricated in 2008. This first design is intended as a proof-of-principle and would not be suitable for use at the ILC. The sensor has an epitaxial layer thickness of 15μm and a pixel size of $50\times50$μm$^2$. The sensor will be $10\times10$mm$^2$ in size and will contain a $168\times168$ array of sensitive pixels, corresponding to approximately 28k pixels total. Four different versions of the pixel electronics have been implemented within each sensor to allow studies for further optimisation in the second fabrication round. The individual pixels operate asynchronously as distributing a clock across the whole sensor surface would have taken a significant amount of power. The discriminated output of each pixel is stretched to the bunch crossing period and tracked to a centralised memory bank column where pixels above threshold are timestamped with the bunch crossing number. The list of pixel locations and timestamps is available for readout following the bunch train. In the first sensor, the memory banks occupy the equivalent width of 5 pixels and occur every 47 pixels (approximately 2mm). Since they are not sensitive, this is a dead area and so contributes to the sensor inefficiency. Figure 4-7 shows the submitted design schematic diagrams for the whole sensor and one of the pixels.

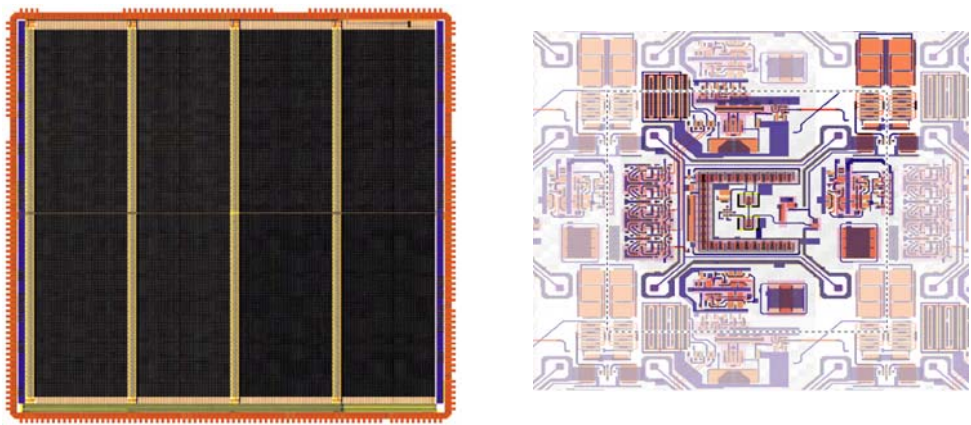

**Figure 4-7 Left: Schematic of complete MAPS sensor submitted for fabrication. The logic and memory storage areas are clearly visible as the four vertical stripes. Right: Schematic of a pixel layout. The electronics associated with a single pixel is highlighted in colour. The pixel boundary is shown by the dashed line.**

A detailed sensor-level simulation of charge diffusion in a MAPS pixel and its immediate eight neighbours has been done using CENTAURUS, with the design GDS file being used to define the pixel geometry. This ensures the simulation is identical in structure to the real pixel. This has allowed optimisation of the sensor during the design as well as making predictions of the sensor performance. As an example, Figure 4-8 shows a simulation of the charge collected by the pixel for three collection diode sizes as a function of the distance of the particle position within the pixel, specifically along a diagonal near the pixel corner. (The corner is the most difficult region as, by symmetry, the average charge collected in each pixel around the corner is at most ¼ of the total liberated.) The 1.8μm diode size is seen to give the highest signal/noise (S/N) and this was the size chosen for the first sensor. Even for the case or a particle right at the corner, the S/N for a MIP deposit is predicted to be above 10 for all diodes. This will allow a threshold around half the MIP value, giving good noise rejection and high efficiency for the pixel discriminator. The target rate for noise hits is $10^{-6}$ per pixel per bunch crossing. Note, even at this low rate, there would be $10^6$ pixel hits per event throughout the whole ECAL.

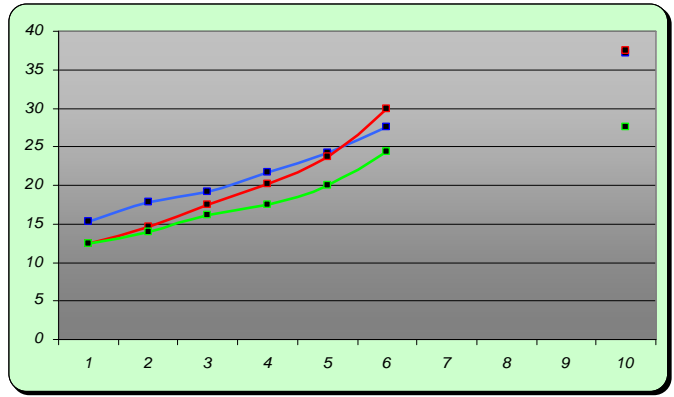

**Figure 4-8: Simulated MAPS pixel signal/noise ratio for MIP deposits for three collection diode sizes as a function of the position of the deposited charge from the corner. The x axis values correspond to increasing distance from the corner of the pixel, with x=1 being at the corner and x=10 being the centre of the pixel. The diode sizes are 0.9μm (green), 1.8μm (blue) and 3.6μm (red).**

### 4.7.2 ECAL performance

The performance of a MAPS digital ECAL has been studied using the Mokka simulation followed by a digitisation step. The latter models the charge diffusion, noise and pixel discriminator output and so gives a list of pixels above threshold for each event, equivalent to the information which would be available in a real calorimeter. Following digitisation, a clustering step is performed to attempt to group neighbouring pixel hits which resulted from a single particle in the sensor. The resulting resolution of EM showers has been studied using this simulation. The effects on this resolution of various contributions are shown in Figure 4-9 as a function of the threshold used.

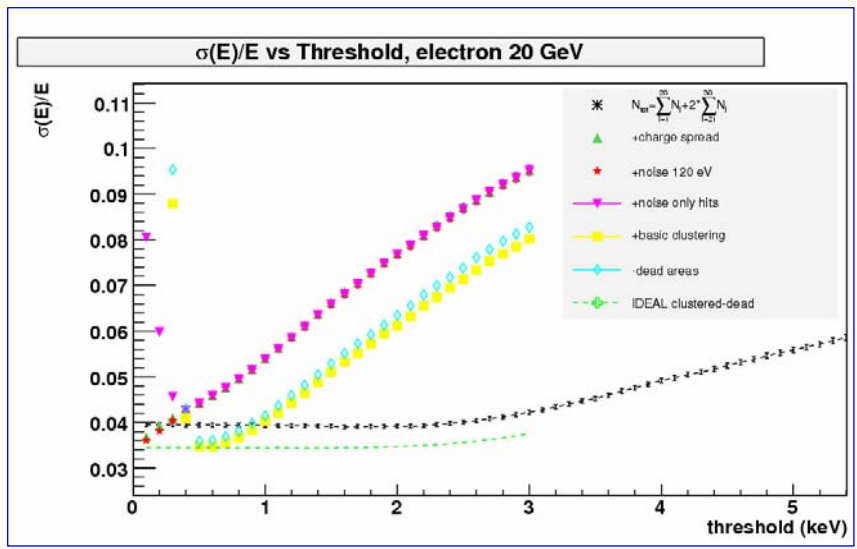

**Figure 4-9**: **For the given sampling fraction used in the simulation, the ECAL resolution obtained from the MAPS with the sensor effects included is seen to be equivalent to that obtained from an ideal digital calorimeter and is around that obtained with the diode pads ECAL.**

PFA studies using a MAPS digital ECAL are ongoing. The much finer granularity resulting from MAPS may allow a significant improvement in two shower separation, particularly in the front part of the ECAL before the shower has spread to the Molière radius. These studies are currently at a very preliminary stage and no quantitative results are available yet.

It is assumed that the response of particles to the MAPS silicon will be identical to that of the diode pads wafer silicon and so a physics prototype for data-simulation comparison studies is not required. However, as it is a novel design, it will be important to acquire operational experience of a MAPS calorimeter. The second fabrication round in 2008 will produce larger sensors of size $20\times20\text{mm}^2$ (limited by the size of a standard CMOS reticule without using stitching) in a dedicated run. Assuming a reasonable yield, this will give a sufficient number of sensors to produce PCBs for one (or possibly two) layers of the ECAL silicon-tungsten technical prototype. This could be swapped into the prototype to replace a diode pad layer at various depths into the structure during the technical prototype beam tests. This will allow a direct comparison of the response of the two types of ECAL at various positions in the shower depth and so allow a meaningful comparison of the performance of the two concepts. The MAPS PCB readout will use the same DAQ system as for the diode pads. Hence, the technical prototype and DAQ for the silicon-tungsten ECAL will be shared between the two designs.

## 4.8 References

1. Calorimeter section, in Tesla Technical Design Report, DESY Report 2001-011, March 2001
2. A calorimeter for Energy Flow, J.C. BRIENT at LCWS2000, SITGES http://www.thep.lu.se/~torbjorn/Pythia.html
3. CALICE collaboration, http://polywww.in2p3.fr/flc/calice.html

# 5 The Tile HCAL

## 5.1 Introduction

In the particle flow paradigm, the HCAL must be an imaging detector: it must allow separating the energy depositions assigned to charged particles from those generated by neutral hadrons and thus eliminating the dominant part of hadronic jet energy fluctuations which is attributed to the charged hadrons. It must then measure the energy of neutral hadrons (mostly neutrons and kaons) and that of not resolved charged hadrons with very good precision. To optimise the particle flow measurement the HCAL should contain the hadronic showers, have a sufficiently high granularity both in the transverse and in the longitudinal direction to resolve the shower substructure, and provide very good hadronic energy resolution. In addition it would be desirable to measure the time of events precisely, to search e.g. for exotic signatures with long lived particles as they are predicted in some SUSY theories, or to help in the rejection of cosmic rays. It has also been suggested to use timing information in the shower reconstruction.

### 5.1.1 The scintillator HCAL concept

With the advent of multi-pixel Geiger mode photo diodes - so-called SiPMs - the high segmentation required for PFLOW reconstruction can be realized with scintillators at reasonable cost. SiPMs have only recently become available in larger quantities from Russian and Japanese industry. With typically 1000 independently quenched pixels on a common load they provide a signal proportional to the number of pixels fired by impinging photons and a gain comparable to that of vacuum photo-tubes. The technology is investigated by several electronics companies around the world and is also driven by non-HEP applications, e.g. astrophysics or medical imaging.

The energy response of scintillators allows to trade amplitude resolution versus granularity and thus to optimize the cost of the readout electronics. In addition to the classical analogue readout, semi-digital concepts with few threshold bit information or a purely digital approach are also followed. A scintillator HCAL has thus become a promising candidate for all PFLOW based detector concepts.

The AHCAL is conceived as a sampling calorimeter, with a material of low magnetic permeability ($\mu$<1.01) like stainless steel, brass, lead or tungsten as absorbers, and scintillator plates as the active medium. The scintillator plates are subdivided into tiles. The millimeter-size SiPM devices can operate with moderate bias voltage in high magnetic field and are mounted on the tiles. The light is collected either directly from the tile or via embedded wavelength shifting fibres (WLS). This concept is different from existing tile calorimeters with long fibre readout; it allows integrating not only the photo-sensors but also the front end electronics into the detector volume.

### 5.1.2 Sampling structure and segmentation

The basic sampling structure consists of 20mm thick absorber plates interleaved with 6.5 mm deep gaps into which the 5 mm thick scintillator plates are inserted. The sampling structure is the same for barrel and for end cap regions; the total depth is still subject to optimization. In the LDC, for example, 38 layers are foreseen for the barrel 53 layers in the end cap. Each layer corresponds to 1.15 $X_0$, or 0.12 $\lambda$ at normal incidence. The layers are subdivided into tiles of transverse dimension 3 × 3cm$^2$ which is considerably smaller than the hadronic shower size. Because of the possibility to implant the SiPM devices into the tiles, all longitudinal layers can be read out individually, which has been shown to significantly enhance the performance.

The segmentation has been optimized to allow for the separation of partially overlapping showers by reconstruction of their internal tree-like structure of sub-clusters, and was also found suitable for the semi-digital approach. Recently, in a full detector simulation and reconstruction study using the Pandora PFLOW algorithm, it could be shown that with such a calorimeter, integrated in the LDC detector environment, the performance goals of the ILC (in terms of W and Z dijet mass separation) can be met. Variation of the tile size between 1x1 cm$^2$ and 10x10 cm$^2$ has shown little impact on the performance for events at the Z resonance, but revealed that the choice of 3x3 cm$^2$ is close to optimal for jets at higher (180 GeV) energy, as shown in Figure 5-1. Further studies are needed to fully optimise the transverse cell (tile) sizes, possibly as a function of the depth in the calorimeter. However, this granularity is considered a reasonable starting point to address integration issues. The studies also indicate that a larger depth of about 50 layers in the barrel would yield better containment and resolution. However, possibilities to estimate leakage from the measured shower shapes and the use of the muon system as tail catcher still need to be evaluated.

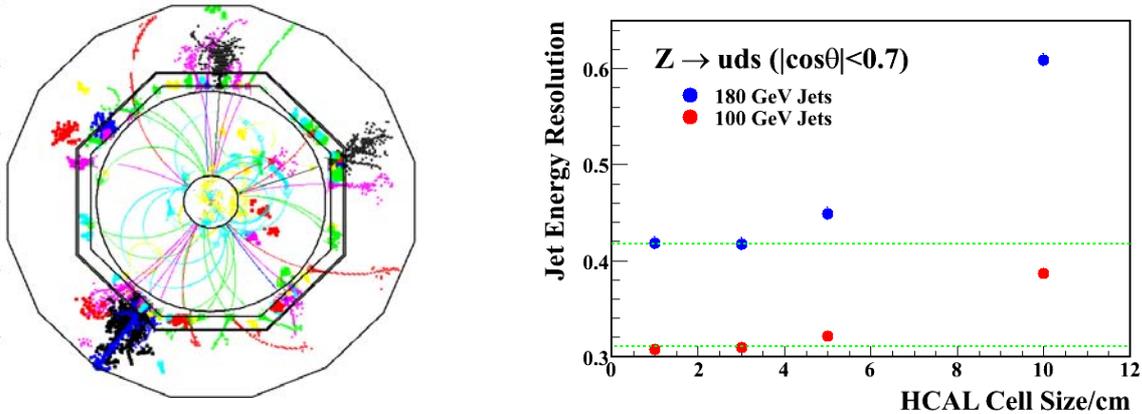

**Figure 5-1** Left: Simulated multi-jet event in a tile HCAL with 3x3cm$^2$ cells. Right: Dependence of jet energy resolution on calorimeter cell size for scintillator tile.

### 5.1.3 Readout and calibration

The scintillator HCAL being an analogue device like the silicon based ECAL, with a channel density about 10 times smaller, we follow an integrated approach to the ECAL and HCAL readout electronics, with many components in common and only the input stage of the analogue front end adapted to the different sensor types. With a scintillator based ECAL the read-out could be even further unified.

Thanks to the high granularity of the HCAL, which leads to of the order of 100 hits for a 10 GeV pion shower, the requirements on the statistical precision of the single cell calibration are rather moderate; 10-20% will be sufficient. However care must be taken to control systematic effects, which coherently may affect many cells, to an accuracy of a few percent, in order to keep the constant term in the energy resolution small. The absolute reference scale for the energy measurement will be set by MIP signals which can be obtained from cosmic rays, beam halo muons or from MIP-like parts within hadronic showers. The time needed to record such events in sufficient quantity is days or weeks. The sensitivity of the SiPM response to temperature and bias voltage fluctuations (typically 5% per 1 K or 100 mV) requires a monitoring system to follow more short-term variations by means of reference signals as provided by PIN stabilized LEDs or radioactive sources. The possibility to measure the SiPM gain directly via the single photo-electron signal separation provides an interesting means to auto-calibrate such variations. The calibration in terms of single photo-electron signals is also needed to apply corrections for the non-linearity of the SiPM response arising from the finite number of pixels.

## 5.2 R&D programme

The test beam effort plays a key role for the validation of the simulations and the further development of the particle flow algorithms which form the basis for the optimal choice of the HCAL granularity. It will also help to identify the crucial operational aspects and guide the development of the calibration and monitoring systems.

Genuine R&D needs to be performed for the optimization and consolidation of the tile photo-detector system. This must be matched with the development of the electro-mechanical concept of the active layer. At present there is no example of a scintillator calorimeter with integrated photo-sensors and front-end electronics. The R&D of the highly integrated front end components is a very ambitious program in itself, which is expected to largely benefit from the synergies with the ECAL electronics development.

### 5.2.1 The test beam prototype calorimeter

The goals of the AHCAL test beam prototype are twofold: on the technology side, the aim is to gain large-scale, long-term experience with a SiPM readout detector and to identify the critical operational aspects for further system optimization. On the physics side, the purpose is to collect the large data samples (order of $10^8$ events) needed to explore hadron showers with unprecedented granularity, validate hadronic shower simulation models and develop energy weighting and PFLOW reconstruction algorithms. Due to the current large model dependence of predictions for PFLOW-relevant shower properties it is indispensable to base the final detector optimization on real beam data. In that respect the term "prototype" may be misleading: the aim is a "proof of principle". Technical solutions scalable to a full detector design are only partially addressed and will be a subject for future R&D.

### 5.2.2 Detector structure

The AHCAL prototype is a 38-layer sampling calorimeter made of a plastic-scintillator steel sandwich structure with a lateral dimension of about 1 m$^2$. Each layer consists of 1.6 cm thick steel absorber plates and a plane of 0.5 cm thick plastic scintillator tiles housed in a steel cassette with two 2 mm thick walls. The total thickness corresponds to 4.5$\lambda$. The tile sizes vary from 3x3 cm$^2$ for 10x10 tiles in the centre of the module, to 6x6 cm$^2$ in an intermediate region and 12x12 cm$^2$ in the outer region, to limit the channel count and cost. In the last eight layers, the granularity is decreased to 6x6 cm$^2$ in the central region.

Each tile is coupled via a wavelength-shifting (WLS) fibre inserted in a groove to a SiPM via an air gap (Figure 5-2). The tile faces are covered with reflector foil; the edged were matted to provide diffuse reflection. Each tile has a 1 mm diameter wavelength-shifting (WLS) fibre inserted into a 2 mm deep groove. The fibre is coupled to a SiPM via an air gap. To increase the light yield, the other firer end is covered with a mirror (3M superradiant foil). The grooves have a quarter-circle shape in the smallest tiles and a full-circle shape in the other tiles. The sides of each tile are matted to provide a diffuse reflection and suppression of optical cross-talk between adjacent tiles, which is then limited to <2%. The tile faces are covered with 3M superradiant foil.

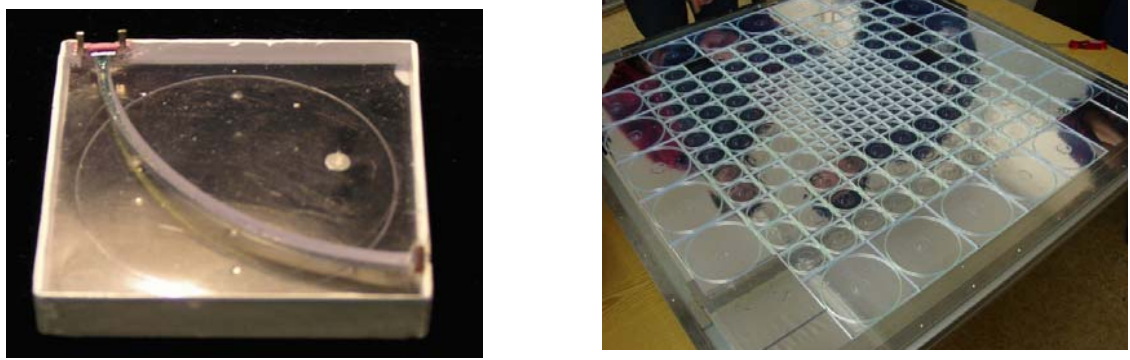

**Figure 5-2 Left: Scintillator tile (3x3cm$^2$) with SiPM. Right: Scintillator tile layer.**

Figure 5-2 shows the tile array laid out in the steel cassette. The size of the active area is 90×90 cm$^2$. In the next step of module assembly, the tiles are covered with a plastic (FR4) board which serves as a support for calibration light fibres and for the 1 mm micro-coax readout cables, which are connected to the SiPMs via small flexible PCBs. A fully assembled module with front end electronics connected is shown in Figure 5-3.

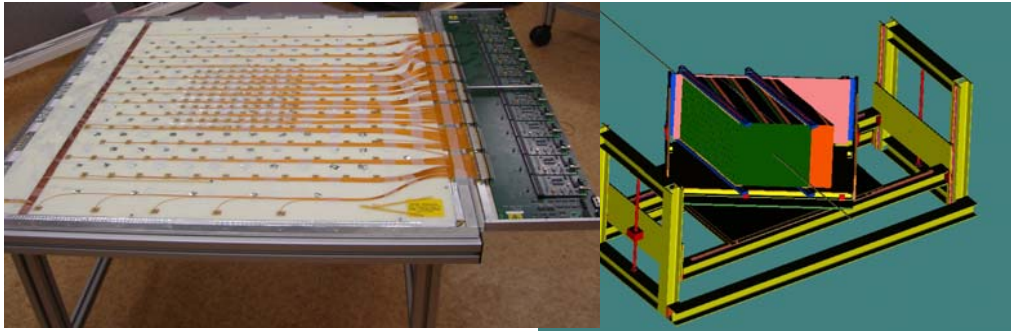

**Figure 5-3 Left: Complete module with front end electronics. Right: HCAL absorber stack on its movable table**

The complete modules are inserted into the absorber stack structure, which is shown mounted on top of a moving stage in the design drawing in Figure 5-3. The drawing shows the configuration set up for inclined beam incidence; the construction ensures that the beam still passes through the high granularity core in all layers for angles up to 35°. The stack and its support have been designed in a modular and flexible way which allows it to be adapted for beam tests with other active modules, for example with resistive plate chambers for tests of the DHCAL. A photograph of the fully assembled set-up is shown in the test beam section of this report.

### 5.2.3 Readout and calibration systems

The readout electronics consists of 18 channel front end ASICs, developed at LAL, with preamplifier shaper and sample-and-hold chain mounted on the cassettes, multiplexed and read-out by an off-detector VME based ADC system which was common for ECAL and HCAL. The ASICs provide a high gain operation mode for auto-calibration of the SiPM gain by observation of single photo-electron signals (Figure 5-4), and include DACs for the channel-by-channel adjustment of the SiPM bias voltage. More details on the ASIC are given in the electronics section.

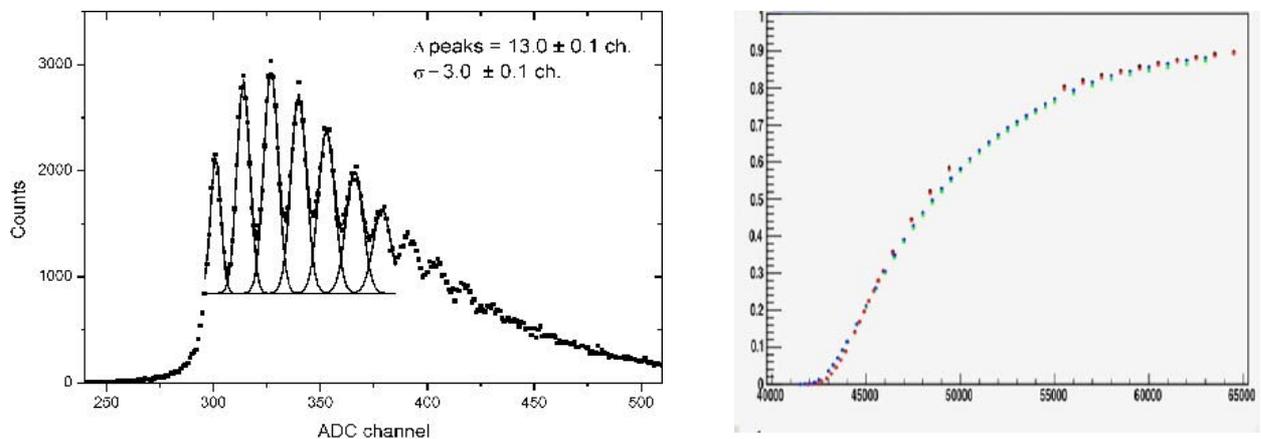

**Figure 5-4 Left: SiPM pulse height spectrum with single pixel signals. Right: SiPM response to LED light as function of LED voltage (arbitrary units).**

The test beam set-up has strongly benefited from the unified readout electronics concept for electromagnetic (ECAL) and hadronic calorimeter, which reduced the system integration effort to an absolute minimum and resulted in a combined and unified calorimeter system from the very beginning. This experience has lent strong support to the approach for future test environments and the final detector system.

An alternative option, based on fast digitization, is being developed at Dubna; its basic functionality was tested with a small scintillator cassette during the CERN test beam run.

A versatile LED calibration system (electronics developed at Prague) provides light signals up to an equivalent of 200 Minimum Ionizing Particle (MIP) to each tile (Figure 5-4). Low light intensities are used for gain calibration; intermediate intensities provide a PIN diode controlled reference for stability monitoring.

## 5.3 Silicon photo-multipliers

The SiPMs were developed, manufactured and tested in Russia. The photosensitive area (1.1x1 mm$^2$) holds 1156 pixels, each with a quenching resistor of a few MΩ. The detectors are reversely biased with a voltage of ~50 V and have a gain of ~10$^6$. SiPMs are insensitive to magnetic fields; this was tested up to 4 Tesla, For the HCAL prototype, more than 10000 SiPMs have been characterized in an automatic setup, with calibrated LED light. The bias voltage working point was chosen as the one that yields 15 pixels for a MIP-like LED signal. We measure gain, relative efficiency, dark rate, inter-pixel cross-talk, noise above a threshold of ½ MIP and the non-linear response function of fired pixels *vs* light intensity over the full dynamic range. Some distributions are shown in Figure 5-5.

Radiation hardness tests of SiPMs have been performed using a proton beam at the ITEP synchrotron. The dark current was observed to increase with the accumulated flux, such that individual pixel signals could not be observed anymore after an irradiation with about 10$^{10}$ protons/cm$^2$. At the ILC, fluxes above this value are only expected very close to the beam pipe. Ageing effects of long-term low-dose irradiation still need to be studied.

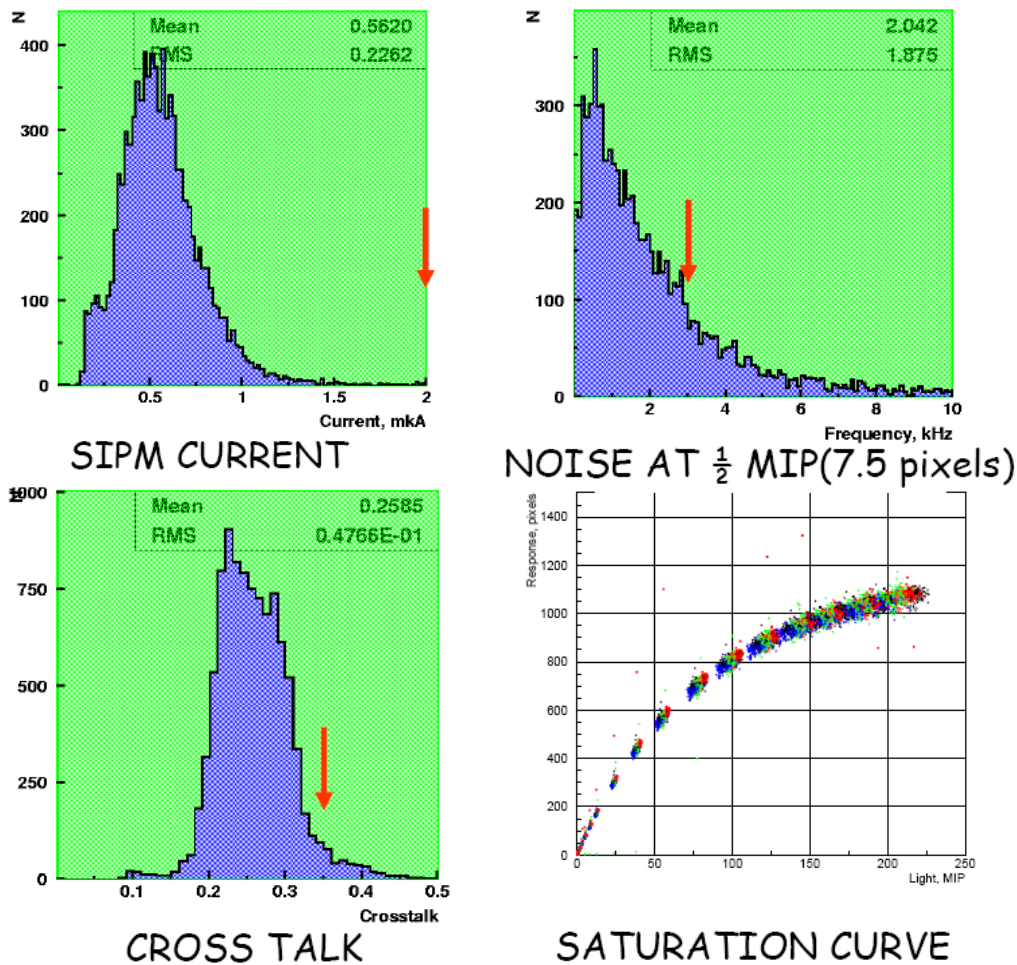

**Figure 5-5: Distributions measured during SiPM characterization: current, inter-pixel cross-talk and noise above nominal threshold of ½ MIP, response functions.**

## 5.4 Operational experience and calibration

The HCAL was assembled and commissioned at DESY, where also an initial calibration of the active layers was obtained in the electron test beam. In 2006, together with ECAL and TCMT, the stack with 23 instrumented layers was exposed to electron and hadron beams of 6-45 GeV and 6-120 GeV, respectively, in the H6 beam line at the CERN SPS, In addition high intensity muon beams were available for calibration. Detector operation proved very stable, with up-times above 90%, and more than 70 million events were collected in 2006.

All calorimeter cells have been calibrated with muons. The MIP signal $A_{MIP}$ is used as a scale for the deposited energy and to set the noise suppression threshold of ½ MIP, which yields a MIP hit efficiency of about 95%. The noise hit occupancy is then about $10^{-3}$, corresponding to ~0.5 GeV on the electromagnetic (em) energy scale. The gain $A_{pixel}$ is measured with low intensity LED light and used for non-linearity correction. The energy per cell in units of MIPs is then obtained from the formula

$$E\,[MIP] = A/A_{MIP} * F\,(N_{pixel}) \quad \text{with} \quad N_{pixel} = A/A_{pixel}$$

where A are amplitudes measured in ADC counts. The non-linearity correction $F$ depends only on the amplitude in units of pixels and is 1 for small amplitudes (all scale factors are absorbed in the MIP calibration factor). The function $F$ is the inverse of the normalized response function and can be approximated as $F = -N/N_{pixel} * \log\,(1-N_{pixel}/N)$ for a total of N active pixels on the SiPM. In practice $F$ is obtained from the test bench measurements. The conversion from MIPs to deposited energy depends on the incident particle type and is taken from simulations or using the known beam energy as reference.

MIP and pixel scale are subject to temperature variations of a few percent per Kelvin. The redundant monitoring system offers various possibilities for correcting these effects, using gain, LED reference signals or direct temperature measurement. The procedures are under development and not yet applied for the first analyses which use data from sufficiently stable periods.

The em shower data are used to validate the calibration and the detector understanding in terms of simulations. For this reason dedicated electron runs have been taken with the AHCAL alone. After correcting the non-linear response of the SiPM using the test bench data and the *in-situ* channel-by-channel gain calibration, the calorimeter response is shown to be linear up to 45 GeV, which is by far sufficient for the hadron shower measurement, see analysis section.

The hadron analysis has just started, data show reasonable qualitative agreement with expectations. Thanks to the high granularity a rich substructure can be observed, which can e used, for example, to determine event-by-event the electromagnetic energy fraction for weighting procedures, and to test particle flow reconstruction algorithms.

## 5.5 Future test beam programme

We presently envisage a continuation of the AHCAL prototype test beam programme including the following steps:

- In 2007 at CERN, the prototype ECAL and AHCAL will be fully instrumented and mounted on the movable stage for scans of the beam incidence angle; data taking will continue in the same beam line as used in 2006.

- In 2008, the same configuration will be exposed to test beams at FNAL, which will extend the energy range down to values as low as 1 GeV and provide the reference sample for comparison with gaseous active layers, e.g. RPCs, which can be installed in the same absorber structure later-on.

- The AHCAL can be tested with the scintillator ECAL in 2008. Production of the necessary additional AHCAL electronics components is under preparation.

- It is foreseen to exchange the steel absorber plates against lead sheets in order to test a compensating scintillator HCAL option, as proposed in the GLD concept.

- The time-sensitive very front end electronics under development for the next generation technical prototype can be used to also equip the present test beam module in order to test simulations of the time dependence of the response to neutrons in hadronic showers.

## 5.6 R&D towards a realistic tile HCAL for the ILC

In the next years, the technology needs to be developed and optimized further, in order to proceed from the pioneering test beam experience towards a realistic and cost-effective proposal for an ILC detector, which should be prepared by the end of the decade when first physics results will become available from the LHC. This challenge goes beyond a straight-forward engineering effort, since no example for a scintillator-based detector with embedded photo-sensors and front end electronics exists so far.

To meet the goal, further R&D is needed both for the active readout layer electro-mechanical design which must integrate the microelectronics readout chips and minimize the cost-sensitive thickness of the readout gap, as well as for the optimization of the scintillator SiPM system. Operational experience and analysis results from the ongoing test beam programme are expected to significantly influence this development.

### 5.6.1 Mechanical structure with integrated front end electronics

The goal is to develop a realistic mechanical structure and calibration concept, which together with the electronics and DAQ activities provide a scalable detector architecture for embedding upcoming new sensor technologies. This work is being pursued in the framework of the EUDET initiative.

Given the proof-of-principle purpose of the test beam prototype, its design is rather conservative, apart from the SiPM. A number of features cannot be translated into a full detector concept and need to be re-addressed in this task:

- The very front end (VFE) electronics components are mounted outside the detector volume. In order to minimize dead space and signal path lengths, the VFE ASIC chips need to be integrated into the layer structure.

- The thickness of the readout layer and the scintillator itself has not yet been minimized. The overall HCAL thickness has significant cost implications, as it has to fit into the bore of the superconducting solenoid of the ILC detector.

- The assembly is quite labour-consuming due to the signal routing via micro-coaxial cables. Printed circuit board technologies amenable at large scale mass production need to be applied instead.

- The prototype has a very versatile and redundant calibration system, in order to ensure reliable data taking with a not yet proven sensor technology, and to allow for comparison of different calibration and monitoring concepts. For a full-scale system, it is too complicated and must be simplified.

For the detector layout we presently limit our considerations to the barrel section. As an example, the subdivision into modules foreseen in the large detector concept (LDC), is shown in the figure below. The hexagonal structure ensures good filling of the volume with absorber material. The longitudinal division into only two half-barrels is preferred with respect to a larger number of barrel "rings" because of the reduced amount of dead material and the easier accessibility of readout electronics from the two end faces of the barrel sections. In this design the total thickness of the HCAL is about 1.1m, the modules are about 2.2 m long. For the scintillator transverse segmentation we assume that square tiles with a size of 3cm as optimized for the test beam prototype are also typical for the full detector. About 1000 to 2000 tiles per layer should thus form a realistic starting point for the expected channel density.

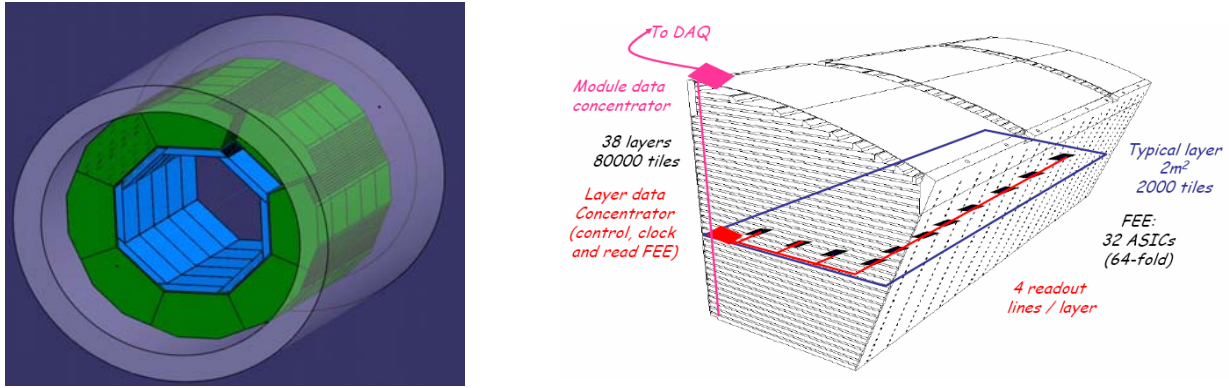

**Figure 5-6: Barrel section of the electromagnetic and hadronic calorimeter in the large detector concept (left), a half-octant HCAL sector (right).**

We consider these dimensions as indicative. The prototype structure must not exactly reproduce the dimensions of the final detector, but it should demonstrate that technical solutions exist for a number of issues:

- The structure must be compact, have minimal dead regions (cracks), be rigid and self-supporting in any of the required orientations and respect the tolerances for easy insertion of active layers. It should be strong enough to support the heavy silicon tungsten ECAL modules mounted on the first absorber layer.

- Electrical readout and control signals as well as optical calibration signals and electrical power must be routed reliably over the up to two metre long active layers, thereby respecting constraints from assembly procedures as well as requirements for maintenance accessibility.

- The connectivity issues at the barrel end (layer-to-layer connection and communication with off-detector components) can be realized within the very limited spatial constraints at the barrel end face, thereby incorporating readout and calibration systems, power supply and cooling.

- Conditions for heat dissipation must be realistic.

For the electronics architecture we again follow the ECAL concept. The VFE ASICs include the analogue stage (pre-amplifier and shaper, sample-and-hold circuit) and the ADC. The front end electronics which controls the VFE and communicates with off-detector components – see DAQ section – will be located at the barrel end. We foresee two stages of concentration, first layer wise, then module wise. The FE must also control calibration electronics and power cycling with respective components integrated on the layer FE board.

The barrel HCAL will contain about 2.5 million scintillator tiles on a surface of about 2500 m$^2$, the corresponding numbers for the two end caps together are about the same. The active layer structure must be compact and amenable at mass production techniques. For reasons of optical coupling stability and motivated by quality control chain considerations we regard the scintillator photo sensor system as a unit. This approach was successful in the test beam prototype effort. The issues to be addressed are

- The coupling between the scintillator and the photo-sensor is a research field in its own, driven by advances in the sensor technology. Important simplifications are possible with improved SiPM performance. We aim at 3mm scintillator thickness and an active layer design which leaves more than one option open for the choice of the photo-sensor and its coupling to the tiles.

- The tile – or tile array - positions must be controlled in a manner which accommodate mechanical tolerances resulting from cost-effective tile production techniques but minimize inefficient regions and still respect the precision requirements of the printed circuit board with its dense electrical signal lines connected to the SiPM.

- The surface of one module layer needs to be sub-divided into several PCBs; the interconnection must be realized in a way which allows exchange of PCBs in case of need for repair and still ensures high reliability. The subdivision must take the many different layer sizes into account.

- Optical or electrical calibration signal routing must be integrated.
- The thermal coupling between ASICs and absorber structure must ensure efficient heat dissipation.

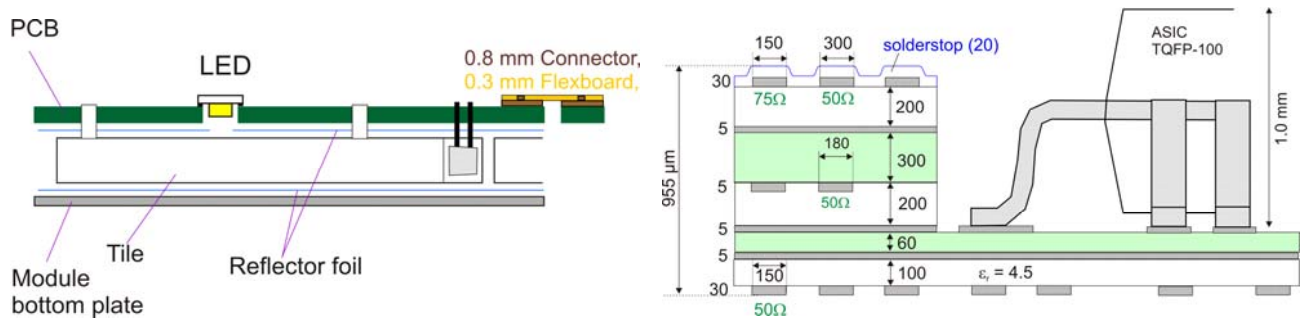

**Figure 5-7: Active layer cross section, with details for PCB, inter-connection, calibration light injection.**

A possible solution is sketched in cross-section above. The heat dissipation in such a structure has been estimated with analytic methods. With power-cycled VFE ASICs but permanent bias voltage supply to the SiPMs, a power consumption of 40μW per channel induces a gradient of 0.3K over the 2m length after typically 6 days of operation. This means that no active cooling is necessary for the bulk of the detector. The currents – 3A per layer during the bunch train – also seem to be manageable.

In an alternative approach, pursued by NIU, the photo-sensor is first integrated into the electronics PCB and only later optically coupled to the scintillator. This has the advantage that it can be included in the automated PCB assembly and soldering procedure.

The calibration system of the test beam prototype is highly redundant and provides a variety of possibilities to monitor the SiPM parameters. It is presently assumed that a much simpler system can be built which relies on the auto-calibration properties of the SiPMs alone. Since the response to individual photo-electrons provides the relevant scale, the light intensity does not need to be precisely stabilized or monitored with PIN diodes. Furthermore it requires only small and fixed light intensities to be injected within a comfortable amplitude range, thus relaxing the constraints on the optical or electrical signal distribution.

The development of small pre-prototypes for the validation of the design approach has started, this includes PCBs to test the LED pulsing near the SiPMs, a long PCB structure to test power pulsing power pulsing in realistic geometry, and a prototype front end board to test the interplay between SiPMs, the new SPIROC chip and the DAQ.

### 5.6.2  Scintillator photo-sensor system optimization

The sensor performance needs to be improved, in order to provide leeway to arrive at a simpler and at the same time more compact design. The performance limitations of the present SiPM, which are most relevant for calorimetry, are the dynamic range and the noise above threshold, where the emphasis is different for ECAL and HCAL applications. The noise depends on the dark rate, the inter-pixel cross-talk and the threshold determined by the light yield of the coupled SiPM scintillator system. The latter can be increased with larger or more efficiently coupled optical signals, but at the expense of dynamic range, or with larger sensor area, which increases the dark rate. Better intrinsic performance of the photo-sensors is needed and actually appearing as result of industrial R&D.

In the test beam prototype, the noise is just small enough to keep the occupancy low and isolated neutron signals distinguishable. Therefore for each channel the bias voltage working point must be exactly adjusted and the threshold calibrated. While this posed no problems in principle, larger operational tolerances would be desirable. The demand for thinner scintillators, to reduce detector volume and cost, sharpens the requirements further.

Simplified optical coupling, omitting, for example, the WLS fibre ("direct coupling"), also reduces the light yield. The blue-sensitive MPPCs (type 1600) were directly coupled to scintillator tiles from the prototype production and found to yield about 7-9 pixels / MIP, less than in the scintillator ECAL, which uses different scintillator material and geometry. This value is too small to ensure good MIP efficiency and leaves no room for thinning the scintillator. However, given the excellent low noise performance of the MPPC, larger area sensors would be possible in principle. Direct coupling also introduces considerable non-uniformity, the more, the thinner the tile (Figure 5-8), which must be compensated by optimizing the geometry. The exact requirements will be studied in simulations.

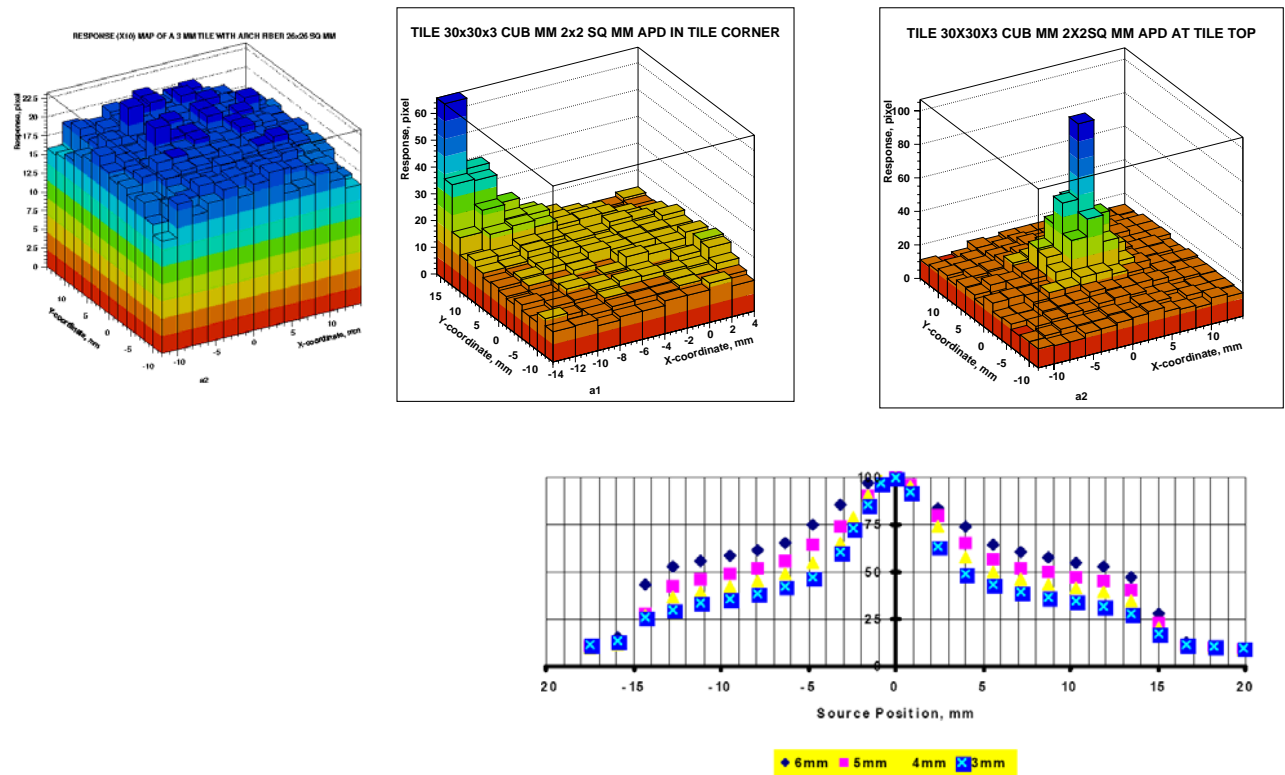

**Figure 5-8: Top: response uniformity of 3 mm thick scintillator tiles with photo-sensors coupled via an arch fibre, or directly at the corner or to the face of the tile (left to right, ITEP). Bottom: scan measurements through the centre for different tile thicknesses, photo-sensor in the centre (NIU).**

## 5.7  Conclusion

The novel multi-pixel Geiger mode photo-diodes open revolutionary detector design options with highly segmented scintillators. A scintillator steel HCAL with 3x3 cm$^2$ tiles was shown to meet the ILC performance goals in full simulation and particle flow reconstruction studies. A prototype with 8000 channels has been built and successfully operated in the CERN test beam. The results will be used to test the simulations and guide the further optimization of the detector concept. In the next years, more R&D is needed to further consolidate and optimize the technology, and to proceed towards a realistic detector design.

# 6 The Tail Catcher and Muon Tagger (TCMT)

## 6.1 Introduction

The CALICE Collaboration has successfully completed the construction and operation of a cubic-meter sized scintillator-steel device, which serves as both a tail-catcher and muon tracker (TCMT). The ~ 6λ thick prototype, constructed in partnership between DESY, Fermilab and Northern Illinois University, was designed with this dual purpose in mind. The TCMT prototype has a fine and coarse section distinguished by the thickness (2 and 10 cm respectively) of the steel absorber plates. The fine section sitting directly behind the hadron calorimeter and having the same longitudinal segmentation as the HCAL will provide a detailed measurement of the tail-end of the hadron showers which is crucial to the validation of the hadronic shower models, since the biggest deviations between models occur in the tails. The following coarse section serves as a prototype muon system for any design of a Linear Collider Detector and will facilitate studies of muon tracking and identification within the PFA framework. Additionally, the TCMT will provide valuable insights into hadronic leakage and punch-through from thin calorimeters and the impact of the coil in correcting for this leakage.

## 6.2 TCMT Prototype

The active layers of the TCMT consist of 1 m long, 5 cm wide and 5 mm thick extruded scintillator strips. The extruded scintillator strips were produced at the Scintillator Detector Development Lab extruder facility operated jointly by Fermilab and Northern Illinois University (see Figure 6-1). A 1.2 mm diameter Kuraray WLS fiber is inserted into the co-extruded holes that run along the length of the strips. Not only was the performance of this novel fiber-coextruded-hole configuration better than the conventional fiber-machined-groove geometry it was also significantly efficient for assembly since no machining, polishing or gluing was involved. The fiber ensconced in the coextruded hole is then mated to a Silicon Photomultiplier sitting at one edge of the strip. The strips and their associated SiPMs in each layer are enclosed in a light tight sheath or cassette (see Figure 6-1). The top and bottom skins of the cassette are formed by 1 mm thick steel with Al bars providing the skeletal rigidity. The aluminum bars also divide the cassette into distinct regions for scintillator, connectors, cable routing and LED drivers such that they can be independently accessed for installation, maintenance or repairs. The cassettes are then inserted, alternately in the X and Y orientation, in the absorber stack. The stack is constructed by welding the steel absorber plates to a frame that also doubles as a lifting fixture. The structure is then placed on a table capable of forward-backward and left-right motion with the help of Hillman rollers riding on steel rails (see Figure 6-2). The electronics boxes are attached to the stack to keep the cable lengths to a minimum.

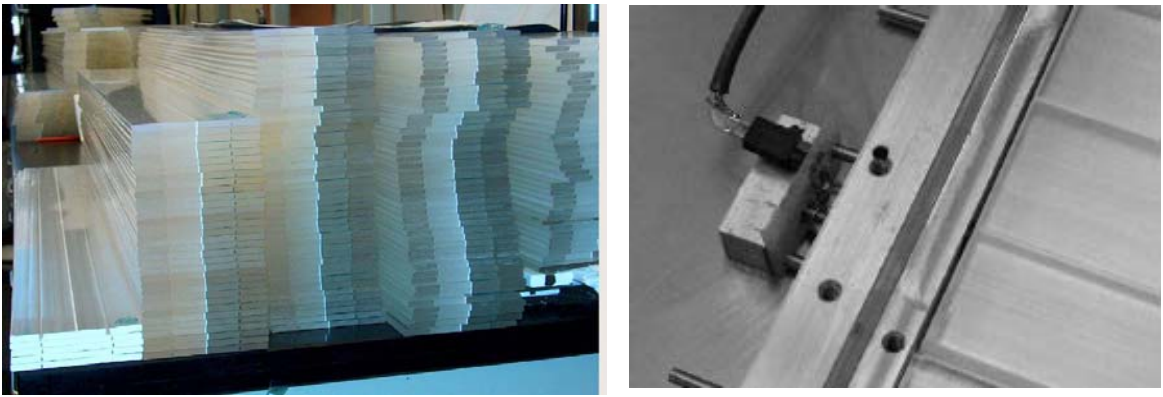

**Figure 6-1 Left: Scintillator strips used in the TCMT. Right: A TCMT cassette.**

## 6.3 TCMT Readout

Since SiPMs served as the photodetectors, much of the electronics developed for the scintillator-based hadron calorimeter could be used for the TCMT. However, the different structure and channel count of the device necessitated the development of some custom interfaces to the AHCAL electronics and DAQ (see Fig. 4). Coax cables from each of the 20 SiPMs in a layer connect to multi-coax connectors at the edge of the cassettes. 1.5m long multi-coax cables then go from the cassettes to adaptor boards which are interfaced to HCAL base boards which in turn communicate with the VME based DAQ.

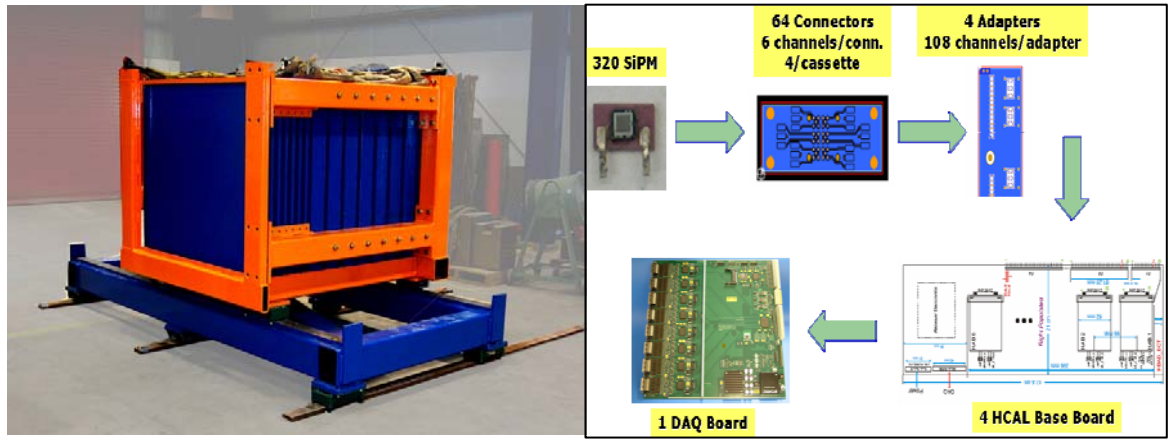

**Figure 6-2 Left: TCMT absorber stack and table. Right: TCMT readout chain.**

## 6.4 LED System

Each cassette carries inside it a LED driver board carrying 20 UV LEDs, i.e. one per strip. The driver boards are placed inside the cassette, on the end opposite to where the strips are read out by the SiPMs. The boards communicate with the outside world with a ribbon cable that is connected to a fanout board. The fanout boards receive the trigger and amplitude signal from the HCAL baseboard and distribute them to four cassettes. Thus four fanout boards are needed to cater to all the sixteen layers of the TCMT. Controlled by the DAQ the LED system provided light of the desired amplitude to provide gain calibration and stability monitoring.

## 6.5 Test Beam

A fully equipped TCMT took beam in the H6B beamline area at CERN as part of the CALICE testbeam in August and October of 2006. Preliminary look at the muon and pion data collected indicates that the detector performed reliably and to specifications in its dual role as a tail-catcher and muon tracker.

# 7 The gas Digital Hadron Calorimeter

## 7.1 Introduction

The Digital Hadron Calorimeter (DHCAL) effort develops gaseous calorimeters with very fine granularity. These calorimeters are conceived to optimize the application of Particle Flow Algorithms to the measurement of hadronic jet energies. The large number of readout channels allows reducing the resolution to a single-bit readout/channel without compromising the single particle energy resolution of hadrons. The granularity of the readout for all developments is currently chosen to be 1 x 1 cm$^2$ within a layer and layer-by-layer in depth.

The group investigates three different active media: Resistive Plate Chambers (RPCs), Gas Electron Multipliers (GEMs) and Micromegas. In addition, two readout schemes, one based on the DCAL chip (developed by Argonne and Fermilab), the other one based on the HaRDROC chip (developed by the French groups of LAL,LLR ,IPNL) are being pursued.

The project initiated with R&D on the active media and the development of a readout scheme capable of handling a large number of channels. Whereas the R&D on RPCs is virtually completed, further characterizations of both GEMs and Micromegas are needed. The R&D on the detector elements will be followed by different experimental setups in order to validate both the active elements and their readout system under test beam conditions. One of these setups is the so-called Vertical Slice Test (VST), being assembled by the US groups. The VST is limited to a small number of chambers, but involves the entire readout chain, as conceived for the readout of larger prototypes (or the HCAL of the ILC detector). Another setup is under preparation by the European groups. It will be made of 1 m$^2$ detector fully equipped by the new electronics readout generation in order to be exposed to a hadronic beam.

Based on the success of these tests the groups will embark on the construction of a 1 m$^3$ Prototype Section (PS), with 40 active layers each with an area of 1 m$^2$, followed by an extensive test program in the Fermilab test beam. The PS will re-utilize the absorber stack and movable stage of the CALICE-scintillator HCAL PS. Detailed measurements in the Fermilab test beam will be essential to prove the concept of a DHCAL, to validate the technological approach to fine granularity gaseous calorimetry, and to provide the data to be compared with the various hadron shower simulation models currently available.

Further R&D on both the detector elements and the electronic readout system will be necessary for the design of a HCAL for an ILC detector. This R&D will lead to the construction of a scalable HCAL prototype wedge with integrated readout electronics and gas & high/low voltage supplies.

In the following we summarize the status of the various parts of the project.

## 7.2 Detector R&D

RPCs, GEMs and Micromegas are being developed as active elements of a DHCAL. While RPCs are cheap to build, reliable to operate (with glass as resistive plates), their single particle detection efficiency decreases substantially with rates of a few 100 Hz/cm$^2$. On the other hand both GEMs and Micromegas can handle high rates, but their construction involves expensive foils (in the case of GEM) and their operation is somewhat more delicate. Since the concept of a DHCAL is a novel approach, the group feels that the development and comparison of performance of different active elements is a worthwhile undertaking.

### 7.2.1 Resistive Plate Chambers

RPCs have been investigated and developed independently by the Protvino and Argonne groups. The two efforts compared their findings in frequent meetings, have come to similar conclusions and propose similar chamber designs for the PS.

The resistive plates are chosen to be glass plates with a thickness of approximately 1.2 mm and a bulk resistivity $\rho \sim 5 \cdot 10^{12}$ Ωm. The glass plates enclose a gas gap of 1.2 mm. The outside of the glass plates is coated by a resistive paint with a surface resistivity $R_\square \sim 1 \div 50$ MΩ/□. The chambers are flushed with a mixture of R134A (93-95%), Isobutane (5%) and Sulfur Hexa-fluoride (0.2-0.5%) and are operated in avalanche mode. Under these operating conditions no aging effects have been reported so far.

The groups measured all relevant performance characteristics, such as the single particle detector efficiency, pad hit multiplicity, lateral charge distribution, performance in magnetic fields (Protvino group), rate capability, mechanical properties (deformations with pressure and electric fields), etc. for a variety of gas mixtures and chamber designs. The tests were performed with cosmic rays and also with particle beams at Protvino and Fermilab. As an example, Figure 7-1 (left) shows the signal charge as function of high voltage for a 2-gas gap chamber. Note that above 8.2 kV the signals include streamers with significantly larger charges, and Figure 7-1 (right) shows the charge radius as measured with 1 x 1 cm$^2$ readout pads and a high-resolution data acquisition system. At a distance of 2 cm from the pad hit, the collected charge is compatible with zero.

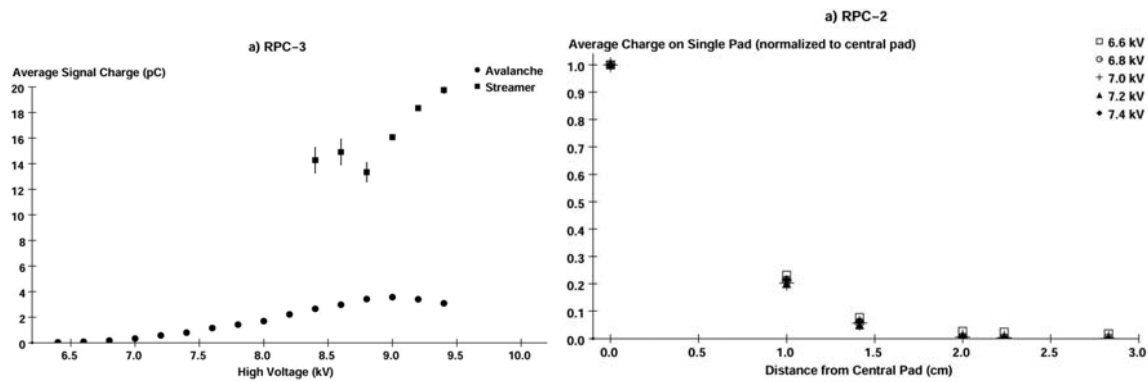

**Figure 7-1 Left: Signal charge as function of applied high voltage for a 2-gas gap chamber. Right: Signal charge as function of distance from the pad hit.**

Figure 7-2 (left) shows the measurement of the single particle detection efficiency and fraction of streamers versus high voltage (as measured by the Argonne group) and Figure 7-2 (right) compares the measurement of the same efficiency with chambers operated in zero and 5 T magnetic fields with different orientations (as measured by the Protvino group). Note that, presumably due to the high electric fields in the chambers, no significant effect due to the magnetic field is observed.

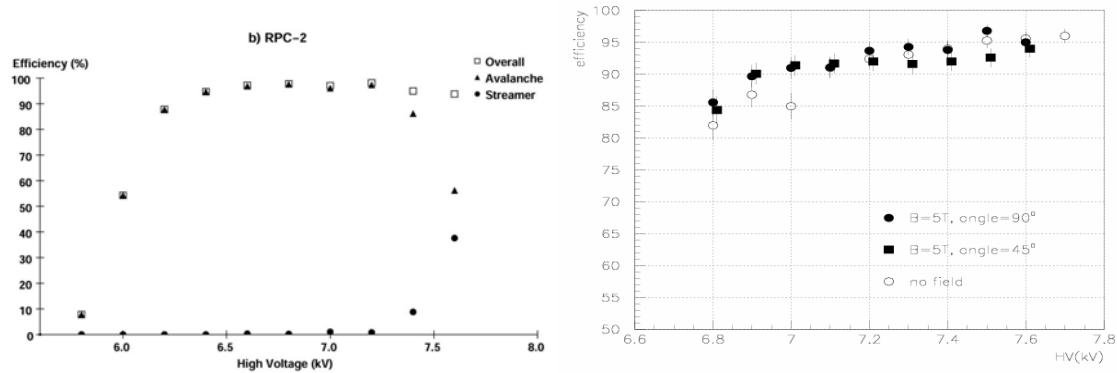

**Figure 7-2 Left: Single particle detection efficiency and fraction of streamers versus applied high voltage. Right: Single particle detection efficiency versus high voltage for chambers operated in a magnetic field of zero or 5 T.**

Figure 7-3 (left) shows the pad multiplicity versus single particle detection efficiency for two different chamber designs. Configuration 1 corresponds to a 2-glass RPC, whereas configuration 2 represents a 1-glass RPC, where on the anode side the gas volume is defined by the pad board. Finally, Figure 7-3 (right)

shows the single particle detection efficiency versus particle rate. In avalanche mode the efficiency is observed to decline for rates exceeding ~300 Hz/cm$^2$.

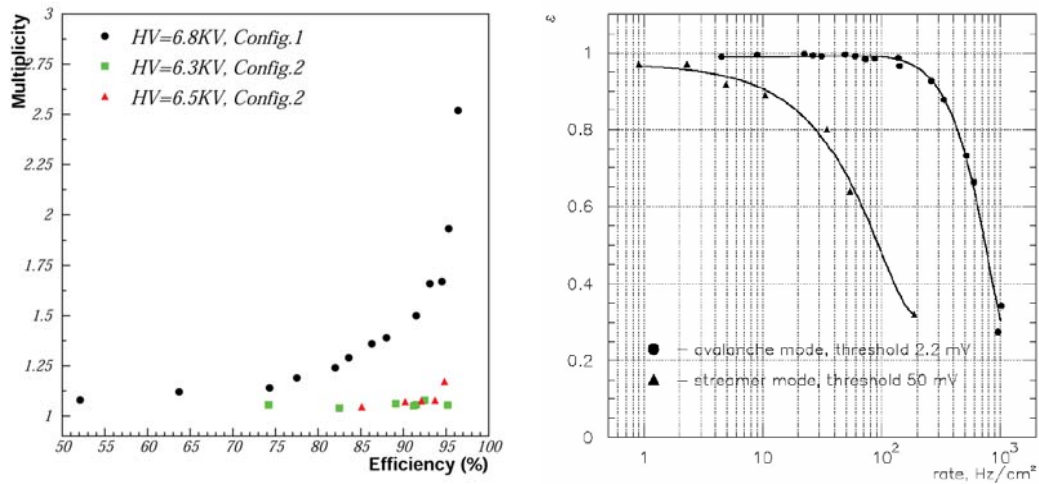

**Figure 7-3 Left: Pad multiplicity versus single particle detection efficiency for 2-glass plate RPCs (configuration 1) and 1-glass RPCs (configuration 2), as measured by the Argonne group. Right: Single particle detection efficiency versus particle rate (Protvino group).**

All results by the Protvino and Argonne groups are in excellent agreement. At this point both groups consider this R&D phase complete, apart from ongoing long-term stability studies for a novel chamber design consisting of only one glass plate. For more information on the long list of measurements performed by the two groups, see references [1-3].

### 7.2.2 Gas Electron Multipliers

The demonstrated high efficiency, high rate capability, robust operation, thin active layer, and flexibility in design, make GEM-DHCAL an excellent choice for an ILC hadron calorimeter. Figure 7-4 shows how a double-GEM structure with on-board analog-to-digital electronics would be configured in a calorimeter.

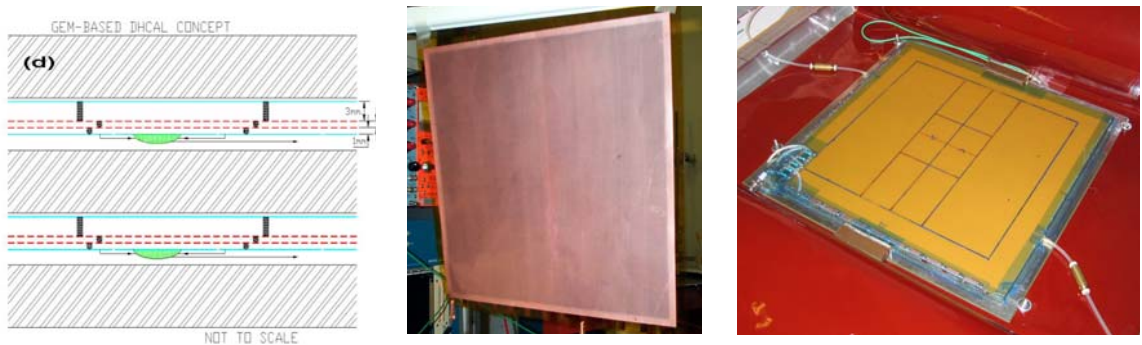

**Figure 7-4 Left: GEM-DHCAL scheme. Middle: 30x30cm$^2$ foil. Right: Completed GEM chamber**

Initial studies used 10cmx10cm GEM foils purchased from the GDD group at CERN. From Sr90 radio active source and cosmic ray, the gain of the double-GEM chamber was determined to be of the order of 3,500, consistent with measurements by the CERN GDD group. The MIP efficiency was measured to be 94.6% for a 40 mV threshold, which agrees with our simulations which also demonstrated GEM based DHCAL performance for PFA. The corresponding hit multiplicity was measured to be 1.27. A gas mixture of 80% Ar/20% $CO_2$ has been shown to give very stable chamber operation, and give an increase in gain of a factor of 3 over the original 70% Ar/30% $CO_2$ mixture. A minimum MIP signal size of 10 fC, and an average size of 50 fC were observed using this mixture. 30x30cm$^2$ foils have been developed for us by 3M Corporation. Figure 7-4 (middle) shows one of these foils which have been subjected to extensive HV testing. Figure 7-4 (right) shows a completed 30x30cm$^2$ chamber. Figure 7-5 shows an anode board with 1x1cm$^2$ pads. We have made a series of such chambers, initially read out using a QPA02 chip from Fermilab PPD. One chamber was exposed to a high intensity, 10 MeV, electron beam at KAERI (Korea) and continued to work well after an exposure of $1.2 \times 10^{-2}$ mC/mm$^2$, a charge density much larger than that expected from 10 years of operations at the ILC.

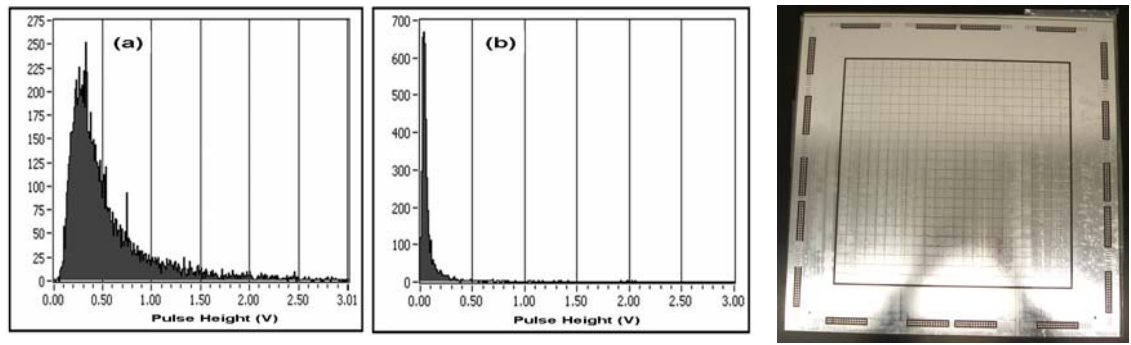

**Figure 7-5: Test beam signal on (left) primary pad, (middle) adjacent pad. Right: Anode pad board.**

Another chamber was tested at the Meson Area Test Beam at Fermilab in April 2007. Figure 7-5 shows the signal from the primary and adjacent pads for 120 GeV protons, respectively. Analysis of data for efficiency, hit multiplicity, cross-talk, as functions of HV and threshold, and rate capability is in progress. Finally, we are assembling GEM chambers to be read out using two newly developed chips: the DCAL chip developed at ANL/FNAL for RPC readout, and the KPiX chip developed at SLAC for the SiD ECal. Both chips have been adapted for GEM use. Each DCAL chamber will use 4 chips and cover a 16x16cm$^2$ area, while each KPiX chamber will use one chip and cover 8x8cm$^2$. Both types of chamber will be tested at Fermilab in the upcoming joint vertical slice test.

Finally, we are working with 3M to produce ~1mx30cm foils to be used in prototype chambers as a precursor to assembling 40 layers of 1mx1m for a 1m$^3$ GEM stack late in 2008.

### 7.2.3 Micromegas

Only recently Micromegas were added on the list of possible active elements for the DHCAL. The Micromegas is similar in principle to GEM but needs lower high voltages and is made of only one stage which means it has a reduced thickness with respect to GEM.

The LAPP and IPNL groups, helped by the CERN Micromegas team and the Saclay group have assembled their first chambers. These 3.5 mm thickness chambers were built using the new bulk technology which allows building big surfaces of such a detector in an industrial way with reduced cost. Both groups are presently testing the Micromegas chambers with cosmic rays.

## 7.3 Electronic readout system

While the development of the active elements is important, the real challenge of the DHCAL project is the development of an electronic readout system, which can handle the large number of channels (~5 · 10$^7$) of an ILC HCAL in a cost effective way. Furthermore, particular care has to be dedicated to the minimization of the digital to analog cross talk in the front-end boards, as the latter are located directly on the chambers. The group is currently pursuing two independent efforts for the development of the readout system, but plans to merge the two into a single effort for the construction of the PS are being discussed.

### 7.3.1 The DCAL-based readout system

The readout system is based around the DCAL front-end chip and utilizes a strategy very similar to the proposed second generation DAQ systems for the EUDET modules, see the relevant section of this report. The system is conceived for the readout of both RPCs and GEMs/Micromegas and consists of six parts, see Figure 7-6.

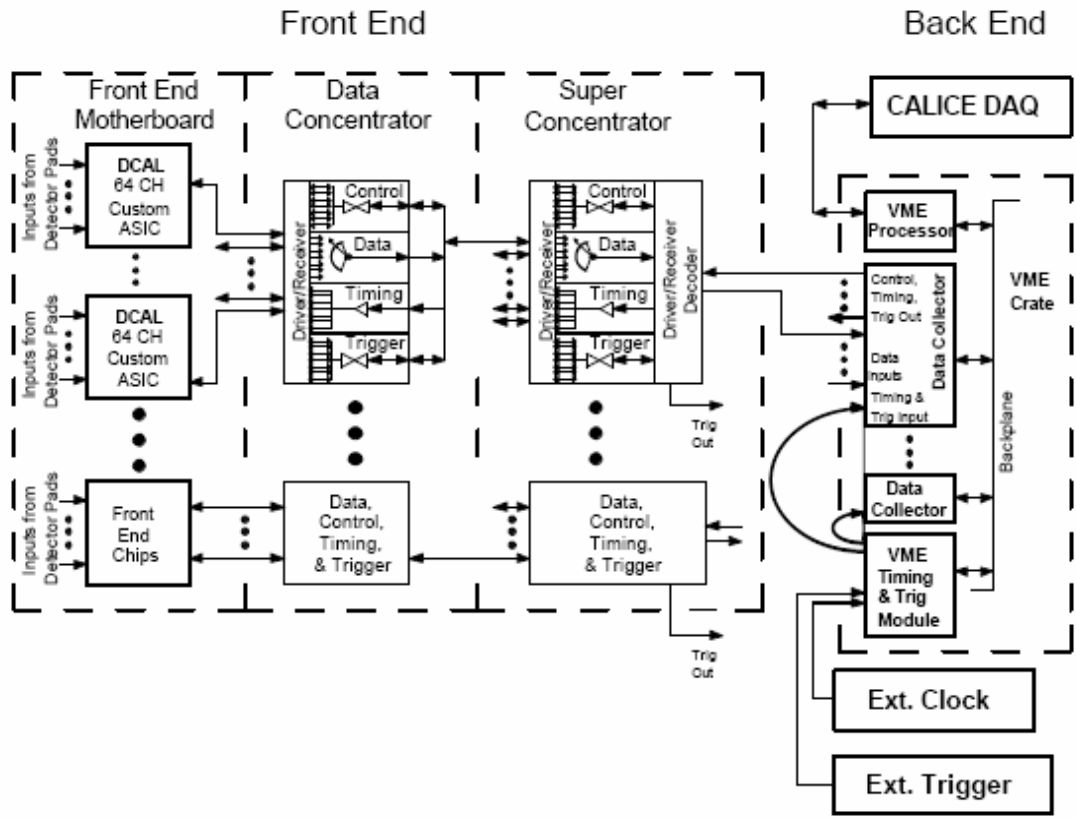

**Figure 7-6: Schematic of the DCAL based electronic readout system for the DHCAL.**

The components included in the dashed rectangle on the left in Fig.5.7 i.e. the front-end ASIC (DCAL), the pad- and front-end boards, the data concentrator boards and the super concentrator boards, are located on the detector, whereas the back-end, i.e. the data collector boards and the trigger and timing unit, can be located off the detector. In the following we briefly describe the various components of the system and the status of their development.

### 7.3.2 The DCAL Chip

The DCAL chip was designed by Argonne and Fermilab, serves 64 channels and provides an output with hit pattern and timestamp. (The design was based on extensive measurements the response to cosmic rays of RPCs with both an analog and a digital readout system.) The chip can be operated in triggered or triggerless mode. The gain of the preamplifiers can be adjusted to accommodate either the large (0.1 ÷ 10 pC) or smaller (0.02 ÷ 2 pC) signals of the RPCs or GEMs, respectively. The threshold of the discriminators is adjustable between about 5 fC and 700 fC and is common to all 64 channels. The chip has been prototyped and underwent extensive testing. Apart from two minor problems, the chip performed as expected. The channel-to-channel cross talk, for instance, was measured to be <0.4% and is expected to have no impact on its readout of RPCs or GEMs. The chip will not need to be prototyped again and is ready for production for the PS.

### 7.3.3 The Pad- and Front-end Boards

The pad-boards contain the signal pads, located close to the outer surface of one of the glass plates in RPCs. The boards count four layers in order to shield the analog signals from cross-talk from the digital lines and from noise pick-up from other sources, but feature no active components. The signals are routed to smaller pads on the opposite side of the board, which are then connected to the corresponding pads of the front-end board (the latter house four DCAL chips). The signal pads on the two boards are connected with conductive epoxy. The split into pad- and front-end board offers the distinct advantage of avoiding buried vias and the possibility of decoupling the size of the two boards. Thus the pad boards can be produced with a large area (32 x 48 $cm^2$ is needed for the PS), while the front-end board can be kept to a smaller and manufacturable size (16 x 16 $cm^2$). The pad boards have been fabricated (for the VST only an area of 16 x 16 $cm^2$ is needed, but the extension to larger areas is highly trivial). The front-end boards contain eight layers, where some layers are connected via blind vias. The boards have been designed by Argonne, fabricated, assembled and successfully tested. Figure 7-7 shows a photograph of the board.

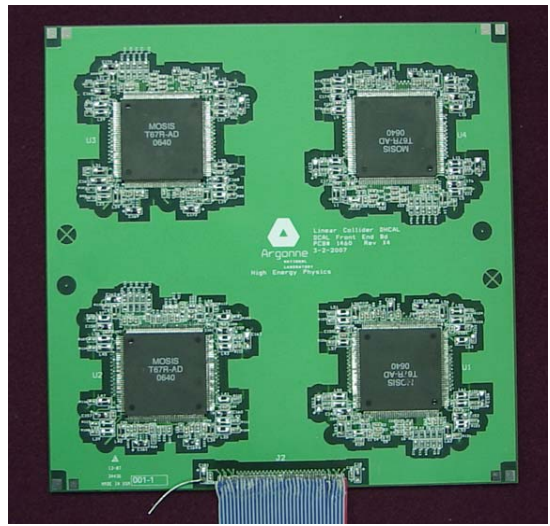

**Figure 7-7: Photograph of the front-end board with four DCAL chips mounted on it.**

The gluing procedure for connecting the pad and front-end boards has been established. For the VST the gluing will be performed by hand, but for the PS an automated gluing procedure will be developed.

### 7.3.4 The Data Concentrators

The data concentrators (DCON) read out 12 DCAL chips and provide the output to the super concentrator. The boards have been designed by Argonne and will be located on the side of the PS. For the VST a DCON with only four inputs has been fabricated and assembled. Standalone tests are ongoing and will be followed by integrated tests with the front- and back-end of the system.

### 7.3.5 The Super Concentrators

The super concentrators read out six DCONs and route the signals to the data collectors. The design of the super concentrators is similar to the DCON. For the VST, involving only of the order of 10 DCONs, the super concentrators have been omitted. The signals from the DCON will be fed directly into the data collectors

### 7.3.6 The Data Collectors

The data collector boards accept inputs from up to 12 super concentrators (or 12 DCONs in the VST). The boards are housed in a VME-crate, which can be located off the detector. The boards have been designed by Boston University, fabricated, assembled and successfully tested in standalone mode.

### 7.3.7 The Trigger and Timing Module (TTM)

The TTM provides the clock and trigger signals to the front-end of the system via the data collectors. The system requires only one such board. The TTM has been fabricated and is being assembled.

### 7.3.8 The DAQ and offline analysis software

The data acquisition software for both the VST and the PS is being written within the framework provided by the CALICE software. The DAQ software is being used for both the commissioning of the different parts of the readout system and for the data acquisition with cosmic rays and in the Fermilab test beam. The offline analysis will provide an event display and the tools to determine both the single particle detection efficiency and the pad multiplicity. Both efforts are well advanced.

In summary, the readout system is well advanced (with prototypes of all but one subcomponent in hand) and first cosmic ray tests with RPCs and GEMs will occur within the next month, followed by a test program of a mini-DHCAL (consisting of approximately 10 layers interleaved with steel-copper absorber plates) in the Fermilab test beam (an MoU between our group and Fermilab is being signed). Assuming the success of the VST, the construction of the PS will commence immediately with the construction of chambers in both Argonne and Protvino and with the production of the various subcomponents of the electronic readout system.

## 7.4 The HaRDROC-based readout system

The French groups under the leadership of IPNL (Lyon) are developing an electronic readout system based on the HaRDROC chip (see Figure 7-8). The latter is an adaptation of the MAROC chip; previously developed and successfully tested for the ATLAS experiment. The chip currently provides both analog (12-bit) and digital (single-bit) information for 64 channels, Thanks to its gain range (0-4), the charge that this chips can read is between (.01 and 6 PC) which covers the different detectors mentioned above. Although this two-output scheme allows to determine efficiency the value for the appropriate digital thresholds, the future versions of HaRDROC will retain only the digital part with most probably a 2-bit output per channel aiming to have a semi-digital readout for better energy measurement. An important feature of this chip is the fact that it is power pulsed, thus eliminating the need for active cooling (less than 20 μW/channel). The chip has been prototyped and successfully tested with a cross-talk from channel to channel less than 2%. An 8-layer board (containing both buried and blind vias) has been designed to minimize the cross talk between digital and analog signals. Prototype boards conceived to house four chips and the readout system, have been produced, see Fig.9. Although, the Board thickness is 0.8 mm, the cross-talks between the different PCB pads were measured and found very low as expected (<.3 %). Tests with prototype RPCs and Micromegas are forthcoming. These prototypes will be readout using FPGA and USB device.

The success of these prototypes will allow to build a 1 m² RPC/Micromegas detectors fully equipped. The comparaison between the two detectors in beam conditions will be of great importance for the detector choice in the future.

To achieve this, the French groups of IPNL,LAPP and LLR have started to work on data concentrator design capable to address 1 m$^2$ detector (10000 channels) and a super data concentrator for the 1 m$^2$ prototype. This study is realized in collaboration with the UK groups of CALICE involved in the future DAQ design (see the appropriate section)

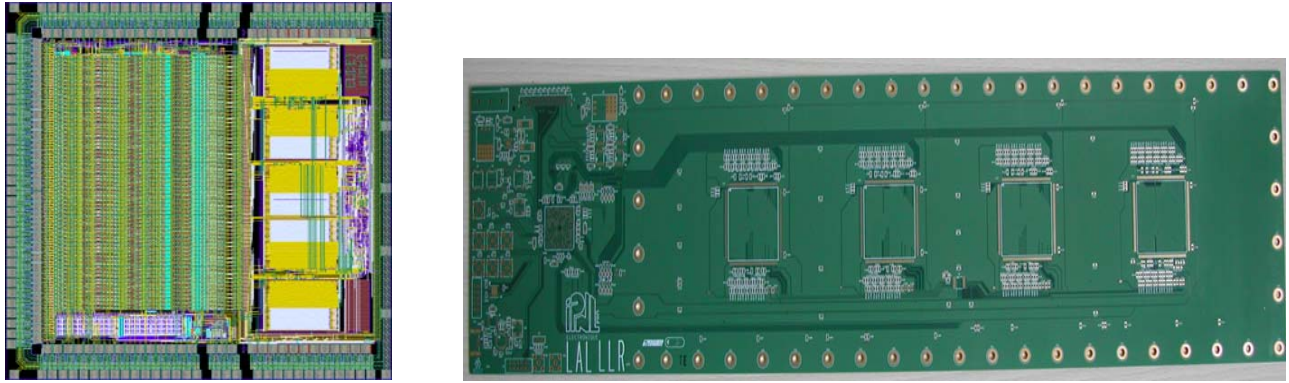

**Figure 7-8: Left: The HaRDROC chip. Right: Front-end board to house 4 HaRDROC chips.**

## 7.5 Developments beyond the Prototype Section

While the designs of the RPCs and the DCAL based readout system are quite close to what could be used in an actual ILC HCAL, further optimizations of both the chambers (the designs of the GEMs and Micromegas are not yet far enough to feature in this part of the report) and the electronic readout system are necessary. The following areas require further exploration, where the experience gained in the mean time with the RPC/GEM stack in the Fermilab test beam will be crucial to define the details of this future research program:

1. Chamber design: test of thin chambers with only one glass plate or of other 'exotic' chamber designs.
2. Long term tests: testing and monitoring of chamber performances over longer periods (several years) to ensure stability over the expected life time of the ILC detector.
3. Neutron sensitivity: explore the possibility of increasing the sensitivity to low energy neutrons (if results from test beam show a need for this).
4. Finer segmentation of the readout (if results from test beam show a need for this).
5. Higher multiplexing of the readout system, to reduce cost and real estate.
6. Thinner readout boards, to reduce the overall thickness of the active element.
7. Exploration of new techniques to eliminate the need for active cooling.
8. Anything else which might be recognized as important from the experience gained in the test beam.

The Argonne group also initiated the development of a concept for a gas-based barrel HCAL with calorimeter wedges containing steel plates of 20 mm thickness, see Figure 7-9. The barrel HCAL is subdivided into three barrels in z, where each barrel contains twelve wedges. The steel plates in a given wedge are held by so-called picture frames located at the end of the modules. These frames provide space for the routing of the gas, high/low voltage and signal lines. The data concentrators (or their equivalents) also will be located in this area.

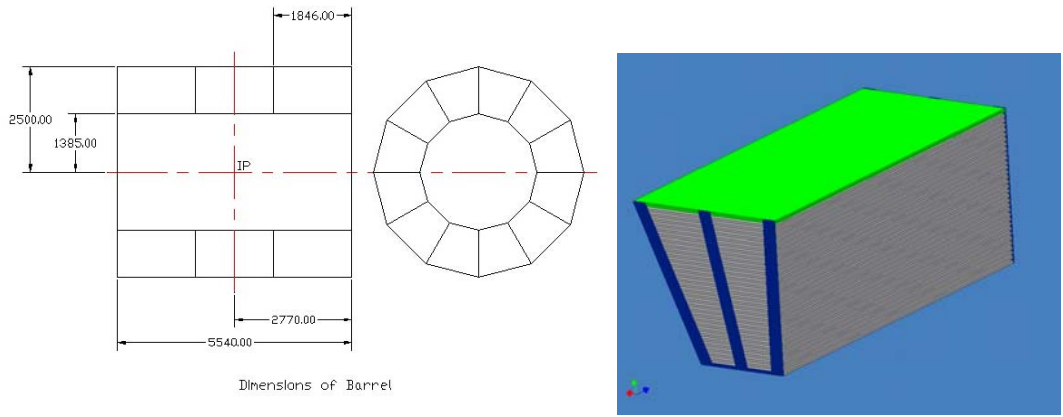

**Figure 7-9: Concept of a gas-based barrel HCAL. Left: The view along the z-axis showing the three barrels, middle) cross section of the barrel HCAL with 12 wedges. Right: A wedge containing steel plates held by 'picture frames'.**

## 7.6 Conclusions

The group explores the novel idea of a PFA optimized, digital hadron calorimeter with a gaseous readout element. RPCs (well advanced and performed independently by two groups), GEMs (ongoing characterization) and Micromegas (new effort) are being evaluated as active elements. An electronic readout system capable of handling large numbers of channels has been developed and prototyped (An alternative front-end is being developed by the French groups). A vertical slice test of the system is ongoing and will lead to measurements with a mini-DHCAL in the Fermilab test beam in June and July 2007. Tests of 1 m$^2$ RPC/Micromegas detectors equipped with French electronics are forseen in few months. Assuming the successful conclusion of those setups, construction of a 1 m$^3$ prototype section with approximately 400,000 readout channels will commence (partial funding for the construction is in hand in both US and Europe). The latter will be essential to prove the concept of digital hadron calorimetry, to validate the technological approach to fine grained gaseous calorimetry and to provide a basis for selecting the best performing hadron shower simulation model.

## 7.7 References

1. V.Ammosov *et al.*, Nucl. Instrum. Meth. **A533**, p. 130-138 (2004)
2. "Resistive Plate Chambers for Hadron Calorimetry: Tests with Analog Readout", G.Drake et al., to appear in Nucl. Instrum. Meth.
3. http://www.hep.anl.gov/repond/DHCAL_US.html and links therein.

# 8 The VFE development

## 8.1 Introduction

The front-end electronics for CALICE is split in two families : one first generation of ASICs to readout the physics prototypes and a second generation for the technological prototype. For the first generation of ASICs, the goal was to emphasize detector characterization and physics performance in test beam and not to use the beam to do electronics developments, therefore several key features have been left aside for the technological prototype, such as low power consumption , high level of integration or sparsified readout. These chips are :
- FLCPHY3 for the Silicon Tungsten ECAL
- FLC_SiPM for the iron-scintillating tiles-SiPM Analog HCAL,
- DCAL for the iron-RPC digital HCAL

The first two chips have been produced in 2003-2004 to equip the 10,000 channels of the physics ECAL and AHCAL prototypes. The third, which is needed in much larger quantity has been delayed for funding reasons and should be produced in 2008. These chips have met the requirements in terms of performance to do efficient testbeam analysis and no major changes are foreseen for the analog front-end for going to the second generation of ASICs.

The second generation of chips has started in 2005 in order to produce ASICs ILC-compatible to equip large scale detectors such as the EUDET modules, close to a module0. There the emphasis has been set on the very high level of integration (see Figure 8-1), which is a new and challenging feature of ILC calorimeter electronics : what used to occupy racks and racks of electronics around detectors will now be embedded inside the detector and only send out zero-suppressed digital data. It is thus essential for ILC to establish the feasibility of this high level integration and operate it on prototypes in testbeam. This constraint also leads to drastic reduction of power dissipation: ILC calorimeter electronics will dissipate 10E4 times less power than LHC electronics! Last, but not least, the scalability or ability to go to mass production must be taken into account at the earliest stage: there is no point in developing electronics that cannot be afforded or mass produced for the final detectors.

Several features of the readout electronics are common to the electromagnetic and hadronic calorimeters, in particular the readout protocol, which led to the simultaneous development of the 3 types of ASICS :
- HaRDROC (Hadronic Rpc Detector Read-Out Chip) for the digital hadronic calorimeter (DHCAL)
- SKIROC (Silicon Kalorimeter Integrated Radout Chip) for the electromagnetic calorimeter (ECAL)
- SPIROC (Silicon Photomultiplier Integrated Read-Out Chip) for the analog hadronic calorimeter (AHCAL)

These ASICs are now in the prototype phase. They are using a more recent and perene technology of Silicon Germanium (SiGe) BiCMOS 0.35 μm from AMS. These chips should be produced in large quantities in 2008 to equip the EUDET modules or module 0s.

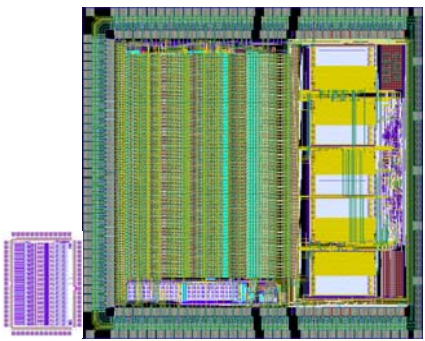

**Figure 8-1 Comparison of size and complexity of first and second generation ASICs.**

## 8.2 Physics prototype ASICS

### 8.2.1 ECAL: FLCPHY3

8.2.1.1 General description

The ASIC described below has been designed to readout the physics prototype in testbeam in 2004-2005 with reduced constraints on integration and power dissipation. The detector is described in [2] and consists of 10 000 channels of Si PAD diodes of 1 cm$^2$, arranged on 2x30 front-end boards containing 6x36 diodes (Figure 8-2).

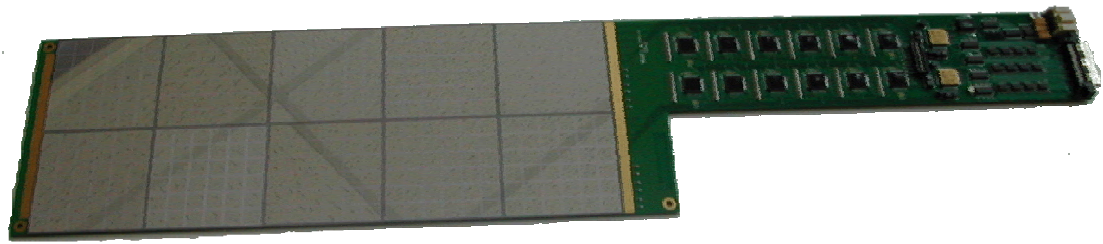

**Figure 8-2 ECAL front end board with Si pad and ASICs**

8.2.1.2 Chip architecture

The chip architecture is built around a low noise variable gain charge preamplifier followed by a by-gain CRRC$^2$ shaper, a Track and Hold and an output multiplexer (Figure 8-3). The chip houses 18 channels, matched to a half detector wafer and is realized in 0.8 μm AMS BiCMOS technology. 1000 circuits have been produced at the end of 2003 and are packaged in TQFP64.

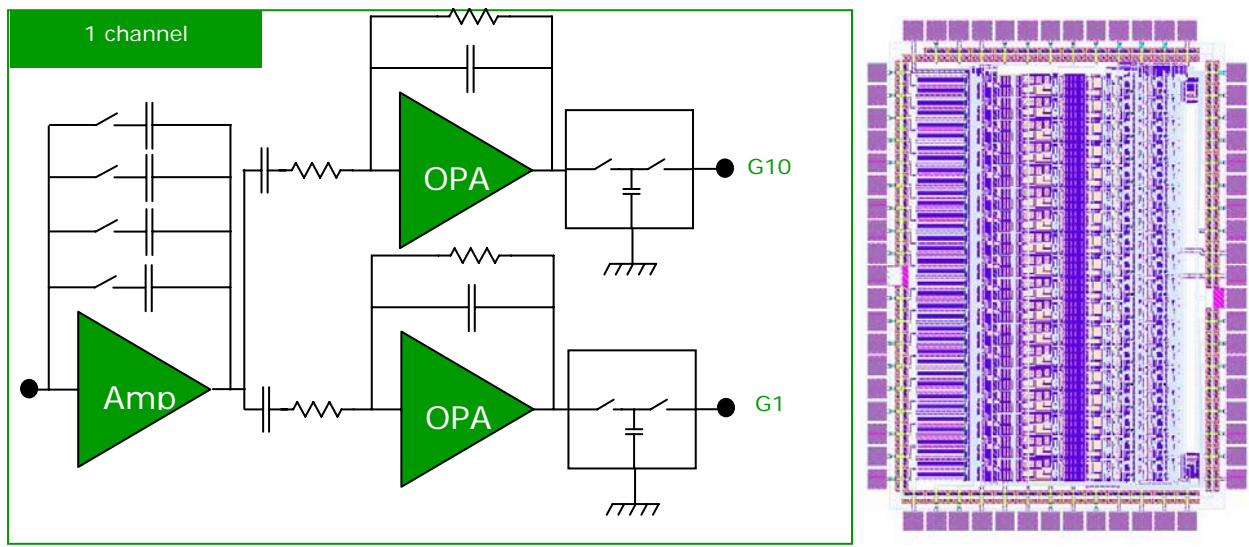

**Figure 8-3 FLCPHY3 block diagram and layout**

The charge preamplifier is a classical folded cascode architecture with a 3000/0.8 input PMOS which exhibits a transconductance of 8mA/V and a noise spectral density of 1.6 nV/√Hz at $I_D$=0.9 mA drain current. The DC feedback is realized with a 36 kΩ resistor multiplied by 625 by a set of current mirrors, achieving an effective feedback resistance of 22 MΩ, while keeping good linearity. The "gain" (feedback capacitance) can be varied from 0.2 pF to 3 pF in 4 bits, the rise time can also be tuned by changing the dominant pole capacitance by a similar ratio.

The following stages are taken from a previous ASIC (OPERA_ROC) developed for the multi-anode PMT of the OPERA experiment [2]. The shaper uses a differential configuration in order to reduce the pedestal dispersion at the output and is built around a Sallen-Key architecture with a peaking time of 180 ns. The track and hold is a simple 1 pF capacitor and a CMOS switch followed by a Widlar differential buffer for low offset, it is read-out sequentially by an OTA in follower configuration.

8.2.1.3    Measured performance of FLC_PHY3 (0.8μm)

The preamp noise has been measured as function of the detector capacitance and yields a series noise spectral density : en=1.6 nV/√Hz and an input capacitance Ca=25 pF from which the test board accounts for 2/3. The 1/f noise has been measured as 12e-/pF.

At the selected shaping time of 200ns, ENC=1000 + 30 e-/pF. Thus with a detector capacitance of 70 pF (25 pF for the Si diode and 20-50 pF for the PCB line) the total noise is expected to be around 3500e- < 1/10 MIP. The preamplifier maximum output voltage is 3 V, corresponding to 600 MIPS with a feedback capacitance of 1.6 pF.   The non linearity is of +/- 0.1% (Figure 8-4).

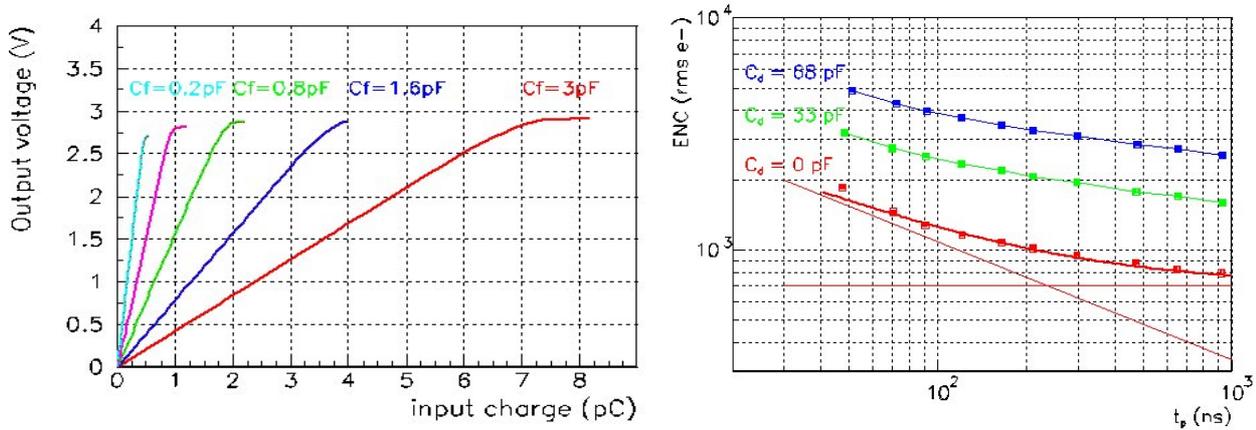

**Figure 8-4 Linearity and noise measurements of FLCPHY3**

Thanks to the differential architecture, the uniformity is good, with a pedestal dispersion of 4.8mV rms, a gain uniformity of 3% rms and a peaking time is 189 ns with a rms of 2 ns.  The crosstalk is below 0.2%.
The noise in this nominal configuration, with a total detector capacitance of $C_d$=68 pF (including PCB contribution of 47pF) is 400 μV in G1 and 200 μV without detector capacitance.. The dynamic range is thus around 6500 (12.5bits) with the detector and 13000 without (13.5 bits).  Using the bi-gain readout with the gain of 10, the dynamic range can be extended to 14-15bits.
The MIP signal is clearly visible with the detector, with a signal to noise around 9 through the full readout.

### 8.2.2    Analog Hadronic Calorimeter: FLC_SiPM

8.2.2.1    Chip architecture

The ASIC for the AHCAL (Figure 8-5) has been developed in less than one year, re-using a large fraction of the chip for the ECAL. In order to read the signal from the SiPM, the charge preamp has been converted into a voltage preamp and the rest of the chip has been kept identical. An 8 bit DAC has been added on the input in order to adjust the SiPM operating point.
The shaper has been made variable through four bits to make two modes of operation : a fast one with high signal to noise ratio for the calibration and a slow one with large dynamic range for physics data taking.

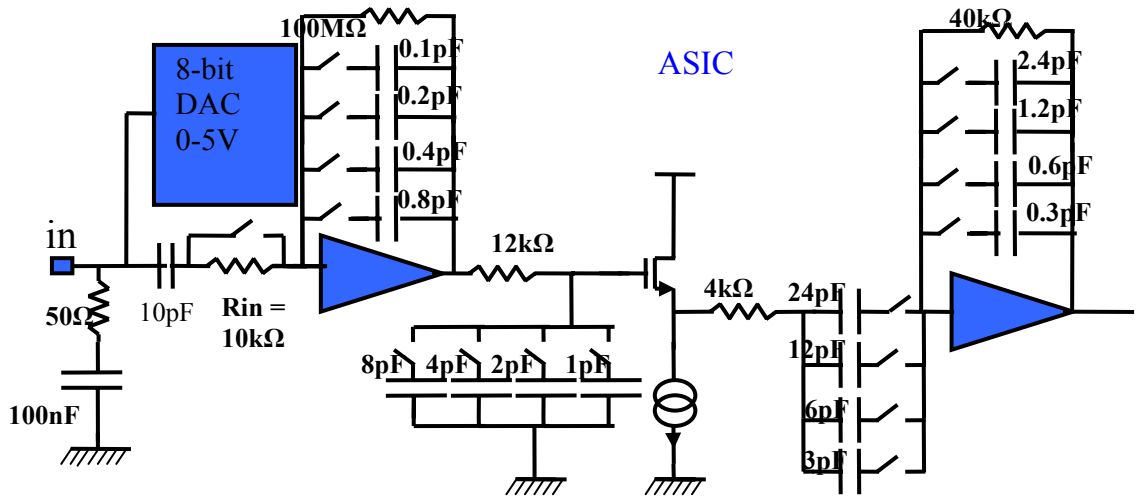

**Figure 8-5 One channel block diagram of FLC_SiPM**

#### 8.2.2.2 Performance

The chip has shown good performance (Figure 8-6 )as shown in the photoelectron peaks below, which are used for the calibration. The input DAC has turned out to be very useful to adjust the operating point of the SiPM.

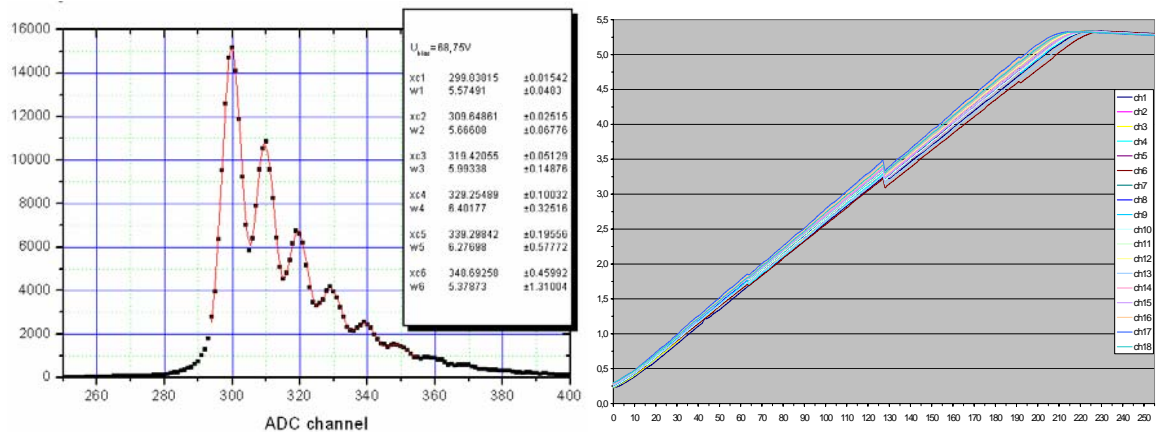

**Figure 8-6 Low light intensity pulse height spectrum with single photoelectron peaks, as measured with FLC_SiP;boas adjustment DAC linearity.**

### 8.2.3  Digital Hadronic calorimeter : DCAL

The Digital Hadron Calorimeter will be read out by the DCAL chip and a system of data concentrator and super concentrators. Due to the large number of channels of the 1 m$^3$ prototype section (~400,000), channel multiplexing has been introduced at all levels of the readout chain. The digitization occurs inside the DCAL chip, the remainder of the readout system only handles digital data.

The DCAL chip is a 64-channel ASIC located on the detector front-end boards, which are glued on top of the detector pad boards. The chip has been realised by the FNAL microelectronics group in 0.25μ CMOS. It is derived from a family of chips developed for Si detectors of the BTeV experiment and has shown good performance for this type of application. The design of the analog part of the circuitry is based on detailed measurements with RPCs read out with both an analog and a digital readout system (see the DHCAL section and the references therein for more details).

Figure 8-7 shows a schematic of the DCAL chip. The signals from 64 detector pads are shaped and discriminated, where one adjustable threshold serves all channels in a given chip. The gain of the input

preamplifier can be selected to be either low (for the readout of RPCs) or high (for the readout of GEMs/Micromegas). The resulting hit pattern (64 bits) traverses a 20-stage pipeline, corresponding to 2 μs, before being written to the output buffer (FIFO), thus providing ample decision time for the trigger.

The chip is clocked with 10 MHz and the clock pulse is distributed to the front-end, the pipelines and the serial data transmission circuitry. The chip works both in triggered and in triggerless mode, with no dead time up to event rates of several kHz. Figure 8-7shows the chip as mounted on its test board.

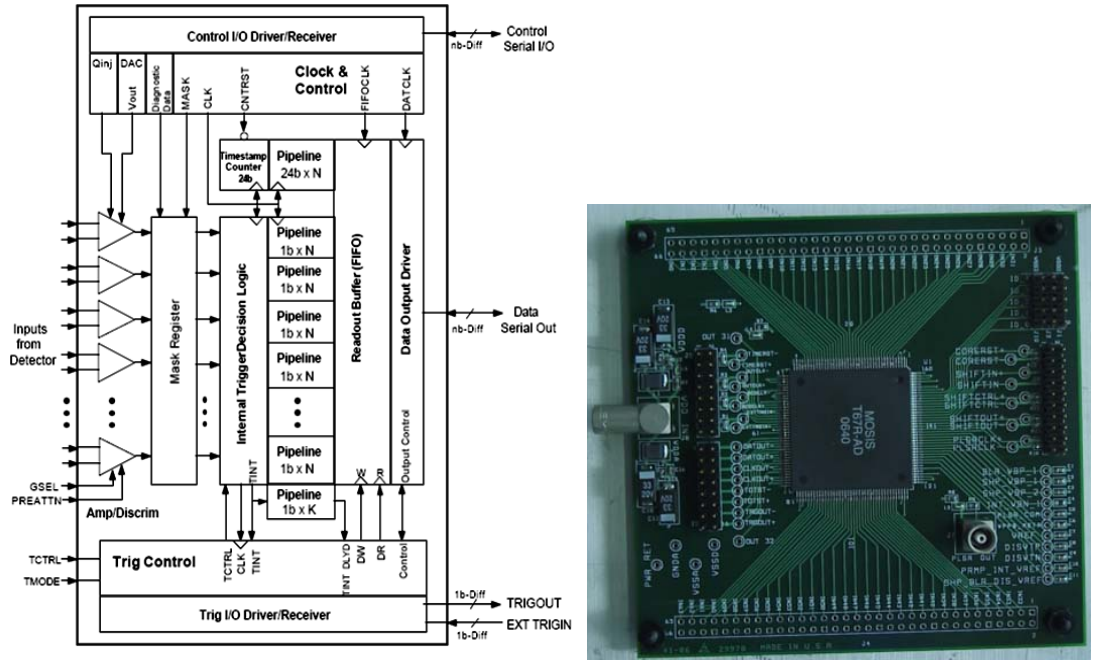
.

**Figure 8-7 Schematic of the DCAL chip; Photograph of the packaged DCAL chip as mounted on its test board**

The functionality of the chip has been extensively tested. Apart from minor problems with the packaging (not affecting the performance in low gain mode) and the self-trigger mode the chip performed entirely successfully. The latter problem will have no effect on the Vertical Slice Test. It has been corrected and its correction has been successfully simulated. No further prototyping is required before initiating the production of large quantities.

Figure 8-8 shows the response of the chip operated in low gain to externally injected charges (in the range between 35 and 700 fC) versus discriminator threshold. For display purposes the curves corresponding to different injected charges are offset vertically with respect to each other. The noise floor is estimated to be well below 10 fC (performance on the actual front-end boards appears to be even better). Figure 8-8 shows the points of 50% efficiency versus injected charge. The response is linear for charges up to 300 fC. The maximum threshold corresponds to approximately 700 fC, well matched to the range of signals (100 ÷ 10,000 fC) of RPCs operated in avalanche mode.

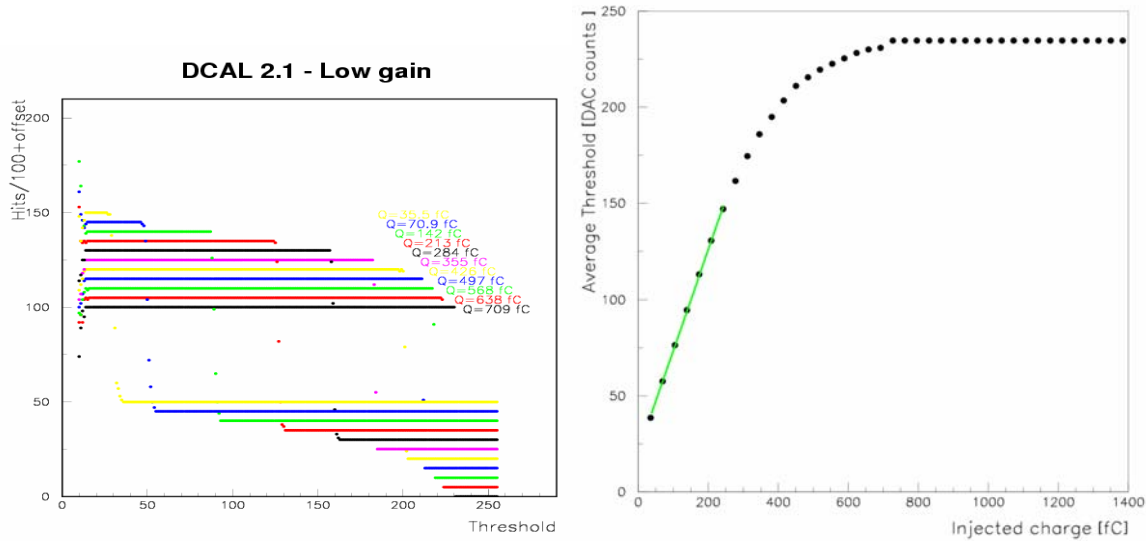

**Figure 8-8 Response versus threshold for a range of injected charges; 50% efficiency versus injected charge.**

The channel-to-channel cross-talk was measured by measuring the response of neighbouring channels to injected charges up to 20 pC. Figure 5 shows the response of channel 54 (with an injected charge of 10 pC) and its neighbours. At very low threshold settings channels 53 and 55 are seen to record hits. The threshold for these spurious hits corresponds to approximately 0.4% of the injected charge and is well below the nominal threshold of 100 fC to be applied to the readout of RPCs in avalanche mode.

In summary, the DCAL chip is ready for production for the DHCAL prototype section.

### 8.2.4  Readout

The readout is performed by digitizer boards called CRC (Figure 8-9) initially developed for CMS by UK groups. They house 96 16bit ADCs that read out 18 multiplexed channels, thus one CRC board services 1728 channels. They have been in use since 2004 and perform very well, with about 1% dead channels overall. However, as the system has more electronics channels than are required for the full readout, these can be left unused. 15 CRC boards are available; the silicon ECAL requires 6 CRCs, the AHCAL 6 CRCs and the TCMT 1 CRC..

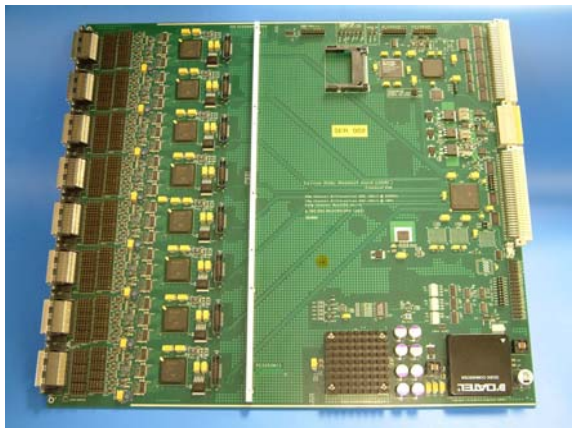

**Figure 8-9 CALICE readout card CRC.**

## 8.3 Technological prototype ASICs

Three main differences characterize the second generation of ASICs :
- Ultra-low power with power pulsing
- Digital data out (on-chip ADC)
- Auto-trigger mode, on-chip data storage and zero-suppress

They also integrate the first part of DAQ to be daisy-chained and minimize the number of lines on the detector PCBs.

### 8.3.1 DHCAL front-end: HaRDROC

HARDROC1 (HAdronic Rpc Detector ReadOut Chip) is the first prototype of second generation readout for the RPC or GEM foreseen for the Digital HAdronic CALorimeter DHCAL. For compactness, the chips must be embedded inside the detector making crucial the reduction of the power consumption to 10 µWatt per channel. This is achieved using power pulsing, made possible by the ILC bunch pattern (1 ms of acquisition data for 199 ms of dead time).

#### 8.3.1.1 ASIC description

HARDROC readout is a semi-digital readout with two thresholds (2 bits readout) which allows both good tracking and coarse energy measurement, and also integrates on chip data storage.
The 64 channels of the ASIC are made of:
- Fast low impedance preamplifier with 6bits variable gain (tuneable between 0 and 4)
- Variable shaper (50-150ns) and Track and Hold to provide a multiplexed analog charge output up to 10pC.
- Variable gain fast shaper (15ns) followed by two low offset discriminators to autotrig down to 10 fC. The thresholds are loaded by two internal 10 bit- DACs.
- A 128 deep digital memory to store the 2*64 discriminator outputs and bunch crossing identification coded over 24 bits counter.
- AMS SiGe 0.35µm technology
- 16 mm$^2$ area
- 3.5V power supply
- 10 µW power consumption/channel
- Package: CQFP240

The block diagram of the ASIC is given in Figure 8-10. Each input signal is first amplified thanks to a variable gain preamplifier which exhibits low noise and low input impedance to minimise crosstalk. It allows accommodating the gain depending of the detector choice, up to a factor 4 to an accuracy of 6% with 6 bits.

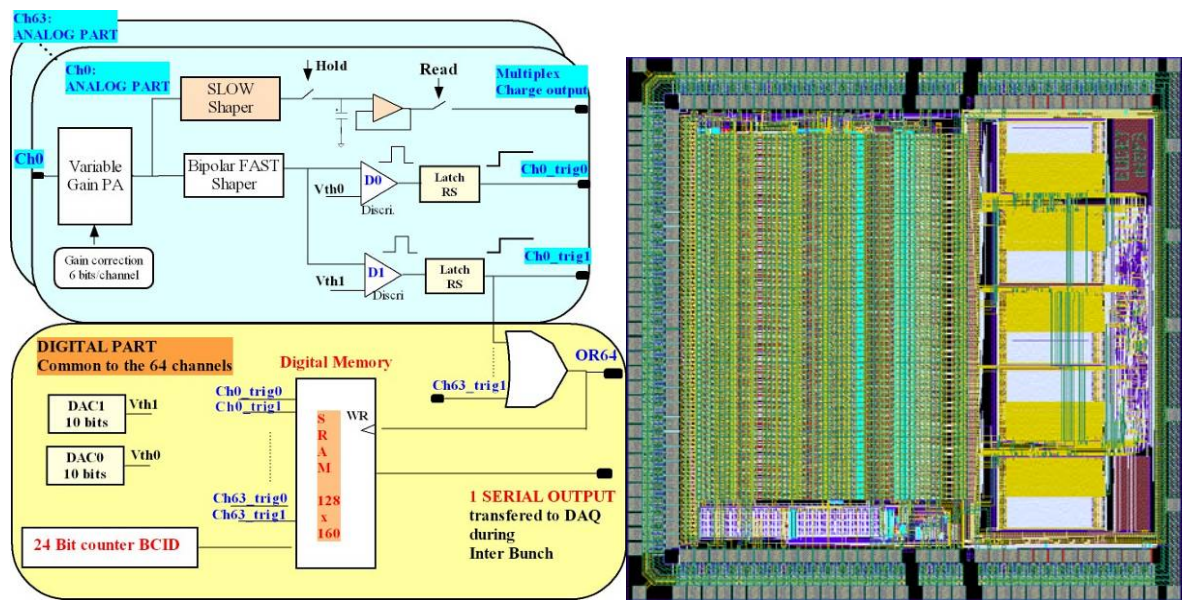

**Figure 8-10 HaRDROC block diagram and layout**

The amplified current feeds then a slow shaper combined with a Sample and Hold buffer to store the charge in 2pF and provide a multiplexed charge output (5MHz) up to 10pC, compatible with the previous generation (analog) DAQ.

In parallel, trigger outputs are obtained via fast channels made of a fast (15 ns) shaper followed by 2 low offset discriminators. The discriminator thresholds are set by two internal 10 bit DACs. Each trigger output is latched to hold the state of the response until the end of the clock cycle. The trigger1 outputs (corresponding to Vth1 <Vth0) are OR wired to generate an internal trigger used to start the memorization of the 128 trigger outputs as well as the Bunch Crossing Identification delivered by a 24 bit counter, needed to associate hits in the DAQ to bunch crossing ID. It is also possible to capture event data using an external trigger provided from outside the chip. All the bias currents are programmed through the Slow Control. The chip is power pulsed to decrease the power consumption. 10μV/channel as targeted with a 1% beam duty cycle

#### 8.3.1.2 Testbench measurements

HARDROC has been submitted in September 2006 and received mid December 2006.

The bipolar fast shaper gain is about 3.5mV/fC and its peaking time is equal to 15ns. The Slow Shaper gain is about 50 mV/pC and its peaking time can be tuned from 100ns to 150ns. The input impedance of the preamp has been measured to be 50-70 ohms, depending on the (tuneable) current flowing in the preamp. The crosstalk has been measured by sending 100fC in one channel and looking to the direct neighbours. This 2% crosstalk is well differentiated and located on the input.

The linearity of the two 10 bits integrated DACs used to generate the thresholds of the discriminatrors, has been measured (Figure 8-11): the residuals of the both DACs are within ±5 mV for a 2.6V dynamic range which corresponds to an Integral Non Linearity of 0.2% (2LSB). The slope is 2.5mV per DAC unit. The s-curve measurement performed on the 64 channels of the chip show good trigger efficiency at 100fC, similar results are obtained down to 10fC . The quite large non uniformity between channels (±25 %) is explained by current mirror mismatch (small size transistor to optimize speed) and can be corrected using the gain tuning of the input preamp.

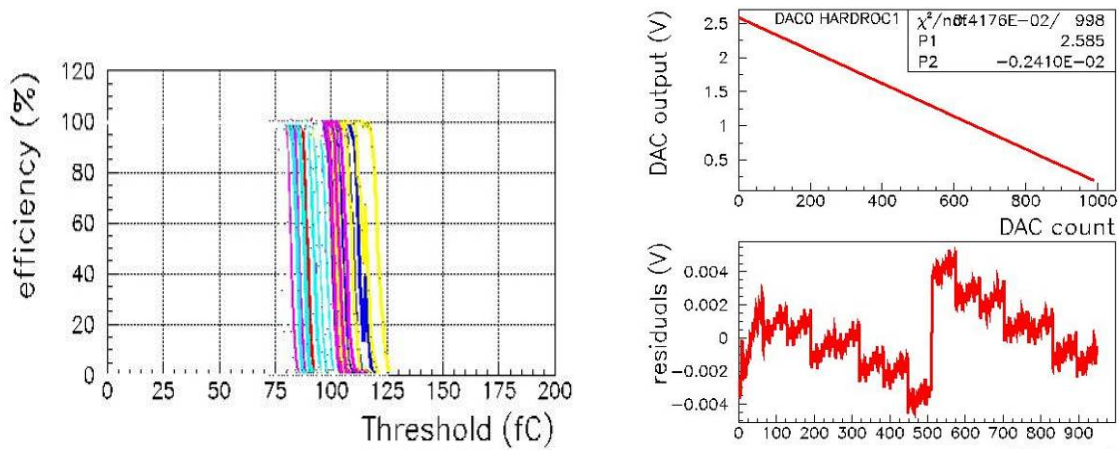

**Figure 8-11 HaRDROC test bench measurtements: trigger efficiency, DAC linearity and residuals**

#### 8.3.1.3 Digital part:

Because of the very high number of electronic channels foreseen in the final detector, chips will be embedded inside the detector and are designed to be daisy chained without any external circuitry, to limit to a bare minimum the number of output lines on the detector. A memory (Figure 8-12) has been integrated in HARDROC to store during the bunch train the 2bits trigger outputs of each channel as well as the BCID, and this for every hit.

The data format is 128(depth)*[2 bits * 64 ch + 24 bits (BCID) +8 bits (header)] =20kbits. There is one serial output which is transferred to the DAQ during the interbunch.

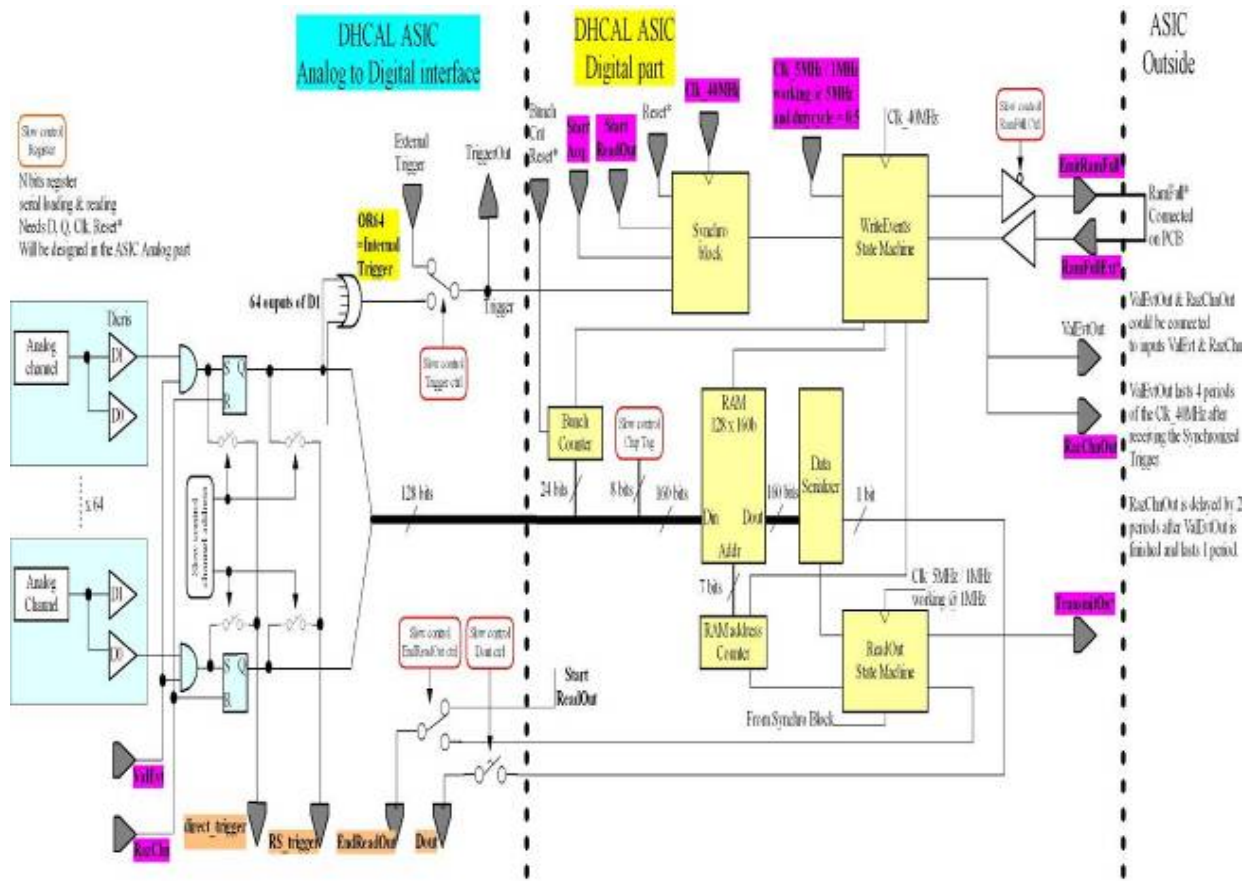

**Figure 8-12 Digital part of the HaRDROC ASIC.**

Figure 8-13 displays a memory frame. The auto trigger mode which is crucial for the chip functionality has been checked successfully down to 10fC.

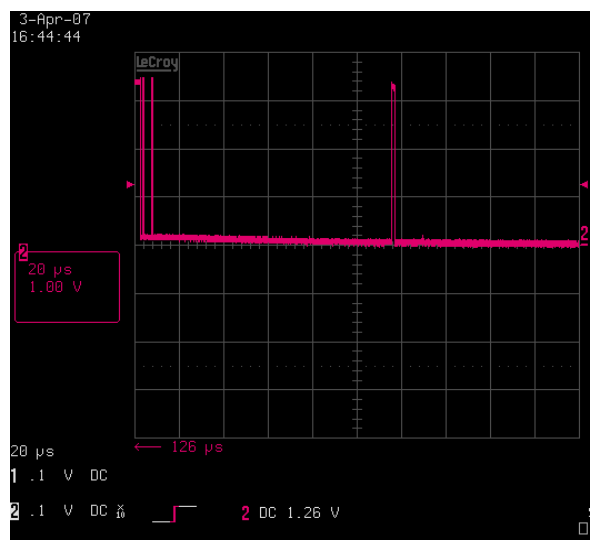
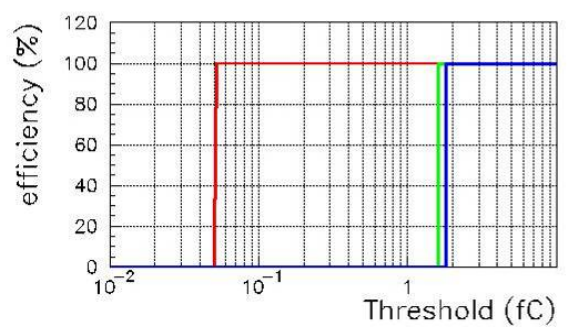

**Figure 8-13 Oscilloscope display of one memory frame in HardROC; trigger efficiency for low threshold settings.**

### 8.3.2 ECAL front-end ASIC : SKIROC

Starting from HARDROC, the second chip developed is for the ECAL, called SKIROC – standing for Silikon Kalorimeter Read-Out Chip – has been designed to read-out the upcoming generation of Si-W calorimeter featuring ILC requirements. The analogue core of SKIROC is based on the front-end electronic designed for that physics prototype. It has been enhanced in many ways using an intermediate prototype called ILC_PHY4. The Maximum input charge has been extended from 500 to 2000 MIP. The number of channel has been doubled – reaching 36 - to fit a pad size reduction in the silicon detector design conducted concurrently. A stand alone working capability comes along with the full power pulsing feature. That means SKIROC does not need any external component such as decoupling capacitance or bias resistor involving a huge room saving. The wake up sequence duration of the power pulsing is around 2μs to ensure a lower than 1% duty cycle in an ILC-like beam structure, involving more than two order of magnitude of power saving.

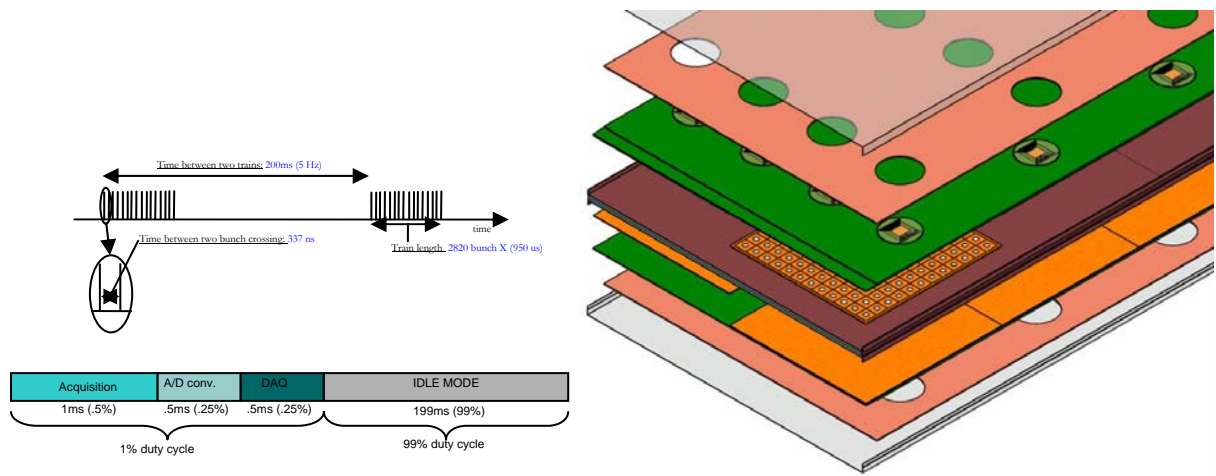

Beyond the analogue core improvement, many features have been implemented in SKIROC. A channel by channel auto-trigger capability has been added allowing a built-in zero suppression. A multi-channel ADC is embedded. The trigger and gain selection threshold is set by an internal dual DAC. Voltage references used in the analogue core use a bandgap reference. A digital core driving all the analogue features and the digital communication with the DAQ has been designed and is implemented in a FPGA to get debugged and improved before being embedded in the next version.

8.3.2.1 SKIROC description

SKIROC (Figure 8-14 )is a 36-channel front-end chip designed to read-out silicon PIN diodes for calorimetry application. It has been designed in a general framework ensuring consistent back-end of different front-end ASIC for several calorimeters (HaRDROC to read out the digital RPC HCAL prototype and SPIROC to read out the SiPM and Sci tiles HCAL prototype are the two others chip existing on that framework)
Its main characteristics are the following:
- AMS SiGe 0.35μm technology
- 20mm² (4mm × 5mm) area
- 3.3V power supply
- Package: CQFP240

Each channel is made of a variable-gain low-noise charge preamplifier followed by both a dual shaper – one with a gain 1 and the other with a gain 10 - to filter the charge measurement and a trigger chain composed of a high gain fast shaper and a discriminator. The measured charge is stored in a 5-depth SCA that can be read either in an analogue way or can be connected to a multi-channel 12 bit Wilkinson ADC. Thresholds are set with a 10-bit DAC for trigger level and for automatic gain selection level. A bandgap ensures the stability versus supply voltage and temperature for all the requested reference in the analogue core.

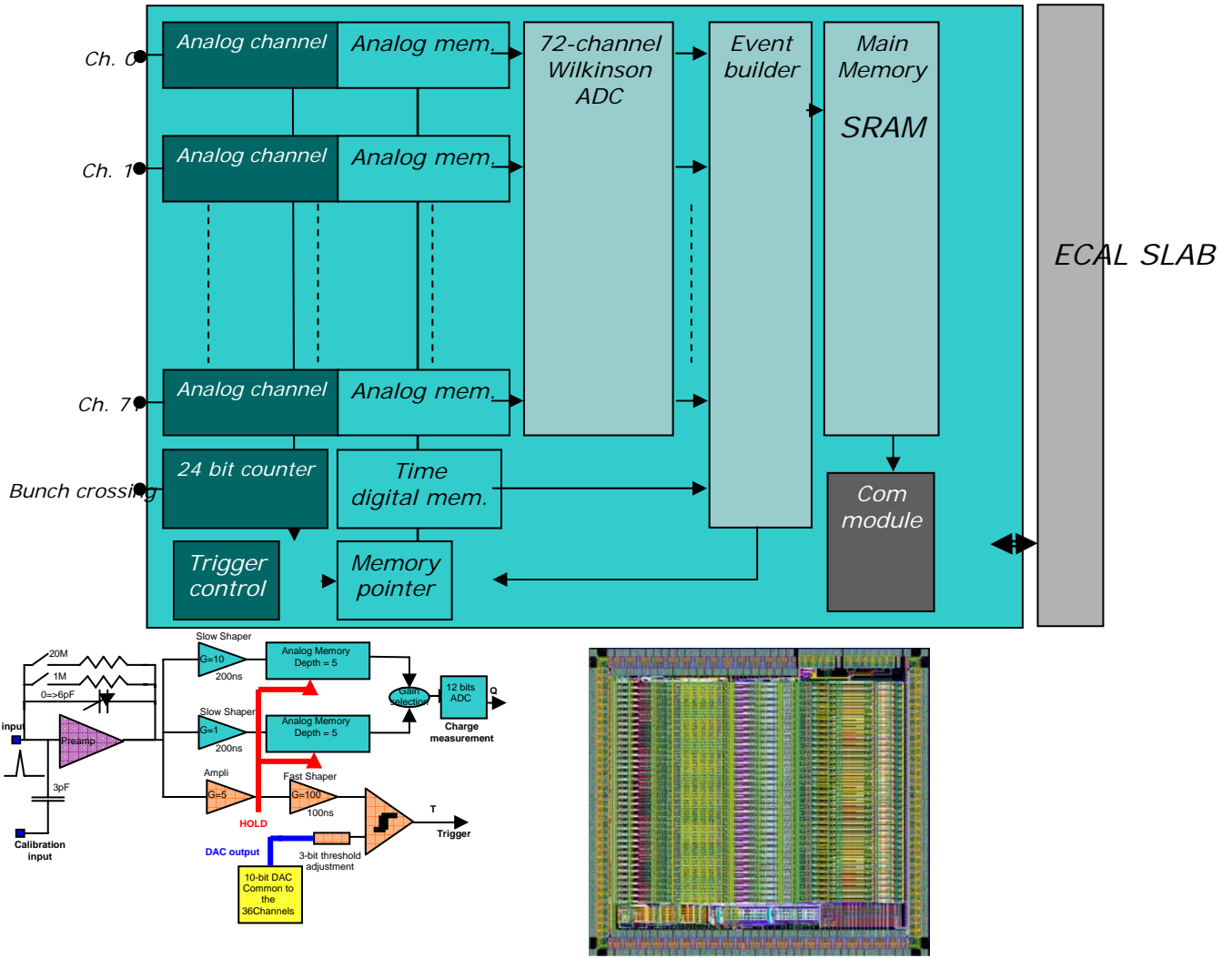

**Figure 8-14: One channel block diagram of SKYROC; ASIC layout**

The digital signals requested for digital and analogue block communications are outputted using a dynamic multiplexing to reduce the pin count while emulating the digital core in a FPGA.

8.3.2.2   Blocks measurements

The blocks used to design SKIROC has been fully characterized before being used in the ASIC. At the moment this abstract is written, results on SKIROC are being processed and final results will be presented during the conference.

The analogue core of SKIROC is mainly inspired from ILC_PHY4 front-end chip prototype. Extensive measurements have been conducted on ILC_PHY4 to validate the stand-alone working capability without any decoupling capacitance or bias resistor. Results below (Fig 4) show that the performances in term of linearity are compatible with calorimetry measurement.

  The equivalent noise charge of the preamplifier is measured around 2000 electrons. After shaping, the simulated MIP to noise ratio is 16 for the trigger line and 11 for the charge measurement. Crosstalk is around the per mil level.

The DAC performances measured on HaRDROC show a linearity in the per mil range and an output voltage swing of 2V covering the whole useable threshold range.

Several points on these blocks has been improved in SKIROC and the coming measurements should validate a fully functional self-triggered front-end chip.

The chip also incorporates 64 12bit Analog to Digital converters (ADCs). Due to the high level of complexity in the analog design and digital sequencing, it has been found too risky to integrate the digital part, which has been left outside in an FPGA. A complete VHDL description of this part has been written and be tested in 2007.

The chip has been realized in SiGe 0.35µm technology by AustriaMikroSystem GmBH. It covers an area of 20 mm2 and has been fabricated in December 2006.

### 8.3.3 SPIROC

The third front-end chip for the AHCAL is called SPIROC (standing for Silicon PM Integrated Read Out Chip) to readout the Silicon PM of the upcoming technological demonstrator foreseen in 2009.

8.3.3.1 SPIROC description

SPIROC (Figure 8-15) has been designed to read out SiPM or MPPC through an ILC beam structure involving a sequence of operation mode as following:
- Acquisition mode: where the charges and the hit time are memorized in the analogue memory each time the chip triggers.
- Conversion mode: where the data are converted from the analogue memory into digital through ADCs. The data are then formatted and stacked up in a digital memory.
- Data transfer: where the formatted event (ie charges measurements and associated triggering time) are outputted to the data acquisition system.
- Idle mode: where the chip consumption is reduced by a factor greater than 1000 to wait for the next machine train

These entire operation modes are fully automatic. SPIROC main characteristics are the following:
- AMS SiGe 0.35µm technology
- 30mm² (4mm × 7.5mm) area
- 3.3V power supply
- Package: CQFP240
- Internal input 8-bit DAC (0-5V) for SiPM gain adjustment
- Energy measurement :
    - 2 gains / 12 bit ADC 1 pe → 2000 pe
    - Variable shaping time from 50ns to 100ns
    - pe/noise ratio : 11
- Time measurement :
    - 1 TDC (12 bits) step~100 ps
    - pe/noise ratio on trigger channel : 24
    - Fast shaper : ~15ns
    - Auto-Trigger on ½ pe
- Analog memory for time and charge measurement : depth 16
- Power pulsing integrated
- Low consumption : ~25µW per channel (in power pulsing mode)
- Calibration injection capacitance
- Embedded bandgap for voltage references
- Embedded DAC for trigger threshold
- 12-bit Bunch Crossing ID
- SRAM with data formatting : 4 Kbytes
- Output & control with daisy-chain

The SPIROC chip will be send to foundry on June 2007. The analogue core is composed of 36 channels embedding an input DAC for SiPM high voltage adjustment on 5V to tune gain channel by channel. Two preamplifiers allow the requested dynamic range and are followed by a trigger line made of a fast shaper and a discriminator. The charge measurement line is made of two variable slow shapers and two 16-depht SCAs. The block scheme of a channel is shown on *Fig 2*.

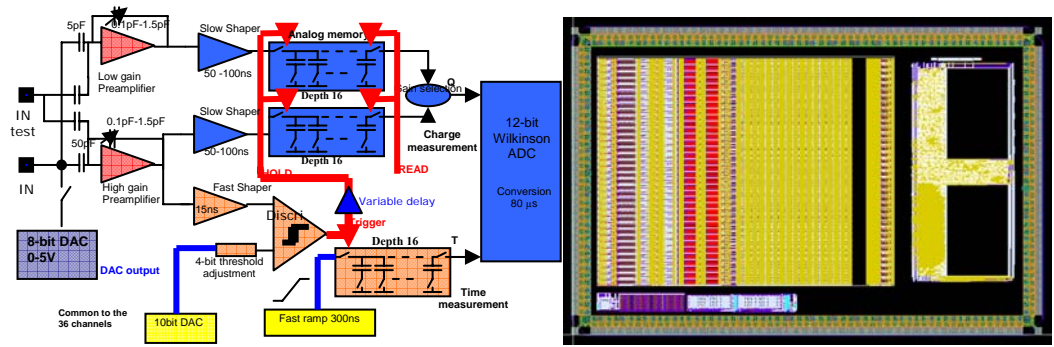

**Figure 8-15 SPIROC block diagram and layout**

#### 8.3.3.2  SPIROC simulation results

The new analogue chain in SPIROC allows the SPE calibration and the signal measurement to be on the same range, simplifying greatly the absolute calibration. An analogue simulation of a whole analogue channel is shown below.
The ADC used in SPIROC has been tested on a functional block before being integrated in the whole system. Integral non linearity of the 12-bit Wilkinson ADC is shown in Figure 8-16. The measurement shows a 11.5 ENOB (Equivalent Number of Bit) fulfilling requirements.

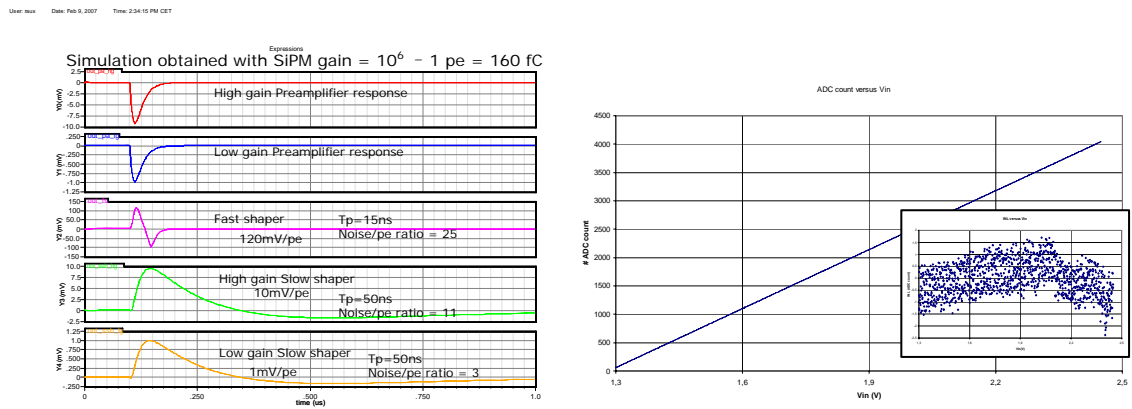

**Figure 8-16: Simulation of the analogue part for a SiPM like input signal;  12 bit Wilkinson ADC integral non-linearity  measurement result. The apparent differential non-linearity shown in the right plot is due to the measurement procedure.**

The SPIROC chip will be submitted in June 07 and will be tested in September 07. It embeds cutting edge features that fulfil ILC final detector requirements including ultra low power consumption and extensive integration. The system on chip is driven by a complex state machine ensuring the ADC, TDC and memories control. The SPIROC chip is due to equip a 10,000-channel demonstrator in 2009.

### 8.3.4  On-going R&D

The second generation of ASICs is not the final one for the ILC and additional R&D is going on, in particular on what concerns ADCs . Pipeline ADCs are the favoured architecture to optimize the power dissipation and a 10bit 10 MHz has been realized by the ClermontFerrand group that exhibits excellent lineanrity while dissipating only 44nW/conversion at 4MHz (Figure 8-17).

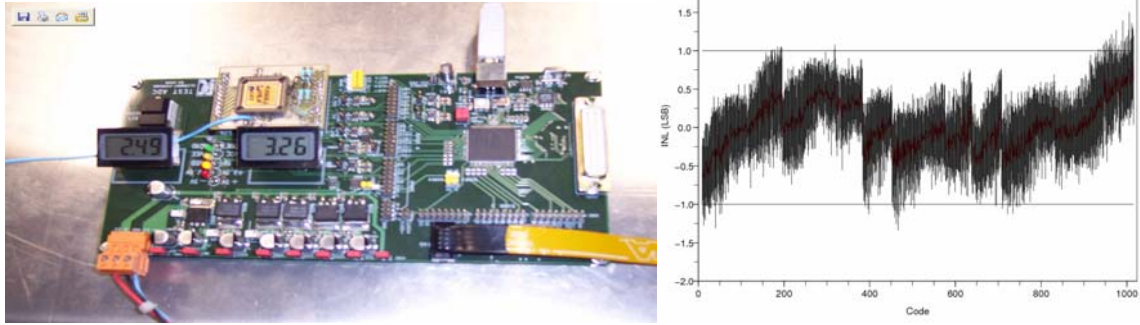

**Figure 8-17 Pipeline ADC on its test board; linearity measurements**

## 8.4 Conclusion

The electronics for the physics prototype have now been in operation for more than 2 years for the ECAL and AHCAL and have shown excellent performance. It is also ready for production for the DHCAL physics prototype.

Two major ASICs have been realized in 2006, which will allow the study of on-detector electronics integration throughout 2007 and the $2^{nd}$ generation DAQ. Good performance of theses ASICs will be a key element for series production in 2008 to equip the EUDET ECAL and AHCAL modules and a DHCAL prototype.

## 8.5 Acknowledgements

Part of this work is supported by the Commission of the European Communities under the $6^{th}$ Framework Programme "Structuring the European Research Area", contract number RII3-026126.

# 9 The DAQ development

## 9.1 Overview

A data acquisition (DAQ) system is described which will be used within CALICE for the next generation of prototype calorimeters for the International Linear Collider and could also be used for the final system. The concept of moving toward a "backplaneless" readout is pursued. A strong under-pinning thread here is to attempt to make use of commercial components and identify any problems with this approach. Therefore the system should be easily upgradeable, both in terms of ease of acquiring new components and competitive prices. The conceptual design of the data acquisition system for the ILC calorimeter will be discussed. DAQ equipment is being developed which attacks likely bottlenecks in such a commercially-based future system and is also sufficiently generic to provide the readout for new prototype calorimeters, such as the prototypes to be built in the EUDET project[1]. Indeed the design[2] is sufficiently generic such that it should have applications elsewhere, be they other ILC detectors or within High Energy Physics in general: e.g. this could be applied to LHC upgrade apparatus. Results and tests already performed will then be shown indicating both the potential and limitations of the approach.

## 9.2 Layout

Although of a generic nature, the final ILC calorimeter provides a test case for the DAQ system under development. A sketch of how the final DAQ system could look like is shown in Figure 9-1 (left).

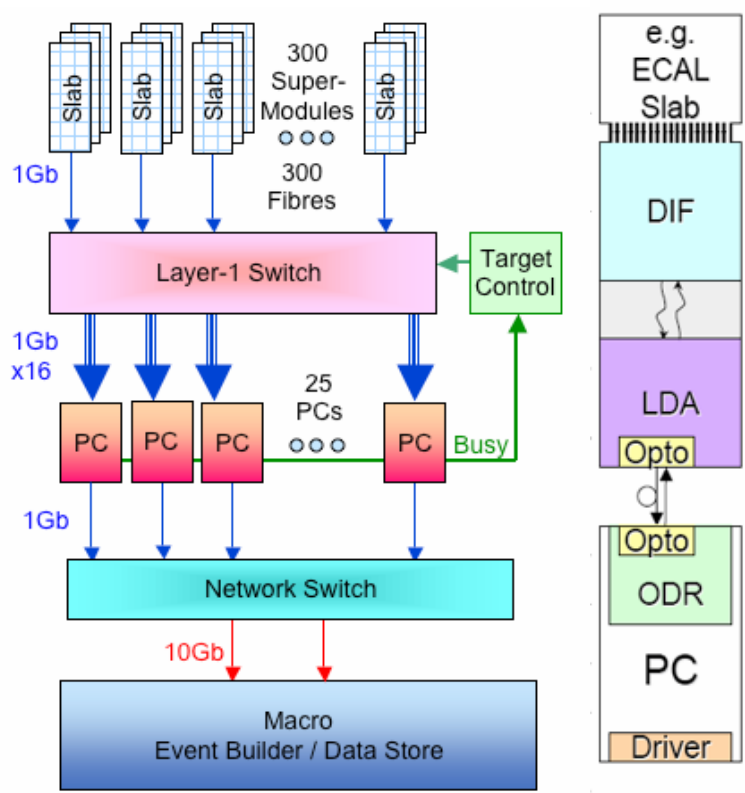

**Figure 9-1:** General outline of expected final ILC calorimeter DAQ system (left) and DAQ chain planned for the EUDET modules (right).

---

[1] www.eudet.org

[2] M. Wing et al., "A proposed DAQ system for a calorimeter at the International Linear Collider", Linear Collider Note, LC-DET-2006-008.

The overall system is shown for the ECAL as an example, but is easily extendable to other calorimeters or detectors. The number of components needed at each stage of the system and the volume of data being transported is also shown. The assumptions for data rates and volume are detailed in the LC note. Clearly these numbers are approximate and also the technology will develop so that the numbers for data rates and hence number of hardware components will change, but this diagram provides a reference for future developments.

Data will be transported off the detector slabs (possibly) via a "Layer-1" switch which can re-route data should the Off-Detector data Receiver (ODR) PC not be available due to a fault or busy signal. These switches are cutting edge technology being used in the telecommunications industry and could improve the efficiency of data taking in high-energy physics experiments. Their performance is one of the projects being pursued accompanied by studies of more conventional, but high-speed 10 Gbit switches. The off-detector data receiver consists of PCI cards housed in PCs; in the above scheme 25 PCs with 50 cards would be needed for the final ILC ECAL. This can be thought of as some upper limit as technology improves, but is already a manageable number. Telecommunications industry is also developing industry-standard crate systems such as ATCA and microTCA, which may be the future direction for high energy physics experiments. Transferring to such a system in the future would be simpler after experience with the current programme where we are using commercial systems and high throughput serial technology such as PCI Express.

In Figure 9-1 (right), the proposed DAQ stream for all calorimeters is shown in more detail and also as it will be set-up for the EUDET prototypes. At the end of the detector slab, the data will be aggregated and transported off the detector via calorimeter-specific electronics, here called a DIF (Detector InterFace). The data will then pass to a Link/Data Aggregator, LDA, which is a generic piece of electronics used by all calorimeter systems, which will serve multiple DIFs. A high-speed optical link will then transfer the data to the ODR.

The PCI cards housed in the PCs, which act as the ODR, are again all based on commercial, off-the-shelf technology. It was originally planned to develop such a card so as best to meet our needs, whilst still using modern commercial components. However, cards, which met all of our requirements, are now available off-the-shelf. The PCI card acquired is shown in Figure 9-2, which has been developed and built by the company PLD Applications[3].

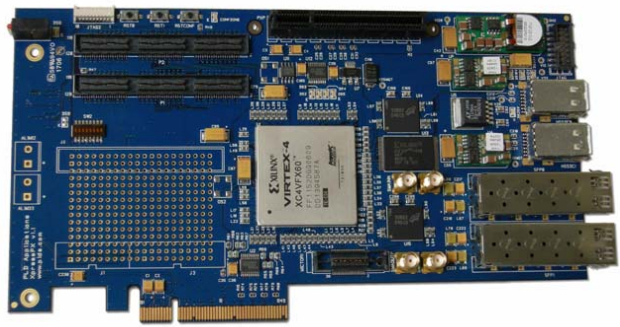

**Figure 9-2: Physical DAQ card bought from PLD applications.**

The card is shown schematically in Figure 9-3, where the major component blocks are highlighted: optical and electrical links; a large FPGA; and a PCI-Express bus. The PCI card will act as a data receiver and also source for the clock, control and configuration data. This card will allow high data rates to be received off the detector, which can then be used by detectors which require this or aggregate large amounts of data and thereby reduce the number of hardware components, and cost.

---

[3] www.plda.com

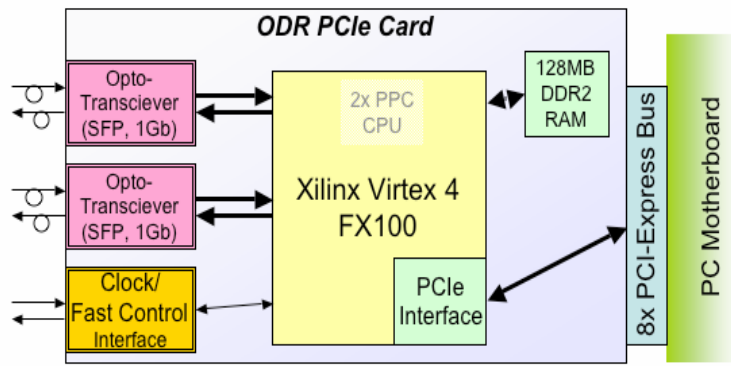

**Figure 9-3: Schematic of DAQ card.**

Several PCI cards have been purchased and are being used for bench tests to determine their ultimate performance and will be used for the EUDET prototype modules. A suite of firmware and test software has been written so that data can now be passed to the PCI card transceivers through the FPGA and into the host computer memory. Measurements of the speed of transfer through the PCI Express bus are underway. Issues being addressed are file-storage systems, storage arrays, data block size, etc..

Moving further back into the detector, studies have been performed of the impact of radiation on the electronics at the end of the slab in the detector volume. Of concern is the rate of single event upsets (SEUs) in the FPGAs in the DIF modules, which will require the FPGA to be reconfigured. This is clearly FPGA dependent and the FPGA to be used in the final ILC calorimeter is unknown. However, a study has been performed to determine the rate of SEUs for current FPGAs. The rate of particles from physics events impinging on the slab electronics was estimated using Monte Carlo simulation. The neutron, proton and pion energy spectra were then converted into a rate for SEUs for FPGAs for which test measurements had been made. The rate varied considerably from about one SEU every 40 days to one every day for the whole ECAL electronics. From this study, we can conclude that this is a feature which needs to be addressed, but does not require the electronics to be reset often on a bunch-train scale. However, the future electronics will need to be tested and their rate of SEUs determined. A method for assessing this has been formulated.

Finally, on-detector data acquisition and integrity is also important and is being studied using a detector model, shown in Figure 9-4. The model currently has readout electronics coupled to a 24 cm PCB, which has the characteristics expected for the EUDET (and final) PCB. Tests are ongoing with the one PCB slab, but will also be extended to seven PCBs connected together, thereby simulating the full-length EUDET slab. The PCB is populated with FPGAs, each of which mimics two ASIC chips. This panel will allow mechanical and thermal tests to be performed such as methods of bonding two PCBs together and power dissipation. However, the main thrust for the DAQ are: signal and distribution options, ultimate speed, output standards, etc. And these options have been included in the model slab to optimise its performance. The results of these tests will feed directly into the design and build of the EUDET module.

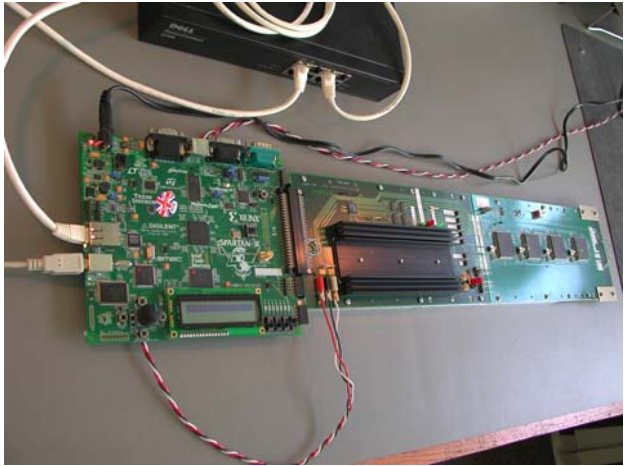

**Figure 9-4: Model test slab showing the electronics board coupled to a short, 20 cm PCB (far right).**

In summary, R&D is ongoing which attacks a new concept in DAQ systems for high energy physics. The applicability of commercial systems is being evaluated with definitive answers to be determined in the coming years. Such a system will also be used for the EUDET prototype modules to be placed in test beams in 2009.

# 10 The software for CALICE R&D

## 10.1 Introduction

The software concept as realized for the processing of the CALICE data is driven by three main guidelines:

1) Application of common ILC Software tools where possible and therefore benefiting from general developments of the ILC software. At the same time the application of these tools allow for the identification of the needs of the ILC Software for 'real' data already at an early stage of the R&D phase.
2) Since test beam data are taken at various locations they have to be available independent of the experimental site. Here, the grid was identified as the ideal tool to meet this goal. At the same time the grid offers significant computing resources needed for simulation and various re-processings of the data..In order to realize the data management and processing within the grid environment the virtual organization (vo) *calice* has been established. As of today this vo counts 42 members.
3) As many users as possible are to get involved in the analysis effort. Therefore entry points for an easy start-up for unexperienced users have to be provided.

In this report, the emphasis will be put on schemes and procedures which were successfully employed during the large scale test beam data taking at CERN in 2006 which included the SiW Ecal the Analogue Hcal and the TCMT.

## 10.2 Data Transfer and Management Scheme

The data transfer and management chain is schematically shown in Figure 10-1.

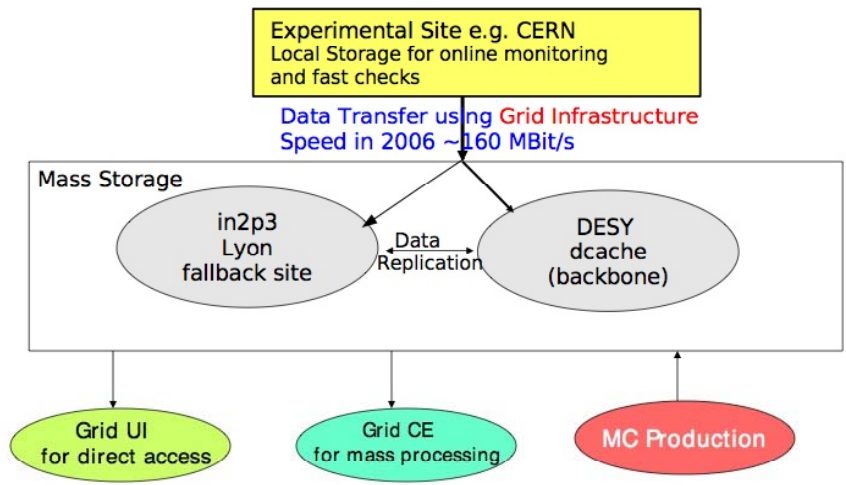

**Figure 10-1: Schematic view on the transfer chain for the CALICE data.**

Data as acquired by the CALICE data acquisition system are stored on a local disk array at the experimental site. Locally, these so called *native* files are used for debugging purposes and online monitoring of the data quality. After run completion a server program running on a Linux PC configured as a grid user interface transfers this data to a mass storage. The files are registered with their logical file names to the grid. By this the data are available to the whole CALICE collaboration approximately 30 Min. after run end without users being forced to log into counting room computers which might lead to interferences with the ongoing data taking.

The primary mass storage is the DESY dcache system. Here, roughly 35 TBytes are available for CALICE. This is the amount of data expected to be by the end of 2007 which include the native data files, simulated as well as files produced during the data processing steps described in the next sections. In addition to the mass storage at DESY similar storage space has been allocated at other sites. Among these sites, there are major computer centers which will hold complete replicas of the data. In total, 13 sites support CALICE with storage and computing resources. Two more sites have expressed their intention of support.

## 10.3 Data Processing Scheme

Note, that the DAQ system is tuned for a maximal acquisition speed. For technical reasons information belonging to one event appears at different places in the data stream. To accommodate for that, these native data files are subject to a first processing step which acts as an event builder and provides first checks on the integrity of the data (see Figure 10-2). This step is also called the 'conversion' step since here the native data are converted into a format compatible with the general ILC software (Details see Section 8.4).

During the conversion, a database is filled with detector configuration data and important conditions data such as e.g. temperatures and voltage settings. This database does hold also other conditions data such as calibration constants and details on the experimental setup.

In 2006, the conversion step was realized by a server program running on a regular Linux PC. For 2007 and beyond, it is envisaged to base this conversion entirely on the grid which allow for a parallel conversion of several runs and therefore for a faster availability of the data. The converted *raw* data are available ~ 1 hour after the end of a given run. These files can be processed using regular ILC Software tools.

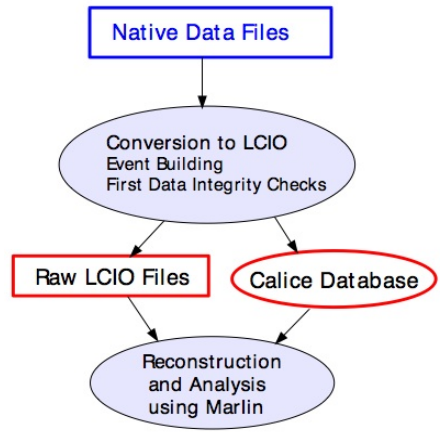

**Figure 10-2: Schematic view of the processing chain of CALICE data.**

In general, the raw data are processed by a reconstruction program of which the main output is calibrated calorimeter hits and, since recently, tracks as measured by tracking devices which are part of the test beam setup. At this stage the data can be directly compared to the output of detector simulations. More on simulations see below.

## 10.4 Software Details

The CALICE software is mainly divided into three packages. The conversion package, the reconstruction package and a user package which holds the interface classes to the raw data plus other utility classes. All these packages are written in the C++ language. The LCIO[1] format , as being standard in ILC detector studies, is used to transport the data between the different steps of the data processing. Where possible, object definitions as given by the LCIO standard are used. An example for this is the *CalorimeterHit* data model. For most of the data types delivered by the CALICE DAQ , however, no such a model is available. Here interface classes are built around *LCGenericObjects* which allows one one hand for an arbitrary, user defined definition of data types but on the other hand have been shown to lead to a penalty in terms of performance. Within the packages the dataflow is realized in form of MARLIN processors. MARLIN is the software framework which is widely in use within the LDC detector studies. The application of LCIO and MARLIN has two important consequences. Within LDC, the latter allows users to perform analyses for CALICE and general detector studies without changing the software framework. The former will allow in addition for the analysis of CALICE data within other frameworks such as in use for ALCPG[2] and GLD[3] studies.

The access to conditions data is realized by the LCCD package. It permits to store the conditions data in different backends and access with the same interface classes. One of these backends is a mysql database. In this case the LCCD package is itself interfaced to the CondDBMySQL package which allows for a structured management of the conditions data. A layer and tagging mechanism assures the reproducibility of results as obtained by a given combination of conditions data. It has to be pointed out that the handling of conditions data is only a first attempt to establish such a software tool which is and will be of vital importance for any running experiment.

Monte Carlo simulations are realized using the MOKKA software package. It provides the geometry interface to the GEANT4 simulation toolkit using a MySQL database. MOKKA is the standard simulation program within the LDC detector study. The various test beam setups are fully implemented into MOKKA. The pure simulation step which creates LCIO files as output is followed by a digitization step which is realized as a MARLIN processor which runs at the beginning of the reconstruction program.

## 10.5 Summary and Outlook on software for CALICE

CALICE has established a full chain of data processing based in ILC Software and grid tools. This way of organizing the software has lead to a wide spread analysis effort not restricted to the main detector experts. The strategy of using ILC Software tools will allow for a easy exchange between test beam results and results or algorithms obtained in general detector studies. This is in particular true for clustering algorithms which will be for the first time tested on data obtained with prototypes of ILC detectors. The CALICE collaboration will pursue this strategy throughout its running in 2007 and 2008.

The experience obtained during the CALICE data taking has clearly revealed the need for a dedicated treatment of data types more related to hardware issues. For the next generation test beams the interface between the DAQ systems and the offline processing has to be better defined. First efforts into this direction are already undergoing.

As indicated above Conditions Data handling is a vital part of every experiment and the way how it is done in CALICE is clearly not optimal. The ILC community is herewith asked to allocate human and financial resources for the development of a well suited system for Conditions Data management.

In total the processing and analysis of the CALICE test beam data delivers important input to the development of the ILC software and computing environment with respect to technical aspects as well as for the development of Particle Flow Algorithms.

## 10.6 Bibliography

1. For LCIO and other software packages see: http://ilcsoft.desy.de
2. N. Graf : *"Simulation and Reconstruction – ALCPG Framework and Toolkit"*. Talk given at the ILC Software Workshop, Orsay (France) May 2007
   http://events.lal.in2p3.fr/conferences/ILCSoftware/ ,
3. A. Miyamoto: *"Software Tools in GLD study"*. Talk given at the ILC Software Workshop , Orsay (France) May 2007.
   http://events.lal.in2p3.fr/conferences/ILCSoftware/ ,

# 11 The test beam effort

## 11.1 Overview

In the four months between July and November 2006 the CALICE collaboration has successfully commissioned and operated, in the H6B experimental area at the CERN SPS, a full calorimeter chain of prototype detectors: ECAL, AHCAL and TCMT (Figure 11-1). The SPS test facility offers the possibility to cover a large portion of the relevant phase-space for ILC studies: positive and negative beam polarities, covering the range 6-120 GeV for electrons and hadrons.

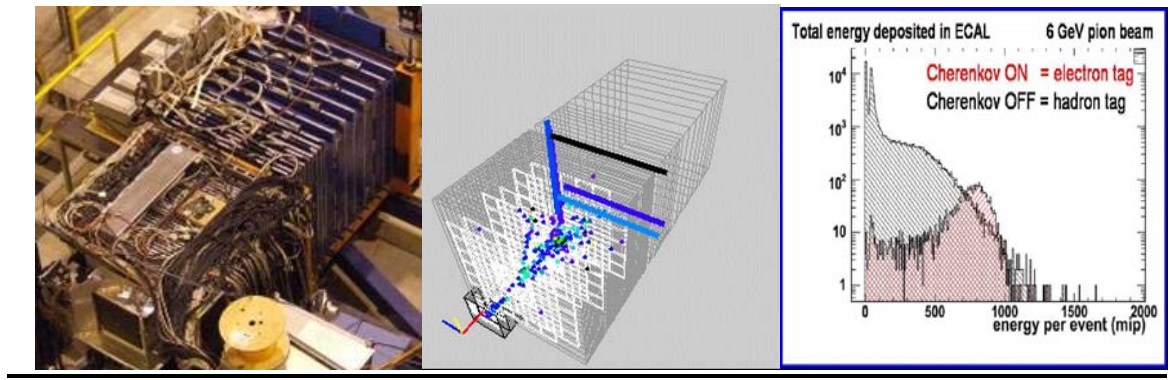

**Figure 11-1: View of the installation area at the CERN test beam (left). A pion event in the on-line display (center). Energy spectrum in the ECAL for 6 GeV pion with and without the Cherenkov bit selected.**

## 11.2 Beamline installation

The CALICE installation at CERN included locally provided beam instrumentation detectors and a custom made trigger system. The trigger to the experiment is provided by the coincidence of two scintillator plates with photo-multiplier readout. Two sizes of plates of have been provided as beam triggers, 3x3cm² and 10x10cm². In addition a 100x100cm² coincidences has been equipped as a muon rejection wall downstream the detector and as muon trigger during calibration. The analog readout of an additional 20x20cm²scintillator plate serves as veto for events with double particles or showers iniciated in the material upstream the detector. All trigger are digitized and recorded event by event by the VME-based data acquisition (DAQ), and can be used offline for data selection.

A threshold Cherenkov counter has been used to discriminate electrons and pions in the range 6-20 GeV. The helium pressure in the 11~m long Cherenkov vessel needs to be adjusted depending on the beam energy. With optimal settings, efficiencies of 90% are obtained, going to 30 % with increasing energy. The discriminated Cherenkov signal is recorded as a trigger bit.

For particle tracking three sets of delay wire-chambers provided by CERN have been included in the CALICE DAQ. Three pairs of x and y planes with two wires each are readout at the event basis by a TDC implemented in the DAQ. The spatial resolution of the tracking system is better than 200~μm.

## 11.3 Calorimeters

The three calorimeter prototypes have been commissioned on the beam-line. The ECAL was equipped with 30 sensitive layers of silicon pads, but each layer was missing a third of the readout channels, thus reducing the active area to 18x12~cm². The total readout channels have been 6480 out of the 9720 foreseen. In the first period of the test-beam only 15 out of 38 AHCAL active layers were available. They have been distributed in the steel stack support to cover a total of about 3.5λ with half of the designed longitudinal segmentation. In a later period a total of 23 modules (4698 readout channels) became available and the installation was rearranged to cover the first 18 layers with fine segmentation and the remaining depth up to 3.5λ with coarser one. The fully equipped prototype with 38 layers will soon cover up to 4.5λ. The TCMT was completely installed with all 16 active layers equipped and a total of 320 readout channels.

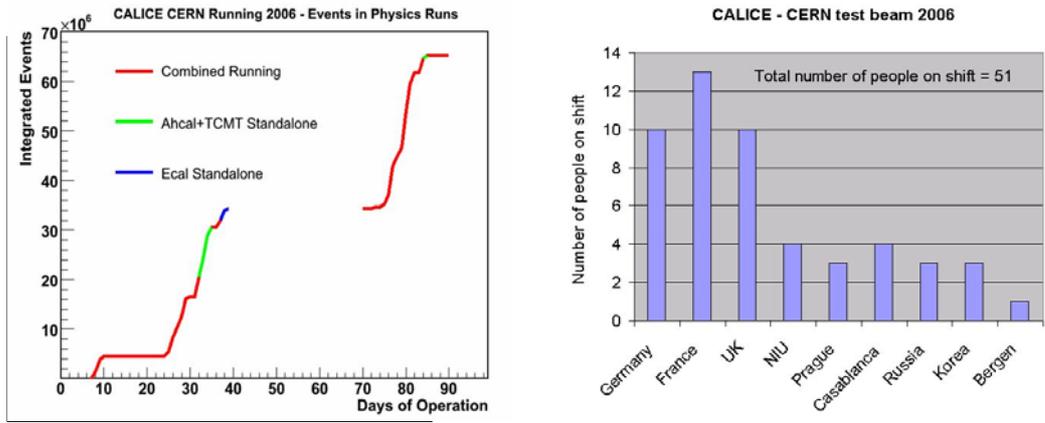

**Figure 11-2: Integrated beam events in the CALICE calorimeter system during the period Jul.-Nov. 2006 (left). Shift sharing among the countries member of CALICE (right).**

The system, with more than 14000 channels and an acquisition rate capability of 120 Hz is a compact HEP experiment in itself. During the data taking period the CALICE detectors had more than 90% up-time and the beam duty cycle was estimated to be about 60%. The collaboration has collected more than 65 millions events (see Figure 11-2), completing the muon calibration of all components, the electromagnetic programs of both ECAL and AHCAL and the first part of the hadronic program for the combined detector at zero degree incident angle of the beam.

The performance of all beam-line detectors as well as that of the 3 calorimeter prototypes has been monitored online during data taking. A special fast analysis tool has been developed to access in real time the relevant beam and detector qualities. This incredible success for the collaboration was only possible thanks to the combined effort of all the institute members. More than 150 physicists have shared shifts at CERN. Experts in place and on-call have been permanently available during the four months of commissioning and running.

In summer 2007, the test beam program will be for the last time at CERN. All detectors will have completed their instrumentation for maximum lateral and longitudinal shower containment. ECAL and AHCAL will be mounted on a motor-controlled movable stage (Figure 11-3) which allows the variations of the beam incidence angle also for the AHCAL structure.

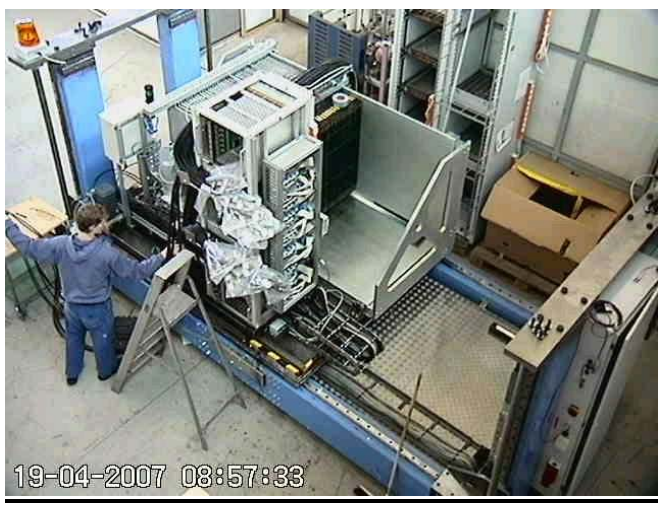

**Figure 11-3: ECAL and AHCAL CALICE prototypes mounted on the motor-controlled movable stage at DESY before shipment to CERN for the summer 2007 test-beam.**

# 12 First preliminary test beam results

## 12.1 Introduction

In this section we provide a very brief outline of the analysis performed on the data recorded in 2006. More detailed results will be presented at LCWS07.

The raw data recorded at the test beams were first converted into LCIO format. These form the basis for all subsequent analyses. After calibration, a reconstruction pass is performed, which provides calibrated hits in individual calorimeter cells for further analysis and comparison with simulation. The validation of Monte Carlo codes such as GEANT4 is one of the main long-term analysis objectives of the CALICE program, in order to establish the reliability of the simulations used for global detector design work

## 12.2 The ECAL data

### 12.2.1 Calibration

During operation of the calorimeters at CERN, a substantial exposure of the detectors to a broad muon beam was carried out, and this provided the main calibration sample used to determine the gains of each channel. Pedestal data (blocks of 1000 events) were also recorded regularly interspersed amongst the beam triggers, so as to monitor pedestals and noise. Pedestals were then subtracted using the locally determined values, and the calibration constants determined from the muon sample were applied to convert ADC counts into MIPS.

The calibration procedure was finally able to calibrate 6471 out of 6480 pads in the ECAL, i.e. just 9 cells were deemed to be dead. The distribution of calibration constants and noise values for all the calibrated cells are shown in Figure 12-1. We see that the gains are almost all in the range 40-50 ADC counts/MIP, with a spread of less than 5%. The small peak at 24 represents a single wafer which was not fully depleted. The average noise is about 0.15 MIP, i.e. signal/noise of about 6.5, with a narrow spread.

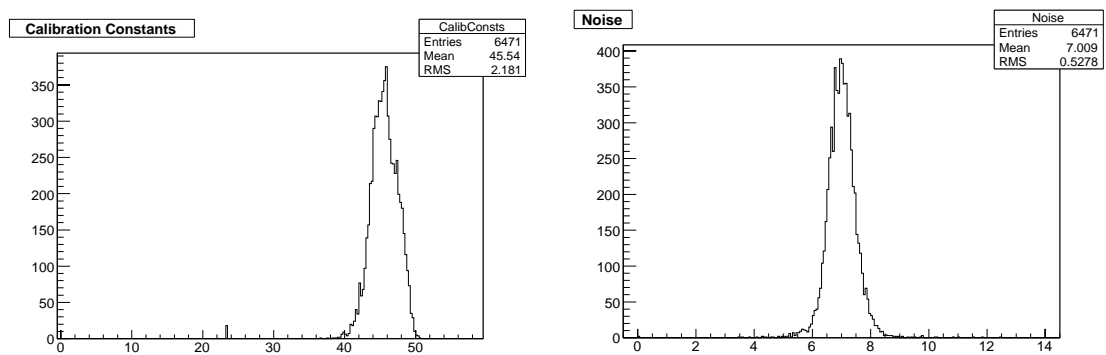

**Figure 12-1: Distributions of calibration constants (ADC counts/MIP) (left) and of r.m.s. noise (right) for all calibrated cells.**

### 12.2.2 Response to electron beams

The prototype ECAL was exposed to electron beams with energies from 1-6 GeV at DESY, and from 6-45 GeV at CERN during 2006. Data were recorded with a variety of beam impact positions, and angles ranging from 0 to 45º. Topics to be studied include the response of the ECAL and its linearity, the energy resolution, their dependence on impact position (because of gaps between wafers) and angle, the longitudinal and transverse profiles of showers, and the position and angular resolution of the calorimeter. All of these should be compared with the predictions of Monte Carlo. Only a handful of results will be shown here as an illustration of the ongoing work.

In Figure 12-2 we show the distribution of energies of cells (with a threshold cut of 0.6 MIP) for two samples: non-interacting 12 GeV pions and 30 GeV electrons, compared with Monte Carlo. The pion sample shows that the MIP peak is well modelled, with a single calibration factor to match Monte Carlo to data. In the electron sample, the tail of high hit energies, up to 350 MIP, in the core of the shower is very well modelled. However, there are discrepancies in the MIP peak region which are still being investigated, but which have little effect on the overall energy response.

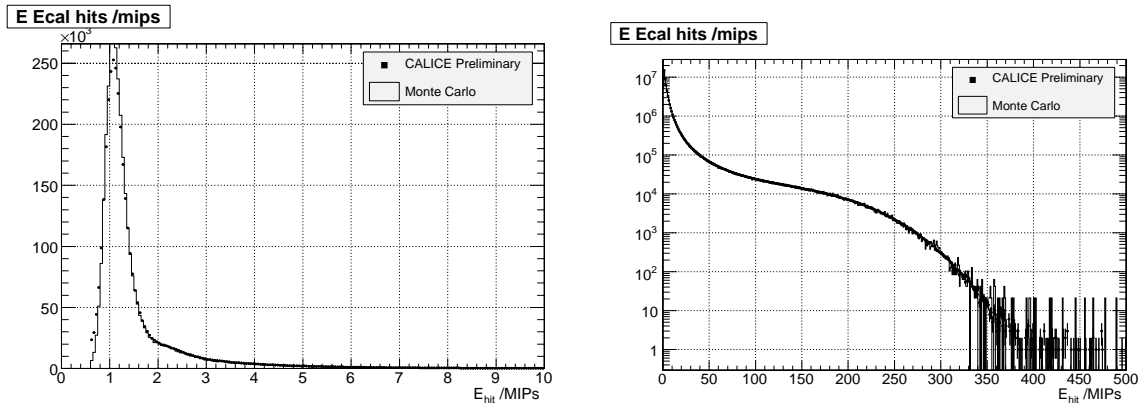

**Figure 12-2: Distributions of hit energies (in MIPs) for 12 GeV pions (left) and for 30 GeV electrons (right). Data and Monte Carlo are compared.**

The total energy in the calorimeter is calculated by summing the energies of all hits, weighting the three detector stacks to account for their sampling fractions, and dividing by an appropriate normalization factor. In Figure 12-3 we show an example of the response of the detector to 30 GeV electrons impinging close to the centre of a silicon wafer at normal incidence. The response is seen to have an approximately Gaussian form. In order to study the average energy response, its linearity and the resolution of the detector, a Gaussian fit is performed, as indicated in Figure 12-3. As an illustration, the fractional energy resolution is also shown in Figure 12-3 as a function of $1/\sqrt{E}$, over the energy range 1-45 GeV. Monte Carlo predictions are also shown, and describe the data well.

These results refer to electrons incident close to the centre of wafers. For electrons impinging close to the inter-wafer gaps, a loss of energy, of up to 19% in the worst case, is seen. Several corrections schemes are under study; these are able to make the response across the calorimeter satisfactorily uniform, though with some degradation of the energy resolution, of order 10% at present.

Studies are also ongoing of the energy profiles within showers. In Figure 12-4 we show an example of the longitudinal shower profiles seen in the energy range 1.5 to 30 GeV. The expected logarithmic growth in the shower energy maximum is observed. In Fig 4 we also show estimates of the shower radius (the radius within which 90% or 95% of the shower energy is confined) using data from 1-6 GeV. This quantity is important in order to achieve good spatial separation of showers in the ECAL. Over this energy range, no significant energy dependence is observed.

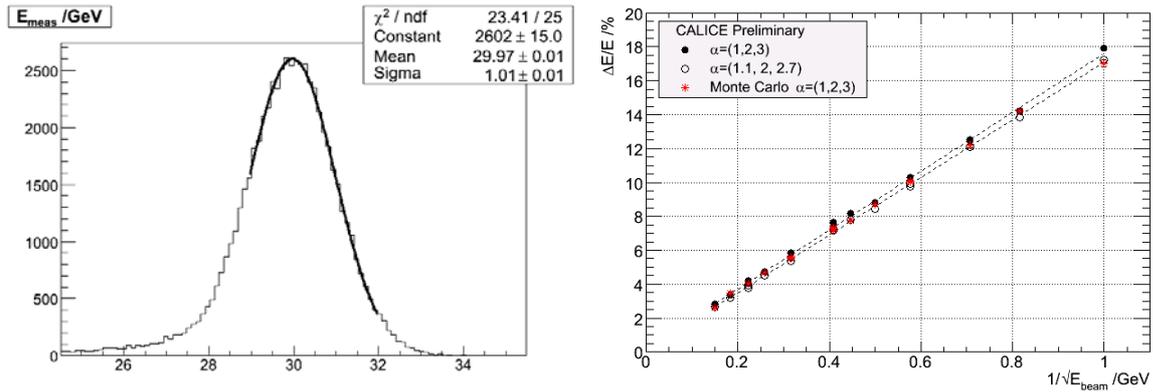

**Figure 12-3: Distributions of total ECAL energy for 30 GeV electrons (left), with a Gaussian fit superimposed; resolution of the ECAL for electrons (right), with data and Monte Carlo compared.**

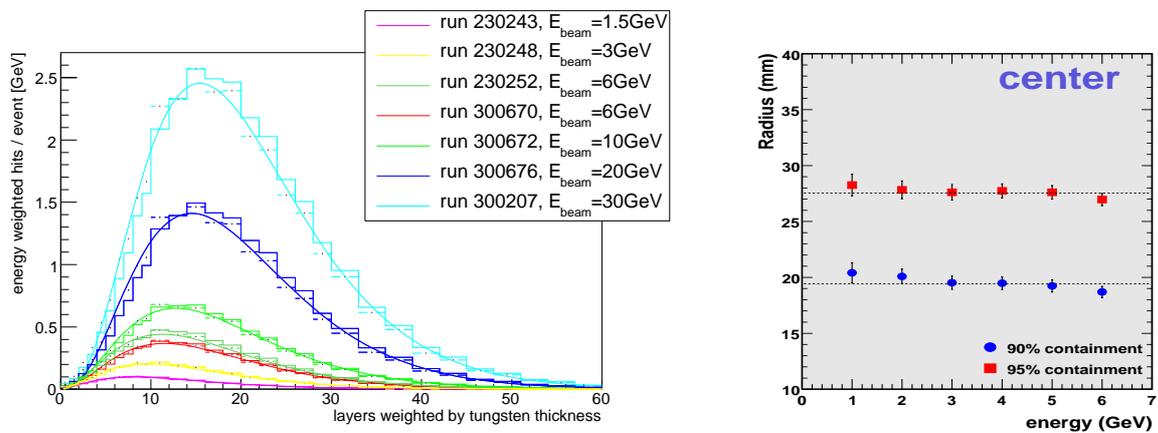

**Figure 12-4 Left: Longitudinal shower profile from runs at CERN and at DESY with several energies measured in the CALICE ECAL; the data are represented by points with statistical uncertainties and the GEANT4 Monte Carlo simulation by the histogram. Parametrisations of the shower profiles by the form $c\, t^{\alpha}\, e^{(-\beta t)}$, where $t$ is the calorimeter depth in radiation lengths, are also shown. Right: Radius for containment of 90% or 95% of the energy of electron showers from 1-6 GeV.**

## 12.3 The AHCAL data

The analogue HCAL (AHCAL) was tested at CERN in 2006 with 15/23 of its intended 38 layers of scintillating tiles during the first/second period of data taking. A calorimeter with roughly 3.8 interaction lengths was exposed to the beam by placing active layers in alternating slots of the steel structure. Seven additional modules have been used to further complete the instrumentation at the beginning of the absorber stack for the second period without increasing the overall depth. Data were taken both with and without the ECAL ($\approx 1\lambda_0$) in front. Many hadronic showers should therefore be contained, albeit with a lower sampling fraction than will be achieved ultimately. The AHCAL analysis strategy is to use muon and electron data, taken without the ECAL, to understand the calorimeter calibration and performance and to tune the simulation if necessary. The hadron data taken with and without the ECAL can then be studied with greater confidence to extract the performance and shower properties. These data are of great interest in comparing with simulations, for which the underlying interaction model is much less well understood than for electromagnetic processes.

The calibration procedure for the AHCAL is more complex than for the ECAL. As well as using muon beams to establish the MIP calibration tile-by-tile, the SiPM photosensors exhibit a significantly non-linear response, and careful calibration of this effect is needed, based both on lab measurements and a light injection system. It is also important to include these effects into the Monte Carlo simulations.

### 12.3.1 Response to electrons

Electron beam data were recorded with the AHCAL in the absence of the ECAL, and these provide an important first check of the robustness of the calibration and reconstruction procedures, since electromagnetic showers yield high particle densities in their cores. Hits in different layers are weighted according to the sampling fractions, as for the ECAL. The summed energy is then converted from MIP to GeV, and a Gaussian fit is used to extract the mean response and energy resolution. Figure 12-5 presents an example of the present state of the analysis, showing the linearity of the AHCAL response as a function of electron energy. Some small residual non-linearity is still seen in the data above 30 GeV, amounting to 5% at 45 GeV. This effect is the subject of active investigation.

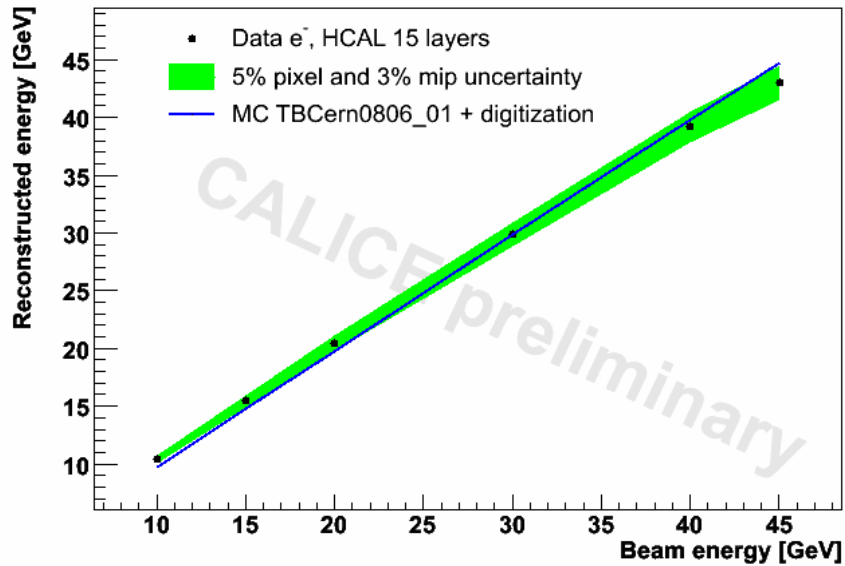

**Figure 12-5: Reconstructed energy plotted against beam energy for electron beams in the Calice AHCAL.**

Studies of electromagnetic shower shape are also ongoing. An example is presented in Figure 12-6, in which we compare the longitudinal distribution of energy in electron showers between data and Monte Carlo, at 10 and 45 GeV. The expected logarithmic growth of shower depth is observed, and the data may be well fitted by the form $at^b\exp(-ct)$ where $t$ is the depth in the calorimeter in units of radiation lengths. We note that the tails of the showers are very well modelled.

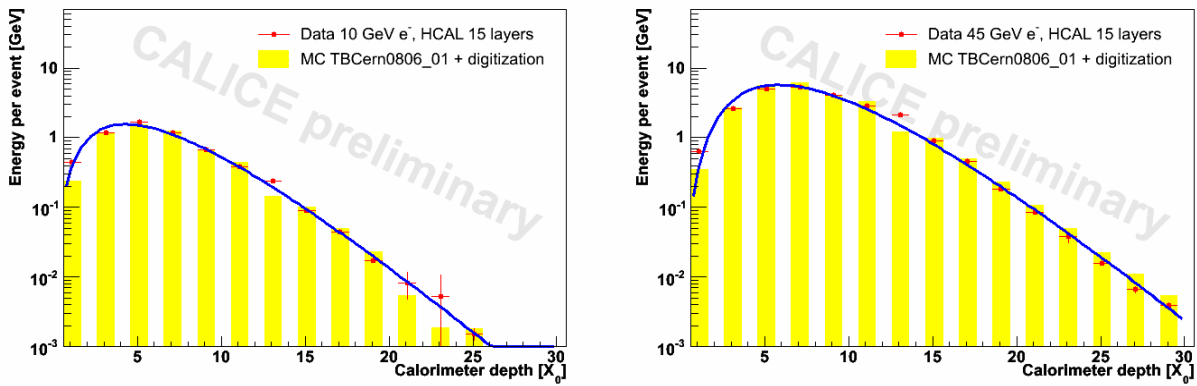

**Figure 12-6: The longitudinal shower profile observed for electron beams in the AHCAL, at 10 GeV (left) and 45 GeV (right). Data and Monte Carlo are compared.**

### 12.3.2 Response to hadrons

The hadron data are in some respects less critically dependent on the calibration procedure, because the hit energies in the core of the showers are typically less than in electromagnetic showers of the same energy. For example, in Figure 12-7 we show the hit energies for several pion energies. Even at 20 GeV, only 3% of the energy is in the form of hits with energy > 15 MIPs, at which point the non-linearity correction is only 10%. The data were collected in combined runs with the ECAL and TCMT, though for the results shown here, events were selected in which no showering occurred in the ECAL, and (optionally) in which there was no leakage into the TCMT. For these samples, in Figure 12-7 we also show the total reconstructed AHCAL energy for pions from 6 to 20 GeV as a function of beam energy. The data show good linear behaviour. The data are compared with two (amongst many possible) Monte Carlo models, indicating that significant differences in predictions may be encountered. The resolution is also found to be in the range of Monte Carlo expectation, but not yet indicative of the final performance, mainly because the instrumentation was still incomplete.

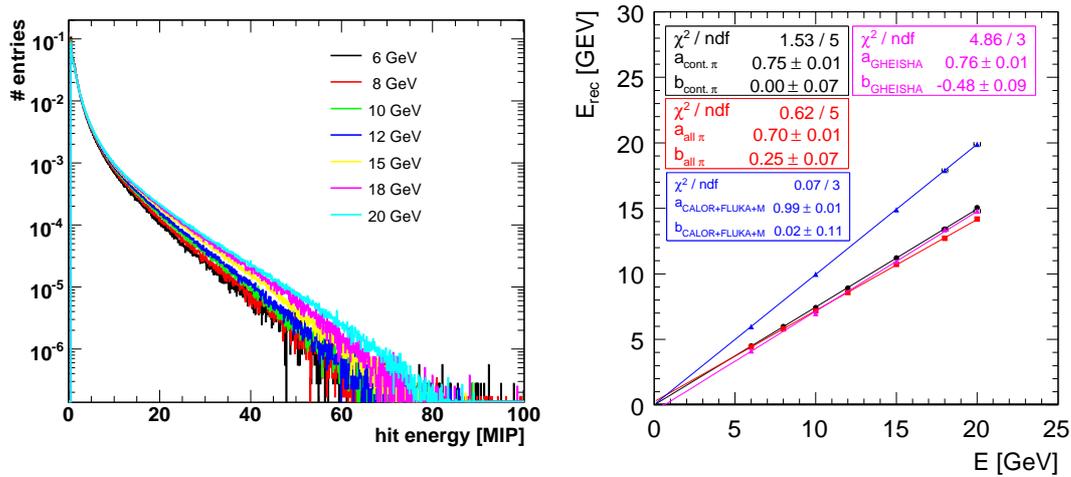

**Figure 12-7: The left hand plot shows the distribution of hit energies in pion-induced showers of various energies. The right hand plot indicates the linearity of the AHCAL response to hadronic beams.**

For the hadronic energy, as well as looking at conventional measures such as shower shape and energy resolution, we intend to pursue a "deep analysis" which explores the shower substructure in some detail, for example dividing the energy deposited into electromagnetic-like, MIP-like, neutron-like components. Understanding of these features will be vital in order to validate the simulations upon which the particle flow algorithms for jet energy reconstruction have been developed.

### 12.4 Combined calorimeter data

The tail catcher (TCMT) placed behind the HCAL is important for detecting leakage of hadronic showers. It is particularly interesting to study correlations between the energy recorded in the ECAL, HCAL and TCMT, because combination of information from all three systems will be needed in order to obtain the best hadronic energy resolution. A preliminary version of this analysis is already under way in parallel with the ECAL and AHCAL analyses. Encouraging results are being seen. For example, in Figure 12-8 we show some results of a study of the correlations between the calorimeter systems, for a 20 GeV pion beam. A more or less linear anticorrelation between the ECAL and AHCAL energies is seen, which can be used as the basis of a simple linear combination. A similar anticorrelation is seen between the TCMT energy and the combined ECAL/AHCAL energy, and the further improvement in the hadron energy resolution on including the tail catcher information is apparent. The peaks at low energies are caused by muons or non-interacting pions. Studies of alternative methods of energy estimating (such as hit counting in a semi-digital mode), and comparisons of all these with Monte Carlo are ongoing.

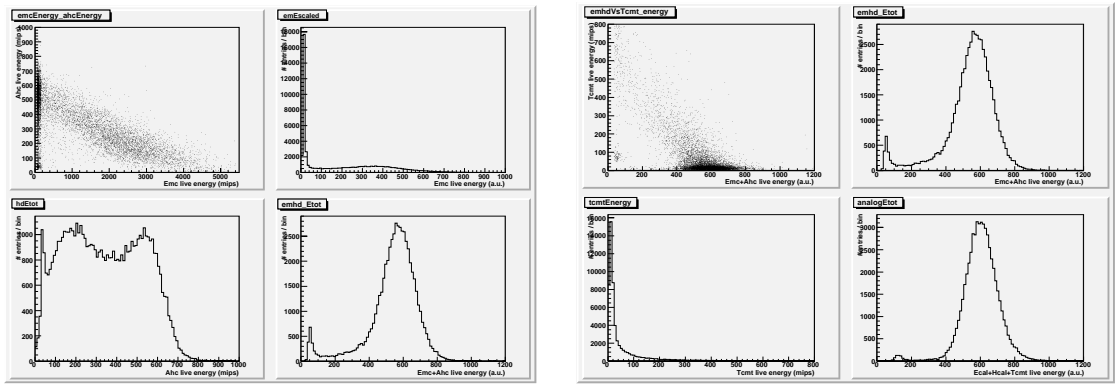

**Figure 12-8: In the left hand set of plots, the upper left shows the anti-correlation between ECAL and AHCAL energies, upper right and lower left the two projections, and lower right the combined energy using the observed correlation slope. The right hand set of plots show the corresponding anti-correlation between the TCMT and the combined ECAL+AHCAL energy.**

# CALICE collaboration

C.Adloff, F.Cadoux, P.Delebecque, R.Hermel, Y.Karyotakis, J.Prast
*Laboratoire d'Annecy de Physique des Particules- Annecy le Vieux*

S.Chekanov, T.Cundiff, G.Drake, B.Haberichter, V.Guarino, A.Kreps, E.May, J.Repond, D.Underwood, B.Wicklund, K.Wood, L.Xia
*Argonne National Laboratory*

A.Brandt, H. Brown, K.De, C. Medina, J. Smith, J.Li, M.Sosebee, A.White and J.Yu
*University of Texas at Arlington*

C.M.Hawkes, Y.Mikami, O.Miller, V. Rajovic, R.J.Staley, N.K.Watson, J.A.Wilson
*School of Physics and Astronomy, University of Birmingham*

J.Butler, E.Hazen, S.Wu
*Boston University*

M.Goodrick, B.Hommels, G.Mavromanolakis, M.A.Thomson, D.R.Ward, W.Yan
*Cavendish Laboratory, Cambridge University*

M.Oreglia
*University of Chicago*

F.Badaud, M.Benyamna, G.Bohner, R.Bonnefoy, D.Boumediene, N.Brun, C.Carloganu, F.Chandez, R.Cornat, M.Crouau, P.Gay, Ph.Gris, J.Lecoq, S.Mannen, F.Morisseau, L.Royer
*Laboratoire de Physique Corpusculaire – Clermont*

G. Blazey, D. Chakraborty, A. Dyshkant, K. Francis, G. Lima, J. Powell, V. Rykalin, M. Smith, V. Zutshi
*NICADD - North Illinois University – De Kalb*

J.P.Crooks ,M.Stanitzki, K.D.Stefanov, R.Turchetta, M.Tyndel , E.G.Villani
*Rutherford Appleton Laboratory  - Didcot*

V.Astakhov, S.Golovatyuk, I.Golutvin, A.Malakhov, I.Tyapkin, Y.Zanevski, A.Zintchenko , S.Bazylev, N.Gorbunov, S.Slepnev
*Joint Institute for Nuclear Research - Dubna*

D.Dzahini, J.-Y. Hostachy, L. Morin
*Laboratoire de Physique Subatomique et de Cosmologie - CNRS/IN2P3 - UJF - INPG - Grenoble*

N. D'Ascenzo, G.Eigen, E.Garutti, V.Korbel, B. Lutz, N. Meyer, V. Morgunov, S. Schätzel, S. Schmidt, F. Sefkow, A. Vargas, N. Wattimena, O. Wendt
*DESY - Hamburg*

M. Groll, R.-D. Heuer
*Hamburg University*

E.Norbeck , Y.Onel
*University of Iowa*

G.Kim, D-W. Kim, K.Lee, S.Lee
*Kangnung National University - HEP/PD - Kangnung*

D.Jeans, K.Kawagoe
*Kobe University - Kobe*

J.A.Ballin, P.D.Dauncey, A.-M. Magnan, M.Noy, H.Yilmaz, O.Zorba
*Department of Physics, Imperial College London*

V.Bartsch, M.Lancaster, M.Postranecky, M.Warren, M.Wing
*Department of Physics and Astronomy, University College London*

M.Faucci Giannelli, B.J.Green, M.G.Green, A.Misiejuk, P.-F.Salvatore, T.Wu
*Physics Department, Royal Holloway University of London*

M. Bedjidian, C.Combaret, J. Fay, R.Gaglione, G.Grenier, I.Laktineh, P.Lebrun,
J-P. Martin, H. Mathez
*Institut de Physique Nucléaire de Lyon*

E. Cortina-Gil, M.C Fouz-Iglesias
*CIEMAT, Centro de Investigaciones Energeticas, Medioambientales y Tecnologicas, Madrid*

D.Bailey, R.J.Barlow, A. Elvin, J.Freestone, R.Hughes-Jones, M.Kelly, S.Kolya, M. Perry, S.Snow,
R.J.Thompson
*The Department of Physics and Astronomy, The University of Manchester*

T.Takeshita, S.Uozumi
*Shinshu University, Matsumoto*

N.Shumeiko, A.Litomin, P.Starovoitov, V.Rumiantsev, O.Dvornikov, V.Tchekhovsky, A.Solin,
A.Tikhonov
*Joint Institute for Nuclear Research - Minsk*

F. Corriveau
*Department of Physics -Mc Gill University - Montréal*

V.Balagura, B.Bobchenko, M.Danilov, R.Mizuk, E.Novikov, V.Rusinov, E.Tarkovsky
*Institute of Theoretical and Experimental Physics - Moscow*

V. Andreev, N. Kirikova, V. Kozlov, P. Smirnov, Y. Soloviev, A. Terkulov
*Lebedev Physics Institute - Moscow*

P.Buzhan, B.Dolgoshein, A.Ilyin, V.Kantserov, V.Kaplin, A.Karakash,
E.Popova, S.Smirnov
*Moscow Engineering and Physics Institute*

P.Ermolov, D.Karmanov, M.Merkin, A.Savin, A.Voronin, V.Volkov
*Moscow State University*

B.Bouquet, S.Callier, F.Dulucq, J.Fleury, H.li, G.Martin-Chassard, F.Richard,
Ch. de la Taille, R. Poeschl, L.Raux, M.Ruan, N.Seguin-Moreau, F.Wicek, Z.Zhang
*Laboratoire de l'Accélérateur Linéaire - Orsay*

M.Anduze, V. Boudry, J-C.Brient, C.Clerc, C.Jauffret, A.Karar, P. Mora de Freitas, G. Musat,
M.Reinhard, A.Rouge, A.L.Sanchez, J-Ch. Vanel, H. Videau
*LLR - Ecole Polytechnique - Palaiseau*

Y.Bonnassieux, P.Roca
*Physique des Interfaces et Couches Minces - Ecole Polytechnique - Palaiseau*

J.Zacek
*Charles University - Prague*

J.Cvach, M.Janata, M.Havranek, M.Marcisovsky, P.Mikes, I.Polak, J.Popule, L.Tomasek, M.Tomasek, P.Sicho, V.Vrba, J.Zalesak
*Institute of Physics, Academy of Sciences of the Czech Republic - Prague*

V. Ammosov, Yu.Arestov, B.Chuiko, V.Gapienko, V.Lishin, A.Semak, Yu.Sviridov, M.Ukhanov, V.Zaets
*Institute of High Energy Physics - Protvino*

M.Barbi, G.J.Lolos, Z.Papandreou
*Department of physics, University of Regina - Regina*

S.W. Nam, I.H.Park, J.Yang
*Ewha Womans University - Seoul*

J.Kang , Y.Kwon
*Yonsei University - Seoul*

Ilgoo Kim, Taeyun Lee, Jaehong Park, Jinho Sung
*School of Electric Engineering and Computing Science, Seoul National University*

I.Yu
*Sungkyunkwan University, Suwon*